\documentclass[acmsmall,nonacm]{acmart}
\usepackage{xcolor}
\usepackage{enumitem}
\usepackage{booktabs}
\usepackage{tabularx}
\usepackage{array}
\usepackage{xltabular}
\usepackage{multirow}
\usepackage{graphicx}
\definecolor{scopeProject}{HTML}{4E7BAE}   \definecolor{scopeSession}{HTML}{6E8253}   \definecolor{scopePersonal}{HTML}{C8643C}  \usepackage{tikz}
\usepackage[most]{tcolorbox}
\definecolor{aptFrame}{HTML}{E4A8A7}\definecolor{aptFill}{HTML}{FAEFEF}
\definecolor{carFrame}{HTML}{FFC9AA}\definecolor{carFill}{HTML}{FFEEE1}
\newtcolorbox{taskbox}[3]{enhanced,colframe=#1,colback=#2,colbacktitle=#1,coltitle=black,boxrule=0.6pt,arc=4pt,left=7pt,right=7pt,top=6pt,bottom=6pt,toptitle=3pt,bottomtitle=3pt,fonttitle=\bfseries\small\sffamily,fontupper=\small\ttfamily,before upper={\raggedright},title={#3}}
\definecolor{chipPeText}{HTML}{B0512F}\definecolor{chipPeFill}{HTML}{FBF0EB}\definecolor{chipPeLine}{HTML}{EECEBF}
\definecolor{chipPText}{HTML}{3B6D9E}\definecolor{chipPFill}{HTML}{EEF4FA}\definecolor{chipPLine}{HTML}{CFE0EF}
\definecolor{chipSText}{HTML}{586B3F}\definecolor{chipSFill}{HTML}{F1F4EB}\definecolor{chipSLine}{HTML}{D8E0C9}
\DeclareRobustCommand{\memchip}[4]{\tikz[baseline=(chip.base)]{\node[draw=#3,fill=#2,text=#1,rounded corners=2pt,inner xsep=2.2pt,inner ysep=1pt,outer sep=0pt,line width=0.4pt,font=\footnotesize\ttfamily](chip){M-#4};}}
\newcommand{\tabgroup}[2]{\multicolumn{#1}{@{}l}{\textit{#2}}\\\cmidrule{1-#1}}
\newcommand{\memP}[1]{\memchip{chipPText}{chipPFill}{chipPLine}{#1}}   \newcommand{\memS}[1]{\memchip{chipSText}{chipSFill}{chipSLine}{#1}}   \newcommand{\memPe}[1]{\memchip{chipPeText}{chipPeFill}{chipPeLine}{#1}} \newcommand{\lucideicon}[1]{\tikz[x=0.042em,y=-0.042em,line width=0.07em,line cap=round,line join=round,baseline={([yshift=-0.5ex]current bounding box.center)}]{\useasboundingbox (0,0) rectangle (24,24); #1}}
\DeclareRobustCommand{\iconMonitor}{\lucideicon{\draw (2,12) .. controls (5.57,2.94) and (18.43,2.94) .. (22,12) .. controls (18.43,21.06) and (5.57,21.06) .. (2,12); \draw (12,12) circle (3);}}
\DeclareRobustCommand{\iconControl}{\lucideicon{\draw (14,4.1) -- (12,6); \draw (5.1,8) -- (2.2,7.2); \draw (6,12) -- (4.1,14); \draw (7.2,2.2) -- (8,5.1); \draw[rounded corners=0.6] (9.04,9.69) -- (20.5,14.4) -- (16.2,15.5) -- (15.1,19.9) -- cycle;}}
\definecolor{circmarkBg}{HTML}{F25A7A}
\DeclareRobustCommand{\circmark}[1]{\tikz[baseline=(c.base)]{\node[circle,fill=circmarkBg,text=white,inner sep=1.2pt,font=\footnotesize\bfseries\sffamily,minimum size=1.1em](c){#1};}}
\DeclareRobustCommand{\stepmark}[1]{\tikz[baseline=(c.base)]{\node[circle,fill=black,text=white,inner sep=1.2pt,font=\footnotesize\bfseries\sffamily,minimum size=1.1em](c){#1};}}
\newcolumntype{L}{>{\raggedright\arraybackslash}X}
\AtBeginDocument{}

\begin{document}

\title{MemoSync: Co-Managing Developer-Agent Metamemory Across Contexts}

\author{Baiqiao Zhang}
\email{bzhangcw@cse.ust.hk}
\affiliation{\institution{The Hong Kong University of Science and Technology}\city{Hong Kong SAR}\country{China}}
\author{Junze Li}
\email{junze.li@connect.ust.hk}
\affiliation{\institution{The Hong Kong University of Science and Technology}\city{Hong Kong SAR}\country{China}}
\author{Chunyang Luo}
\email{lcy007@stu.pku.edu.cn}
\affiliation{\institution{Peking University}\city{Beijing}\country{China}}
\author{Yuying Tang}
\email{yuying.tang@connect.ust.hk}
\affiliation{\institution{The Hong Kong University of Science and Technology}\city{Hong Kong SAR}\country{China}}
\author{Wenshuo Zhang}
\email{wzhangeb@connect.ust.hk}
\affiliation{\institution{The Hong Kong University of Science and Technology}\city{Hong Kong SAR}\country{China}}
\author{Shiyuan Piao}
\email{shiyuan.piao@connect.ust.hk}
\affiliation{\institution{The Hong Kong University of Science and Technology}\city{Hong Kong SAR}\country{China}}
\author{Dingdong Liu}
\email{dliuak@connect.ust.hk}
\affiliation{\institution{The Hong Kong University of Science and Technology}\city{Hong Kong SAR}\country{China}}
\author{Shuai Ma}
\email{mashuai@iscas.ac.cn}
\affiliation{\institution{Institute of Software, Chinese Academy of Sciences}\city{Beijing}\country{China}}
\author{Xiaojuan Ma}
\authornote{Corresponding author.}
\email{mxj@cse.ust.hk}
\affiliation{\institution{The Hong Kong University of Science and Technology}\city{Hong Kong SAR}\country{China}}

\renewcommand{\shortauthors}{Zhang et al.}

\begin{abstract}
AI coding agents increasingly manage what context carries across sessions and projects, but developers have limited visibility into what these agents store, retrieve, and apply. This opacity creates a metamemory problem: the developer's meta-level judgments about what should be remembered are separated from the agent's object-level memory operations, disrupting the monitoring and control that keep the two levels aligned. We conducted a formative study with 12 developers and identified what they want to monitor and control across four dimensions of agent memory use. Based on these findings, we designed MemoSync, a co-management system that treats agent memory as a shared, operable representation: the agent proposes memory candidates and reports which items it uses, and the developer reviews, edits, and scopes them. In a between-subjects study ($N=132$) against two baselines, MemoSync participants achieved significantly higher task performance and the highest mean developer-agent memory alignment. In an exploratory mediation analysis, the alignment advantage was primarily driven by improved monitoring accuracy. Despite performing substantially more monitoring and control actions, participants reported no increase in perceived workload. A week-long deployment with nine developers showed that memory monitoring and control fit into everyday development work. MemoSync is open-sourced at \url{https://github.com/SLEEPYBQ/MemoSync}.
\end{abstract}


\begin{teaserfigure}
  \includegraphics[width=\textwidth]{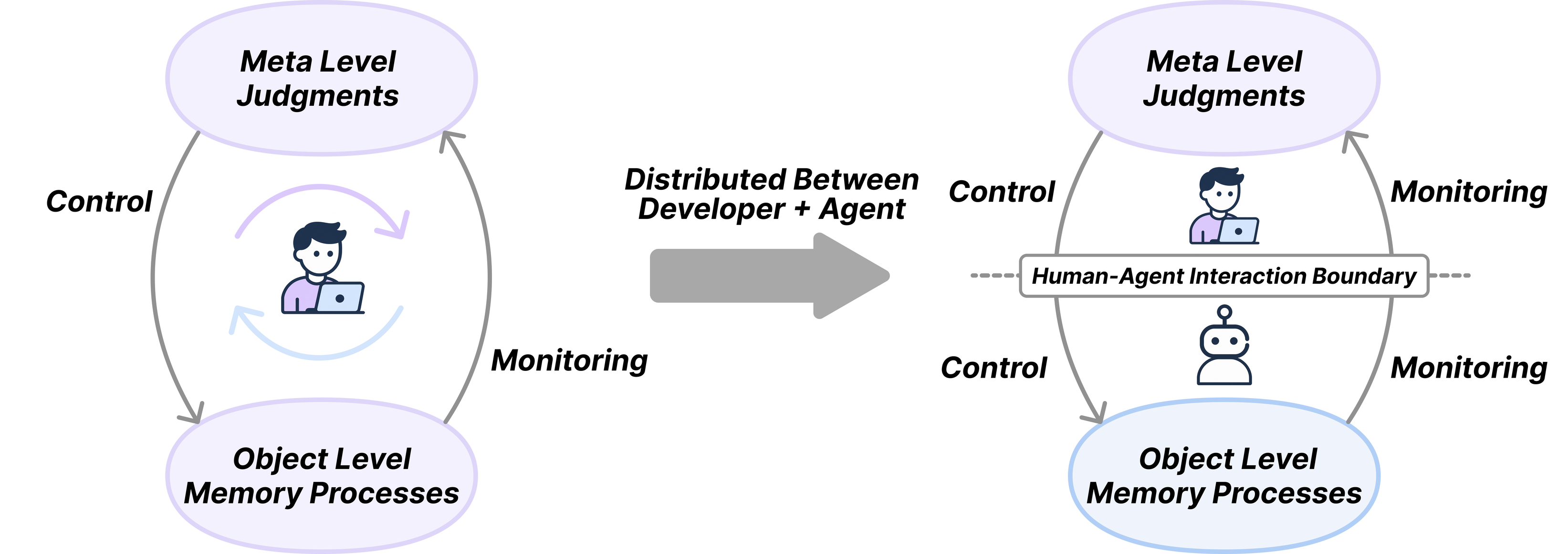}
  \caption{Metamemory regulation in individual cognition versus developer-agent collaboration. \textbf{Left:} In human cognition, meta-level judgments and object-level memory processes reside in the same individual; monitoring and control form a closed loop that keeps both levels synchronized~\cite{nelson1990metamemory}. \textbf{Right:} Developer-agent collaboration distributes this loop across a human-agent interaction boundary. The developer makes meta-level judgments (what to remember, in what scope, when to update), while the agent operates at the object level (storing entries, retrieving context, applying them to code generation). This boundary disrupts monitoring and control, causing memory states to diverge without either party detecting or correcting the mismatch.}
  \label{fig:teaser}
  \Description{Two diagrams side by side. Left: an oval labeled Meta Level Judgments above an oval labeled Object Level Memory Processes, with a developer icon between them. An arrow labeled Control runs from the upper oval to the lower one, and an arrow labeled Monitoring runs back up, forming a closed loop within one person. A large gray arrow between the diagrams reads Distributed Between Developer + Agent. Right: the same two ovals, but a dashed horizontal line labeled Human-Agent Interaction Boundary separates a developer icon above from a robot icon below. The Control and Monitoring arrows each cross this boundary.}
\end{teaserfigure}

\maketitle

\section{Introduction}
\label{s:intro}

As AI-based coding agents become increasingly integrated into software engineering, developer-agent collaboration has transitioned from a series of isolated prompts to a continuous, multi-session, and multi-project workflow~\cite{liang2024large, chen2026codewithme, sergeyuk2026evolving}. Throughout this ongoing collaboration, developers routinely return to the same codebase across multiple sessions and rapidly switch between disparate projects, relying on the coding agent to maintain and carry forward relevant context. This shared history constitutes the agent's working and long-term memories\footnote{We define \textit{memory} broadly as the synthesis of in-session interaction, cross-session experience, and external knowledge that influences agent behavior~\cite{10.1145/3748302}.}, which dynamically shapes its code reasoning, generation, and planning across diverse contexts~\cite{xiong2026memory, wang2025agent, park2023generative}.
Developers expect these memory systems to exhibit precise contextual awareness. Each of their interaction with the coding agent can produce diverse specifications, ranging from global, programmer-level stylistic preferences to local, project-specific architecture conventions~\cite{liang2024large}. Accordingly, when developers establish, scope, or delete these rules, they assume the agent will strictly match their mental model: reliably retaining instructions, applying them only within their valid boundaries, and immediately purging them upon revocation.
In practice, however, essential cross-session context may be dropped, project-specific rules may leak into unrelated workspaces, and obsolete guidelines may continue to shape the agent's outputs. These issues often go undetected because agent memory states are not easily accessible or legible on conventional programming interfaces~\cite{openai2025codexmemories, jones2025users}. We term this invisible gap between what the developer expects the agent to remember and the system's actual memory state \textit{Developer-Agent Memory Misalignment}.

Such misalignment stems from a structural breakdown in metamemory regulation. In cognitive psychology, Nelson and Narens conceptualize this regulation process as a closed-loop system operating across two memory levels: a meta-level that scrutinizes what memory contains, and an object-level that executes information storage and retrieval~\cite{nelson1990metamemory}. In human cognition, this loop is sustained by \textit{monitoring} (informing the meta-level of object-level contents) and \textit{control} (steering object-level actions according to meta-level understanding). Recent work has extended this framework to human-AI interaction, characterizing the monitoring and control demands that generative AI places on its users~\cite{tankelevitch2024metacognitive}. Developer-agent collaboration splits this loop in co-coding practices: the developer makes meta-level judgments (deciding what memory entry to keep, which scope it applies to, or when to obsolete it), while the agent operates at the object-level (updating stored memory entries, retrieving context, and applying them to code generation). We refer to this distributed loop as developer-agent metamemory. Largely opaque object-level states sever both pathways of the loop. Developers can neither clearly inspect stored and used context nor precisely control how memory shapes outputs, leaving much of this misalignment unnoticed and uncorrected across sessions and projects.

Empirical human-computer interaction (HCI) studies show that developers highly value agency and control over AI outputs~\cite{huang2025professional, 10.1145/3803437.3806709}. Many of them currently rely on three common but inadequate practices to manage cross-context coding memory. Manual Context Injection (i.e., manually pasting code, locating files, or repeating rules) preserves user control, but it is restricted to a single session. When entering a new session, developers need to decide what knowledge and context to carry over and manually recreate them, imposing high metacognitive demands~\cite{tankelevitch2024metacognitive} alongside output verification effort~\cite{mozannar2024reading, vaithilingam2022expectation, 10.1145/3772318.3791176}. Static Configuration via files such as \texttt{CLAUDE.md} and \texttt{.cursorrules} allows the developer and the agent to co-edit memory instructions in natural languages~\cite{chatlatanagulchai2025use}, but monitoring is limited to the file itself. Which rules the agent applied or ignored during a generation turn remains hidden~\cite{jiang2025beyond, cai2026rule}. Automatic Memory Extraction employed by tools such as Cursor~\cite{anysphere2025cursormemories}, Claude Code~\cite{anthropic2025claudecodememory}, and Codex~\cite{openai2025codexmemories} autonomously persists conversational history across workspaces, which relieves manual effort but strips user control~\cite{openai2025codexmemories, chen2026relational}. More crucially, bringing static configurations or automatic memories across workspaces unchanged can cause outdated rules to conflict with new project-level conventions~\cite{treude2026context}, adding to the misalignment.
Research on AI agent memory systems tried to address these challenges by making stored items visible~\cite{huang2023memory, 10.1145/3810195, 10.1145/3772318.3791669}, editable~\cite{huang2023memory, yen2024memolet, 10.1145/3810195, 10.1145/3772318.3791669}, reusable~\cite{yen2024memolet, 10.1145/3810195, 10.1145/3772318.3791669}, and conflict-aware~\cite{vaithilingam2025semantic, 10.1145/3772318.3791669}.
Still, developers cannot monitor which memory entries are selected during a generation turn or whether/how they shape code outputs, leaving the critical loop between the developer's mental model and the agent's active memory operations unclosed.
This gap leads to the following research questions:

\begin{itemize}[leftmargin=*,nosep]
\item \textbf{RQ1}: In cross-context collaboration with AI coding agents, what do developers want to monitor and what do they want to control in agent memory?
    \item \textbf{RQ2}: What interaction design can help developers monitor and control agent memory across contexts?
    \item \textbf{RQ3}: How does memory co-management affect developer-agent memory alignment and task performance in cross-context collaboration?
\end{itemize}

To answer \textbf{RQ1}, we conducted a formative study with 12 developers from industry and academia who regularly use coding agents across multiple projects. The study yielded five findings, covering how memory should be represented and what developers can monitor and control across four dimensions of memory use: evolution (how memory changes), selection (what is retrieved for a turn), execution (how it is applied), and impact (its effect on the output). Guided by these requirements, we designed MemoSync, a co-management system that actualizes agent memory as a shared, operable representation jointly navigated and modified by both the developer and the agent (\textbf{RQ2}). To manage boundaries across workspaces, MemoSync associates each memory item with a scope (i.e., personal, project, or session). Upon context switches, the system abstracts specifications from past instances, proposes their provisional scopes for developer review, and transforms them for the destination codebase. MemoSync closes the metamemory loop by pairing explicit monitoring with direct control across each of the four identified memory dimensions. It allows developers to authorize proposed changes, inspect and dynamically adjust the context selected for every code-generation turn, trace rule application in real time, interrupt the process whenever spotting an error, and audit the final output's compliance with memory specifications. To evaluate MemoSync's efficacy on fostering metamemory regulation in developer-agent collaboration (\textbf{RQ3}), we conducted a between-subjects study ($N=132$) comparing the system against an automatic memory baseline (e.g., ChatGPT- and Claude-style memory) and a static configuration baseline (e.g., developer-maintained Markdown memory files). During the study, participants completed two distinct, full-stack projects, each encompassing two development sessions. We also deployed MemoSync with nine developers on their own projects for one week.

This paper makes four contributions:
\begin{enumerate}[leftmargin=*,nosep]
\item A formative study with 12 developers that yielded five findings on memory representation and on monitoring and control across the four dimensions of memory use, leading to five design goals.
\item MemoSync, a co-management system that treats agent memory as a shared operable representation, pairs monitoring with control across the four dimensions, and scopes memory for transfer across contexts.
\item A between-subjects study ($N=132$) in which MemoSync participants achieved significantly higher task performance than two baselines and the highest mean memory alignment, with the alignment advantage driven mainly by monitoring accuracy and no added perceived workload despite more monitoring and control actions.
\item A week-long deployment of MemoSync with nine developers on their own projects, where participants incorporated memory monitoring and control into their everyday development work.
\end{enumerate}
 \section{Related Work}

\subsection{AI-Assisted Programming and Human-Agent Collaboration}

\textbf{Code completion and generation.}
In workflows centered on code completion and generation, developers use model suggestions to accelerate implementation or explore alternative solutions~\cite{barke2023grounded}, while reading, interpreting, and inspecting generated code to decide whether to incorporate it into their implementation~\cite{mozannar2024reading}. Research on programming question answering also shows that users may overlook errors in AI-generated answers, highlighting the need for effective verification support~\cite{aic042}. Prior systems provide anchored explanations~\cite{yan2024ivie}, live execution feedback~\cite{aic050}, and multidimensional code evaluation~\cite{aic089} to help developers understand, verify, and revise generated code. However, this support primarily concerns local code: translating overall goals into specific requests and coordinating successive generations to satisfy task requirements still depend substantially on developers' own efforts~\cite{yen2024coladder}.

\textbf{Conversational programming and structured steering.}
Natural-language and conversational programming enable developers to progressively clarify requirements and refine implementations through continued interaction~\cite{aic015,ross2023programmer}. To connect high-level goals with implementation details, prior systems make task decompositions, plans, and the model's interpretation of user intent inspectable and editable~\cite{yen2024coladder,kazemitabaar2024decomposition,zhang2025neurosync}. Interactive explanations and dataflow visualizations extend this control to intermediate operations~\cite{aic059,xie2024waitgpt,aic087}. For implementation choices that remain open, branching and merging and structured design exploration help users compare and refine alternatives~\cite{aic030,aic066,aic073}, while sketch-based and component-level editing support local revisions~\cite{yen2025codeshaping,aic093}. These mechanisms strengthen understanding and intervention within the current task. However, ongoing revision of task representations does not fully address knowledge reuse across tasks: when goals or projects change, further interaction support is needed to select, adapt, and carry forward established requirements, design decisions, and experience.

\textbf{Collaboration with AI Agents.}
Coding agents further expand the scope of work that developers can delegate.
Developers can specify higher-level goals and allow the agent to choose concrete steps, invoke tools, and carry out sequences of file modifications and code execution~\cite{chen2026codewithme}.
Developers can consequently reduce their direct involvement in individual operations while maintaining control through planning, incremental delegation, and result inspection~\cite{huang2025professional}.
Related systems also explore proactively offering suggestions based on task context, using status cues and edit previews to help developers coordinate these interventions~\cite{aic065,10.1145/3706598.3714002}.
Other systems organize agent-discovered program dependencies into annotated traces to help developers understand the answers provided~\cite{aic086}.
However, inspecting current execution does not ensure that developers know which prior agreements the agent is carrying forward.
As implementation decisions and context selection are delegated to agents, insufficient feedback and control over retained and retrieved information can allow developers' understanding of agent memory to diverge from its actual state.

\subsection{Metamemory in Human-AI Collaboration}

\emph{Metamemory} focuses on people's understanding and regulation of their own memory states and processes. Nelson and Narens describe two interrelated levels: the \emph{object level} carries out memory activities such as learning, storage, and retrieval, while the \emph{meta level} maintains a dynamic model of these activities. \emph{Monitoring} informs the meta level about object-level states, and \emph{control} adjusts object-level activities according to meta-level judgments~\cite{nelson1990metamemory}. Within individual cognition, these pathways connect people's judgments about memory with the memory processes they regulate.

HCI research has drawn on the broader concept of metacognition to examine how users monitor and regulate their cognitive activities during interaction. Chen et~al. incorporate metacognitive experiences, knowledge, and skills into a model of visualization, explaining how users assess their understanding and adjust their exploration strategies~\cite{10.1145/3706598.3714400}. In human-AI interaction, Tankelevitch et~al. characterize prompting, output evaluation, and workflow adaptation as activities that place metacognitive demands on users, and discuss how explainability and customizability can help address these demands~\cite{tankelevitch2024metacognitive}. These accounts provide a theoretical basis for designing metacognitive support, while leaving open how users' judgments can remain connected to an agent's persistent memory operations across contexts.

Building on this perspective, we conceptualize developer-agent memory co-management as a regulatory loop distributed across a \emph{human-agent interaction boundary}: developers judge what to retain, where it applies, and when to update it, while the agent stores, retrieves, and applies memory. We therefore investigate how interfaces can help developers observe these memory operations and translate their judgments into effective interventions, supporting memory alignment across sessions and projects.

\subsection{Memory Management in AI Agents}

Research on automated agent memory focuses on information extraction, organization, retrieval, and updating to support experience accumulation and improve long-term task performance~\cite{10.1145/3748302,wang2025agent,xiong2026memory}. These studies primarily evaluate memory mechanisms in terms of experience reuse and task performance, giving less attention to how users participate in deciding what information should be retained and under what conditions it should be used. In human-AI collaboration, however, memory also carries users' preferences, goals, and constraints, and decisions about its retention and updating need to be grounded in user intent~\cite{vaithilingam2025semantic}. Information accuracy and task benefits alone are therefore insufficient to establish whether memory continues to match users' expectations.

HCI research on interactive memory management further examines how users participate in these decisions. Prior work presents conversational memories as visible, editable objects that users can inspect and revise~\cite{huang2023memory,yen2024memolet}, and uses conflict detection and local revision to help users understand and manage the consequences of memory updates as their intentions change~\cite{vaithilingam2025semantic}. Interactive memory management also helps users decide which experiences to reuse and how to apply them to new tasks. Reusable content includes conversation excerpts, cognitive workflows derived from collaboration, and domain knowledge implicit in historical code~\cite{yen2024memolet,10.1145/3772318.3791669,zhou2026cerebra}. In addition, Knoll allows users to configure accessible knowledge modules and inspect the knowledge sources associated with responses~\cite{zhao2025knoll}; Cerebra links knowledge items, code snippets, and execution results to help users inspect and revise how knowledge is applied~\cite{zhou2026cerebra}.

These studies provide interaction mechanisms for users to inspect and control memory content and its use. An open question is how to organize these interactions around developers' metamemory regulation in cross-session coding, connecting their meta-level judgments with agents' object-level memory operations through a sustained monitoring and control loop that supports memory alignment. \section{Formative Study}

To understand what developers want to monitor and control in agent memory (RQ1), we conducted a formative study.

\subsection{Participants, Procedure, and Data Analysis}

We recruited 12 developers who regularly use AI coding agents across multiple projects (demographics in Table~\ref{tab:appendix_participants}), as they are most likely to encounter the cross-context memory challenges this study investigates. All sessions were conducted remotely and lasted approximately 60 minutes. Before the session, we collected from each participant a recent agent-assisted project with its conversation logs and configuration files (e.g., \texttt{CLAUDE.md}, \texttt{AGENT.md}). Following a contextual inquiry approach~\cite{holtzblatt1997contextual}, the session was divided into three parts:

\begin{itemize}
    \item \textbf{Semi-structured Interview}: We asked participants about their current collaboration patterns with the agent, including session structure, cross-project workflows, and configuration practices ($\sim$15~min).
    
    \item \textbf{Artifact Walkthrough}: The participant and researcher jointly examined conversation logs, configuration files, and the agent's automatically extracted memory. Participants identified information they wished to persist and annotated its scope and portability, revisited their experiences with cross-project migration and long-term maintenance, and pointed out moments where they could not monitor or control the agent's memory ($\sim$30~min).
    
    \item \textbf{Design Expectations Interview}: Participants described what visibility and control they wanted over memory management, including what they needed to see, when to intervene, and how to handle deviations ($\sim$15~min).
\end{itemize}

\noindent Details for each part can be found in Appendix~\ref{app:procedure}. Each participant received $\sim$\$10 for their time. This study adhered to the Declaration of Helsinki and was approved by the Human Research Ethics Committee. We recruited iteratively and stopped after 12 participants when no new codes emerged. Two researchers independently coded the transcripts using an inductive approach and iteratively refined codes through discussion (Appendix~\ref{app:analysis}).

\begin{figure*}[t]
  \centering
  \includegraphics[width=\textwidth]{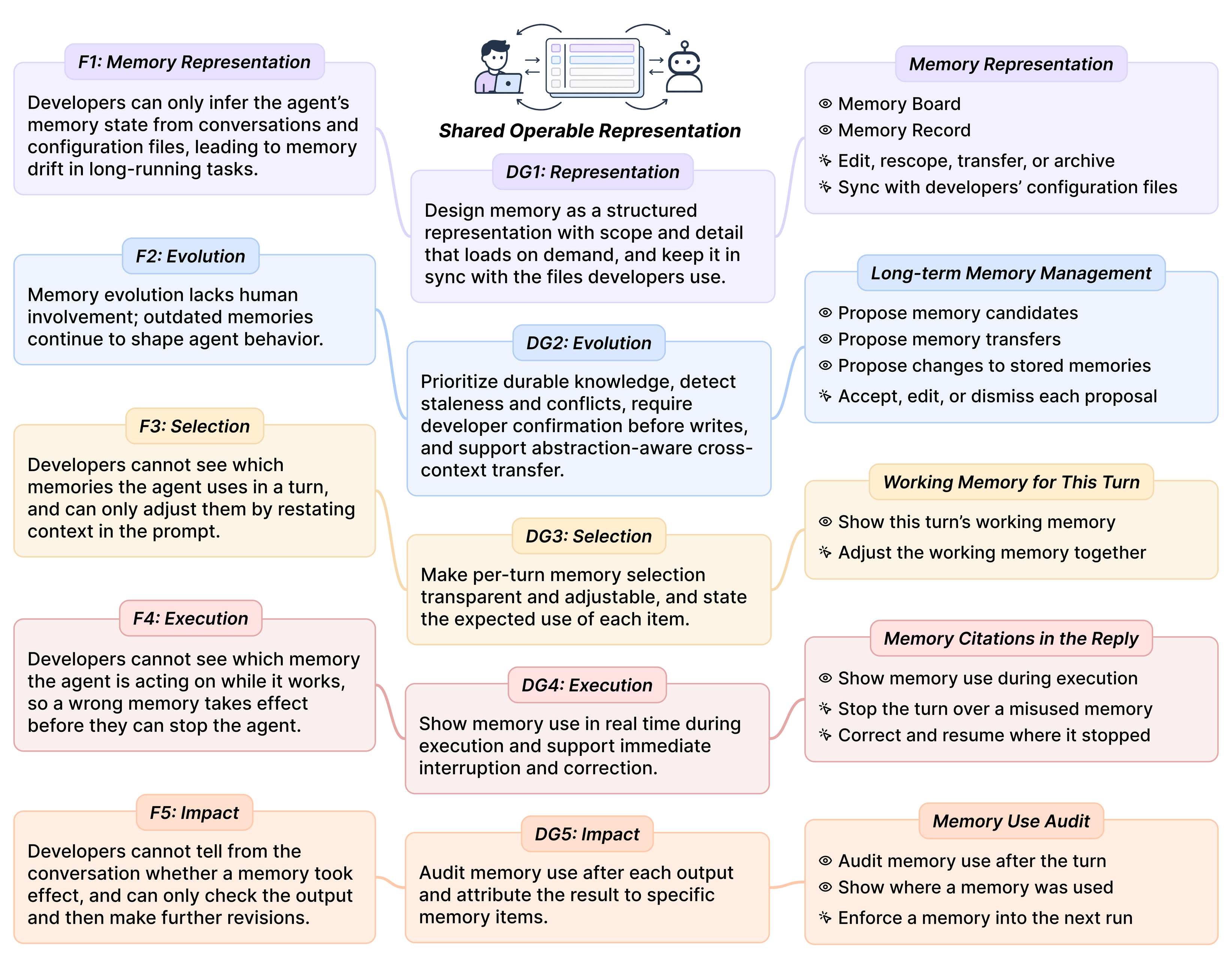}
  \caption{
    Overview of how formative findings inform design goals and system features.
    \textbf{Left:} Five findings: the representation of memory (F1) and the four dimensions of memory use, evolution (F2), selection (F3), execution (F4), and impact (F5).
    \textbf{Center:} Five design goals derived from the findings.
    \textbf{Right:} MemoSync features that realize each goal, named by the part of the interface they live in; \iconMonitor{} marks what the developer monitors and \iconControl{} marks what the developer controls. The features are described in Section~\ref{s:system}.
  }
  \label{fig:findings-to-features}
  \Description{Three columns connected by curved lines, one row per theme, color-coded from purple to orange. Left column: five finding boxes. F1 Memory Representation: developers can only infer the agent's memory state from conversations and configuration files. F2 Evolution: memory evolution lacks human involvement and outdated memories continue to shape agent behavior. F3 Selection: developers cannot see which memories the agent uses in a turn. F4 Execution: developers cannot see which memory the agent is acting on while it works. F5 Impact: developers cannot tell from the conversation whether a memory took effect. Center column: an icon of a developer and a robot on either side of a shared panel, labeled Shared Operable Representation, above five design-goal boxes DG1 to DG5 that restate each finding as a goal. Right column: five feature groups. Memory Representation lists Memory Board and Memory Record (monitor) and edit, rescope, transfer, or archive and sync with configuration files (control). Long-term Memory Management lists proposing candidates, transfers, and changes (monitor) and accepting, editing, or dismissing each proposal (control). Working Memory for This Turn lists showing the working memory (monitor) and adjusting it together (control). Memory Citations in the Reply lists showing memory use during execution (monitor) and stopping the turn and correcting and resuming (control). Memory Use Audit lists auditing memory use and showing where a memory was used (monitor) and enforcing a memory into the next run (control).}
  \vspace{-5pt}
\end{figure*}

\subsection{Findings}
Figure~\ref{fig:findings-to-features} shows how the five findings from the interviews connect to design goals and system features. 

\textbf{Finding 1. Developers can only infer the agent's memory state from conversations and configuration files, leading to memory misalignment in long-running tasks.} Developers and agents primarily share information through three channels: conversations, configuration files, and the codebase. For conversations, information is gradually lost as the agent discards earlier turns during context compression and developers forget specific agreements after days away from a project. When starting a new session, participants either used skills such as \texttt{/handoff} or manually re-explained the project background (N=5). Both approaches had limitations: P7 noted that handoff documents \textit{`dump everything in there''}, while P3 described repeatedly re-explaining the same context as effortful. Static configuration files such as \texttt{CLAUDE.md} persist across sessions. However, as these files grew, participants rarely read them in full and could not confirm which rules the agent actually followed, especially when the files contained conflicting entries (N=11). Participants rarely reviewed agent-generated code line by line and inspected it only when results were wrong (N=10). 

Participants wanted to directly view and edit the memory the agent was actually using (N=12). They viewed memory as a structured collection (N=12), organized along three levels: eight participants explicitly distinguished Personal, Project, and Session scopes, while the remaining four described equivalent distinctions using different terms. Participants also wanted individual memory entries to expand on demand to provide richer details (N=9). Participants also wanted to continue using their existing \texttt{CLAUDE.md} files. When shown existing open-source memory tools, seven participants reported having tried similar tools before but found them difficult to integrate with their existing workflows.

\textbf{Finding 2. Agent memory lacks sustained human involvement; outdated and leaked entries continue to shape agent behavior.} Long-term memory was primarily created and modified by the agent in automatic mode, and participants rarely inspected individual entries (N=11). During the walkthrough, P8 opened the CodeX's memory list and found over a thousand lines, many containing execution details. Participants wanted to retain reusable knowledge rather than execution details (N=11). No participant continuously maintained memory over time. Their primary goal was to complete the current task, so they typically checked memory only when results went wrong (N=10). As projects progressed, entries became outdated, duplicated, or contradictory, with no mechanism alerting developers to these changes. When switching projects, participants typically copied existing context documents with little revision and assumed that the agent would interpret them correctly, but this often led to unexpected behavior (N=10). Participants wanted the agent to ask for confirmation before writing to memory and to surface outdated or conflicting entries (N=12). During cross-project migration, they also wanted guidance on which memories could transfer directly, which needed rewriting, and which should remain in the original project (N=9).

\textbf{Finding 3. Developers cannot see which memories the agent uses in a turn, and can only adjust them by restating context in the prompt.} The agent selected which memory items to use for each turn on its own. Participants could see the memory store but not which items the agent had selected for the turn, so they could not judge whether the selection was appropriate (N=9). Seven participants reported asking the agent to repeat back what it remembered as a way to verify whether it had the right context. This indirect verification depended on the agent's self-report; developers had no independent way to check the completeness of the selection. Participants wanted to see which items the agent planned to use and how it expected to use each one before generation started, with the ability to modify the selection (N=9). P11 suggested: \textit{``Before it starts generating, I want to see what it plans to use and how it will use each one. And I should be able to change its memory.''}

\textbf{Finding 4. Developers cannot see which memory the agent is acting on during execution; a wrong memory may take effect before they can stop the agent.} During execution, the agent might act on an outdated memory the developer did not expect in the current conversation. Participants could not observe which memory the agent was following as it generated (N=12). In most cases the agent did not indicate which memory it was acting on, and errors surfaced only when the developer inspected the finished output. Only when the agent happened to mention a piece of information in its reply, and that information conflicted with the developer's expectations, could the participant catch the problem mid-execution and interrupt the generation. Participants wanted to see in real time which memory the agent was applying, so they could stop and correct the agent on the spot and resume from where it stopped (N=12).

\textbf{Finding 5. Developers cannot evaluate whether their memory took effect in the agent's output.} Eleven participants (N=11) reported that they could not tell whether a specific memory item was used, ignored, or misapplied. P4 described a typical experience: \textit{``I noticed something was off in its reply, went back and checked, and it clearly didn't follow what I had set up. I asked it why, and it just said sorry, I'll fix it. Then it made the same mistake again.''} This inability to attribute results to specific items made repair indirect: developers could only re-emphasize the relevant memory in the next prompt without knowing whether the agent would follow it. Participants wanted each output to show which memory items shaped it and which were violated, so that they could trace the result to specific entries (N=12).

\subsection{Design Goals}

Based on these findings, we propose five design goals (Figure~\ref{fig:findings-to-features}). They converge on a core principle: agent memory should serve as a \textit{shared operable representation} that both the developer and the agent can read, modify, and act on. DG1 addresses the representation of memory itself; DG2 through DG5 correspond to the four dimensions identified in the formative study: evolution, selection, execution, and impact.

\textbf{DG1. Design memory as a structured representation with explicit scope and progressively loadable detail, compatible with developers' existing tools.}
Building on F1, the system should represent each memory item as a structured entry with an explicit scope and detail that the agent loads on demand, and should stay in sync with formats such as \texttt{Markdown}, so that developers can import existing configuration files and export memory for use elsewhere.

\textbf{DG2. Evolution: prioritize durable knowledge, detect staleness and conflicts, require developer confirmation before writes, and support abstraction-aware cross-context transfer.}
In response to F2, the system should propose memory items from the conversation, favoring preferences, constraints, decision rationales, and lessons over execution traces and temporary information, and require developer confirmation before any write. It should detect stale, redundant, or contradictory entries and propose fixes for the developer to review. When a memory item migrates to a new context, the system should abstract it from its source, rewrite it for the destination, and propose a landing scope, so that the developer reviews each form before the item crosses the boundary.

\textbf{DG3. Selection: make per-turn memory selection transparent and adjustable, and state the expected use of each item.}
In response to F3, the system should expose the proposed memory set and the expected use of each item before generation. Developers should be able to remove irrelevant items, add missing ones, or ask the agent to revise the selection, and confirm it before generation starts.

\textbf{DG4. Execution: show memory use in real time during execution and support immediate interruption and correction.}
In response to F4, the system should expose memory use during generation. When developers spot a deviation, they should be able to stop execution at that point, supply a correction that the resumed run must follow, and resume from that point.

\textbf{DG5. Impact: audit memory use after each output and attribute the result to specific memory items.}
In response to F5, the system should report, for each injected memory item, whether it shaped the turn, was violated, was not applicable, or had no visible effect, and it should judge this independently of the agent's own report, since the agent may claim to follow a rule while breaking it. Developers should be able to trace a verdict to the place in the output where the item was applied, and to mark an item as mandatory for the next turn. Audit results should feed back into the store for ongoing maintenance.

 \section{MemoSync}
\label{s:system}

\begin{figure*}[t]
    \centering
    \includegraphics[width=\textwidth]{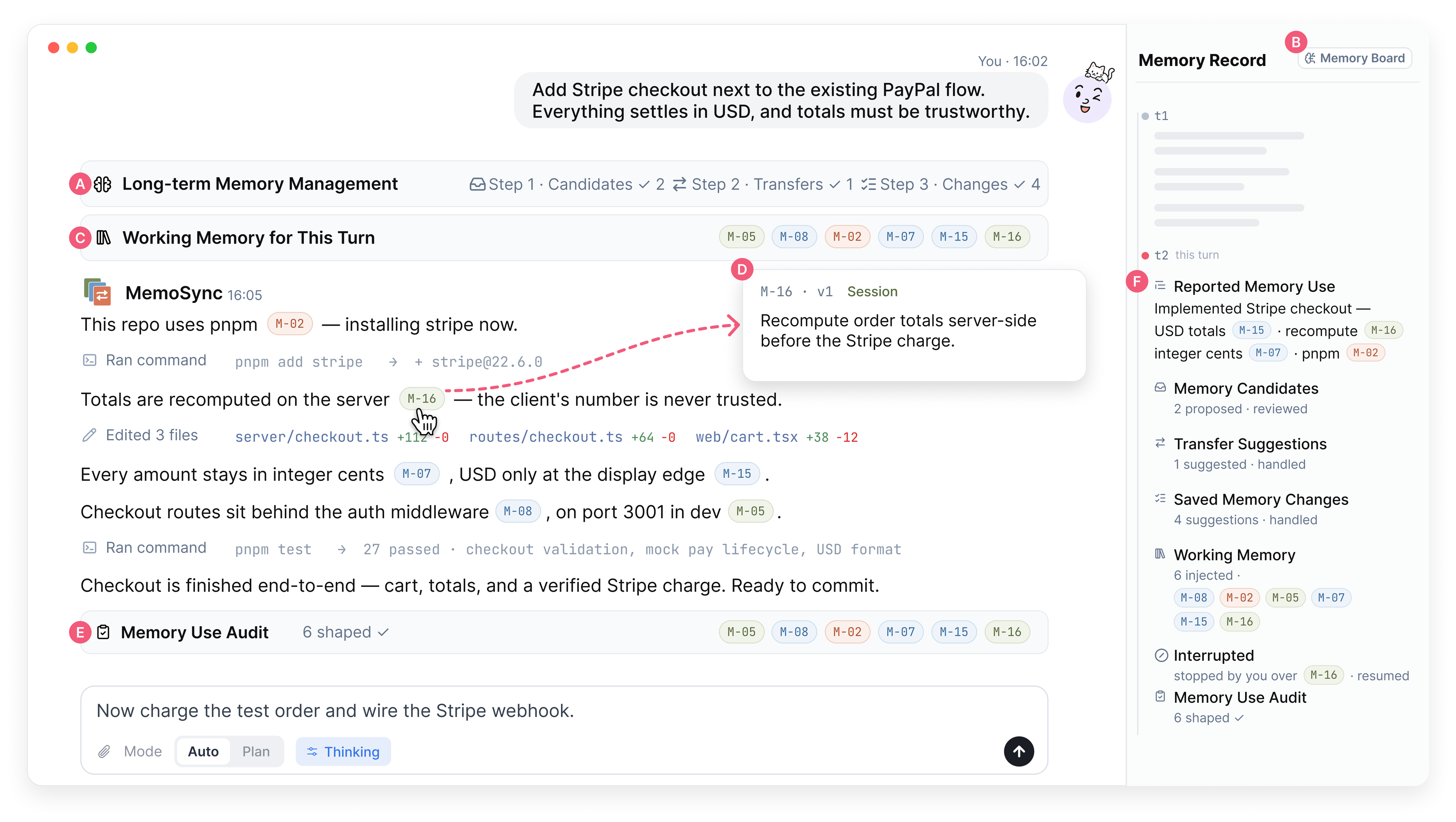}
\caption{MemoSync during Ben's work on \textit{storefront}. \circmark{A}~\textit{Long-term Memory Management} summarizes the review of changes to stored memory. \circmark{B}~The \textit{Memory Board} button opens the shared memory library. \circmark{C}~\textit{Working Memory for This Turn} lists the six confirmed items. \circmark{D}~Inline citations reveal memory items on hover and provide \textit{Stop} controls during execution. \circmark{E}~\textit{Memory Use Audit} reports item-level outcomes after execution. \circmark{F}~The \textit{Memory Record} traces memory activity across the session's turns.}
    \label{fig:main_ui}
    \Description{A chat application with a main panel on the left and a Memory Record sidebar on the right. The main panel contains a developer prompt, a completed long-term memory review, six working-memory items, a coding-agent reply with inline memory citations, a hover card showing a cited item, command and file-edit rows, and a memory-use audit. The sidebar lists proposals, injected items, reported citations, an interruption, and an audit summary for each turn. A Memory Board button appears in the sidebar header. Letter markers A to F identify these regions.}
\end{figure*}

MemoSync integrates developer monitoring and control of agent memory into the coding agent's task execution through a shared, operable memory representation. Around each turn\footnote{A turn is one cycle of coding-agent execution in response to a user prompt, including its tool actions.}, MemoSync supports memory review before execution, intervention during execution, and auditing after completion. Figure~\ref{fig:main_ui} illustrates this workflow. Before execution, \circmark{A} \textit{Long-term Memory Management} presents the memory review preceding the turn, and \circmark{C} \textit{Working Memory for This Turn}  lists the memory items confirmed for the current turn. During execution, \circmark{D} inline citations in the agent's reply indicate relevant memory items. After execution, \circmark{E} \textit{Memory Use Audit}  reports how those memories were used. The right sidebar contains the \circmark{F} \textit{Memory Record} for reviewing memory activity across turns, with a \circmark{B} \textit{Memory Board} button at the top for accessing the persistent memory store for viewing and editing. 

We illustrate these interactions through Ben, who adds Stripe\footnote{\url{https://docs.stripe.com/}} as a second payment provider to \textit{storefront}, an online shop that already uses PayPal\footnote{\url{https://developer.paypal.com/}}. His memory store contains the preference ``Use pnpm, never npm''\footnote{pnpm and npm are package managers used to install project dependencies: \url{https://pnpm.io/} and \url{https://docs.npmjs.com/about-npm/}.} (\memPe{02}) and the project rule ``Prices are stored as integer cents, never floats'' (\memP{07}). It also holds lessons from \textit{acme-payments}, a Stripe billing service from another project.

\subsection{Evolution: Reviewing Changes to Long-term Memory}
\label{s:evolution}

\begin{figure*}[t]
    \centering
    \includegraphics[width=\textwidth]{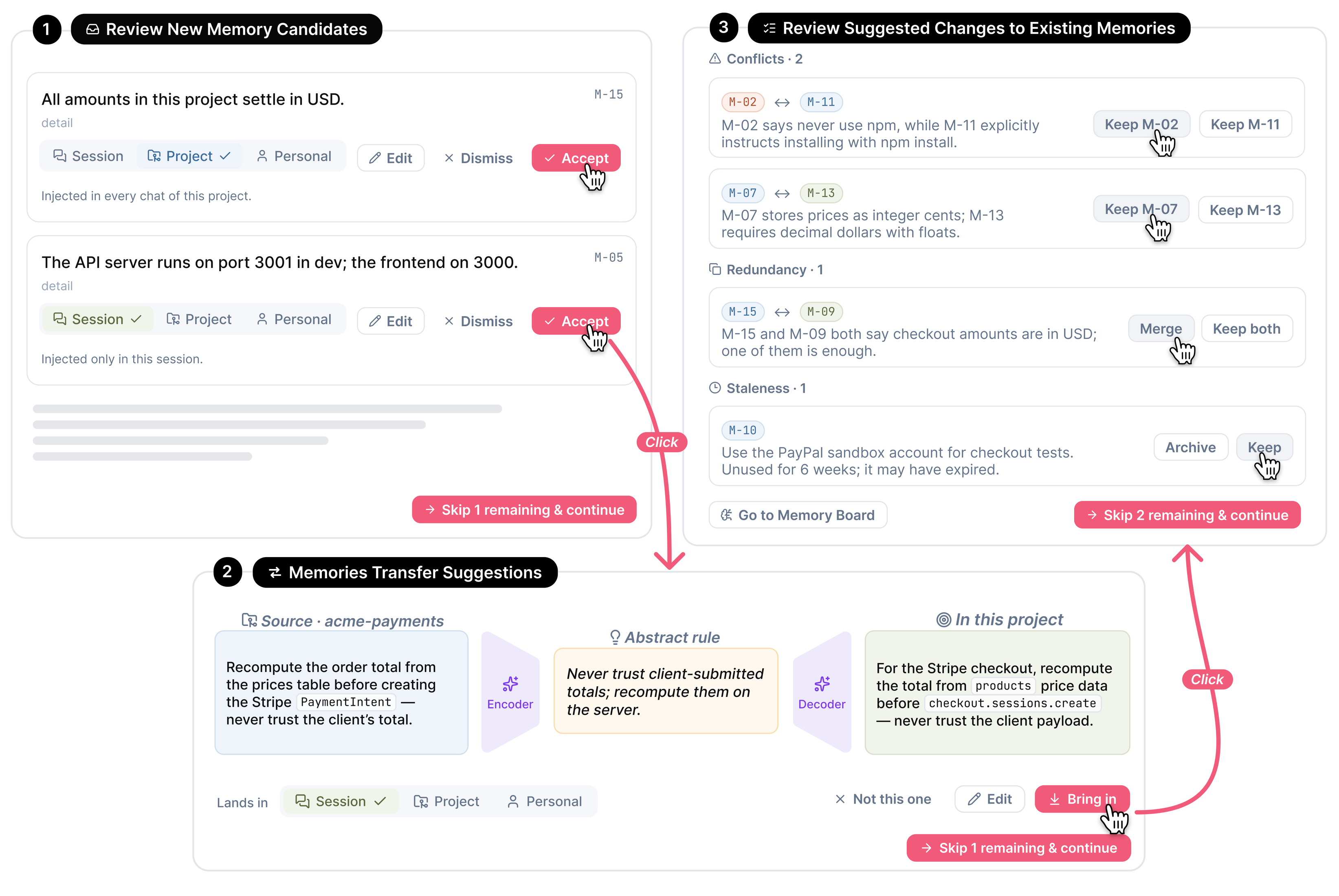}
\caption{Reviewing changes to long-term memory before execution. \stepmark{1}~\textit{Review New Memory Candidates} presents extracted items with scope selectors and editing, dismissal, and acceptance controls. \stepmark{2}~\textit{Memory Transfer Suggestions} presents the source, abstract, and localized forms with a proposed destination scope. \stepmark{3}~\textit{Review Suggested Changes to Existing Memories} presents conflicts, redundancy, and staleness for review. The \textit{Go to Memory Board} button provides access to direct memory editing.}
    \label{fig:evolution}
    \Description{Three numbered panels. Panel 1 shows two memory candidates with scope selectors and Edit, Dismiss, and Accept buttons. Panel 2 shows a source rule from acme-payments, its abstract form, and the version rewritten for storefront, connected by encoder and decoder symbols. A destination-scope selector and Bring in button appear below. Panel 3 shows conflicts between pnpm and npm instructions and between integer-cent and decimal-dollar rules, a redundant pair about USD amounts, and a potentially stale PayPal test-account entry. Controls allow keeping, merging, or archiving items.}
\vspace{-8pt}
\end{figure*}

Ben opens a session under \textit{storefront} and enters the prompt: ``Add Stripe checkout next to the existing PayPal flow. Everything settles in USD, and totals must be trustworthy.'' Before the coding agent executes the task, \textit{Long-term Memory Management} presents three review steps for changes proposed from the conversation and existing memories (Figure~\ref{fig:evolution}). Only changes explicitly approved by Ben are written to the store. 

The first step lists new memory candidates extracted from the conversation, each with a proposed scope (Figure~\ref{fig:evolution}, \stepmark{1}). Ben accepts ``All amounts in this project settle in USD'' (\memP{15}) into the project scope and ``The API server runs on port 3001 in dev'' (\memS{05}) into the session scope.

The second step suggests memories from other projects that may support the task (Figure~\ref{fig:evolution}, \stepmark{2}). MemoSync identifies a rule in \textit{acme-payments} about recomputing order totals before creating a Stripe \texttt{PaymentIntent}, an API object that tracks a payment\footnote{\url{https://docs.stripe.com/payments/payment-intents}}. To avoid carrying source-specific implementation details into \textit{storefront}, an encoder abstracts the rule into ``Never trust client-submitted totals; recompute them on the server.'' A decoder then rewrites it using the destination's existing memory and current task. \stepmark{2} displays all three forms and a proposed destination scope for review. Ben clicks \textit{Bring in} to accept the localized rule as \memS{16} in the session scope.

The third step checks for conflicts, redundancy, and staleness (Figure~\ref{fig:evolution}, \stepmark{3}). Ben keeps the pnpm preference (\memPe{02}) over an npm instruction, and the integer-cents rule (\memP{07}) over a decimal-dollar rule, retiring the contradictory entries. He clicks \textit{Merge} for two USD rules and \textit{Keep} for a stale-flagged PayPal test-account entry that remains relevant.

\subsection{Selection: Composing the Working Memory for a Turn}
\label{s:selection}

\begin{figure*}[t]
    \centering
    \includegraphics[width=\textwidth]{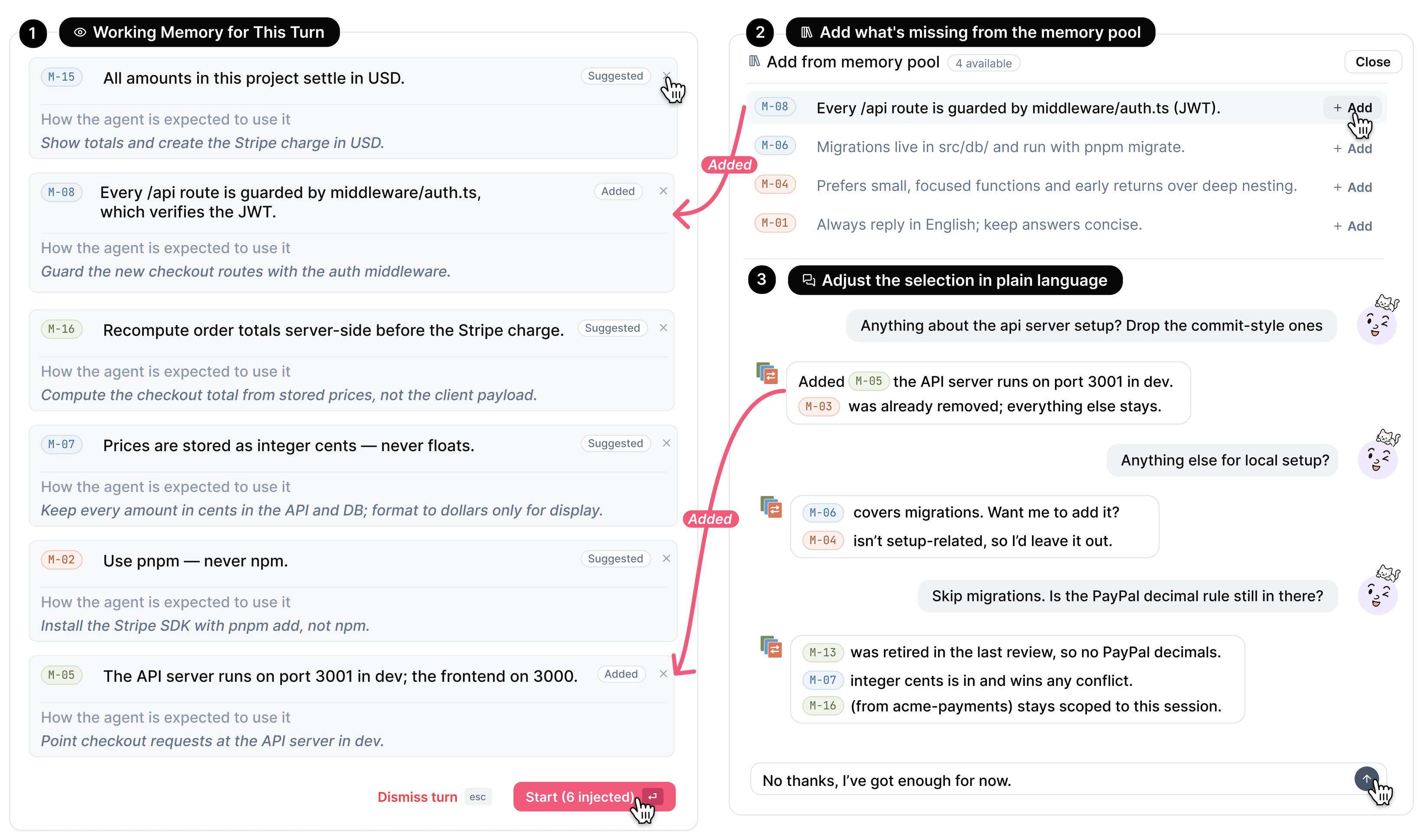}
    \caption{Composing working memory before execution. \stepmark{1}~The preview shows memory items, scopes, expected uses, and removal controls. \stepmark{2}~The memory pool provides \textit{Add} controls for other active items. \stepmark{3}~The selection dialogue supports requests in plain language. Ben confirms the final set with \textit{Start (6 injected)}.}
    \label{fig:selection}
    \Description{Three panels. Panel 1 shows six memory rows with scope-colored identifiers, expected-use descriptions, removal controls, and a Start (6 injected) button. Panel 2 lists other active memories with Add controls. Panel 3 shows a dialogue in which Ben requests the API server setup, excludes commit-style items, and asks about a retired PayPal rule. The replies identify the items added or excluded.}
\end{figure*}

After the approved changes update the store, \textit{Working Memory for This Turn} previews the items proposed for injection into the coding agent's context (Figure~\ref{fig:selection}, \stepmark{1}). Each row shows the item's scope and expected use; for \memS{16}, the latter reads ``Compute the checkout total from stored prices, not the client payload.'' Ben can remove an item using the cross at the end of its row, open \textit{Add from memory pool} to select other active items not currently included, or request changes through the selection dialogue.

Ben clicks \textit{Add} beside \memP{08}, which requires API routes to use \texttt{middleware/auth.ts} (Figure~\ref{fig:selection}, \stepmark{2}). In the dialogue's text input (\stepmark{3}), he enters ``Anything about the api server setup? Drop the commit-style ones.'' MemoSync adds the port fact (\memS{05}) and confirms that the commit-style preference (\memPe{03}) remains excluded. Ben confirms the six-item set by clicking \textit{Start (6 injected)}, which starts the coding agent's execution.

\subsection{Execution: Tracing and Interrupting Memory Application}
\label{s:execution}

\begin{figure*}[t]
    \centering
    \includegraphics[width=\textwidth]{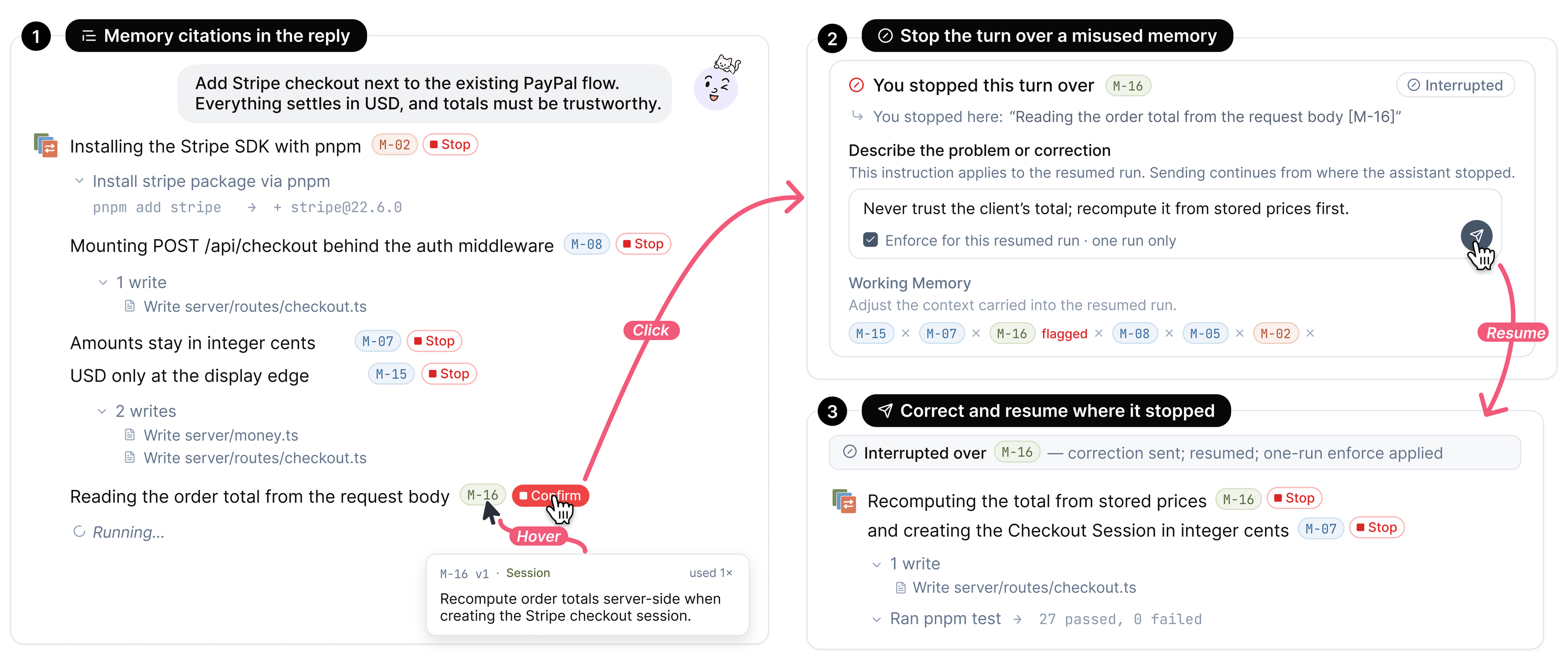}
\caption{Inspecting and correcting reported memory use during execution. \stepmark{1}~Hovering over the citation reveals \memS{16}; \textit{Stop} and \textit{Confirm} interrupt the current run. \stepmark{2}~The recovery card quotes the associated sentence and provides a correction field and an \textit{Enforce this run} option. \stepmark{3}~After Ben submits the correction, the agent continues with server-side total computation and integer-cent amounts.}
    \label{fig:execution}
    \Description{Three panels. Panel 1 shows a coding-agent reply with inline memory citations and Stop controls. A hover card displays the server-side total rule, while the associated sentence describes reading the total from the request body. Panel 2 quotes that sentence and contains a correction text box, the option Enforce for this resumed run, one run only, and a send button. Panel 3 shows the continued reply recomputing totals from stored prices, creating the checkout session in integer cents, and reporting passed tests.}
\end{figure*}

During execution, the coding agent reports memory use through inline citations, which MemoSync displays as the reply streams (Figure~\ref{fig:execution}, \stepmark{1}). Hovering over a citation reveals the item's content and scope. Ben sees the agent write ``Reading the order total from the request body'' while citing \memS{16}. The hover card states that totals must be recomputed server-side from stored prices, revealing a conflict between the cited rule and the reported action.

\begin{figure*}[htbp]
    \centering
    \includegraphics[width=\textwidth]{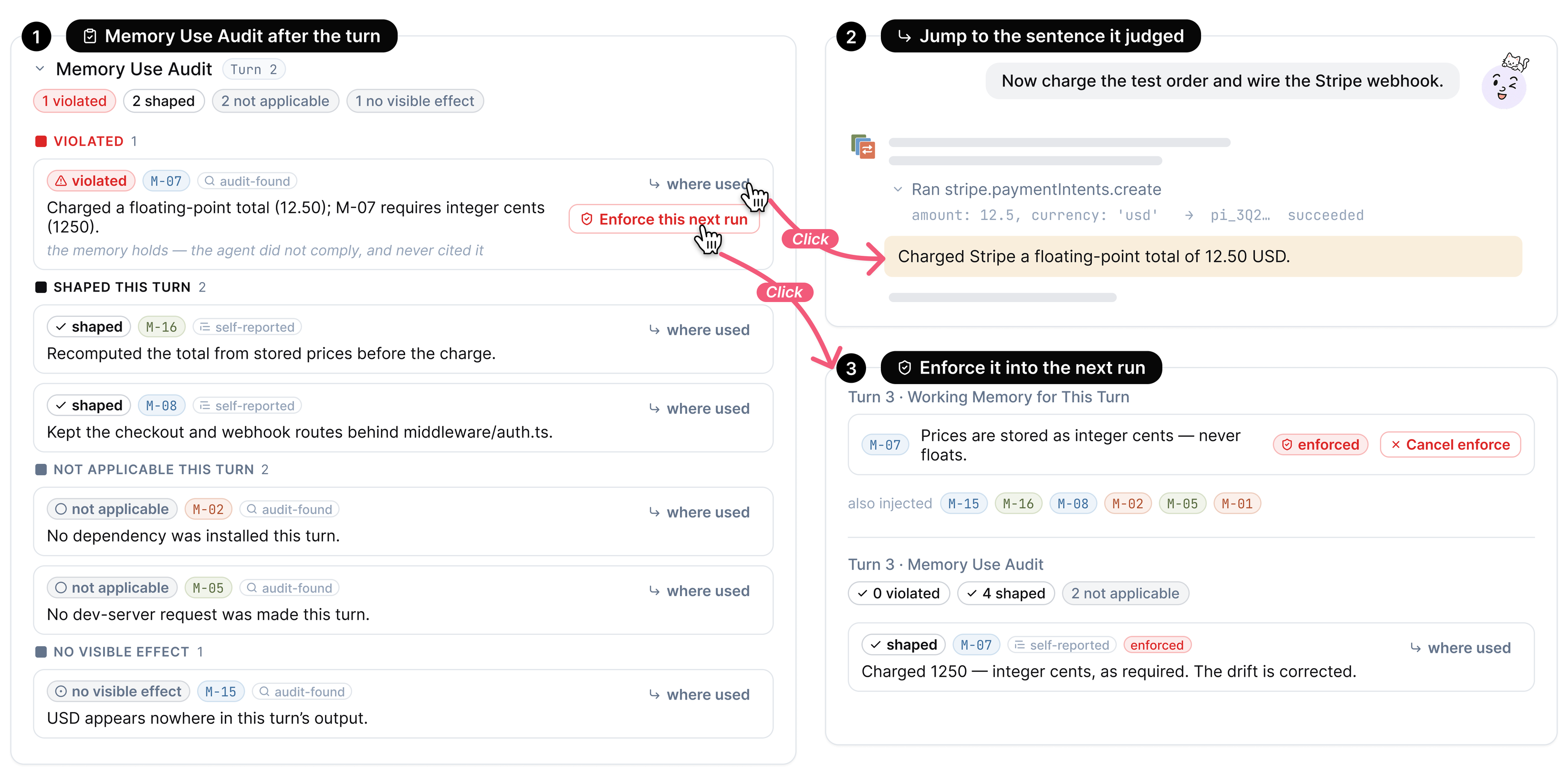}
\caption{Auditing Ben's second turn and carrying a rule into the third. \stepmark{1}~Each injected item has an audit verdict and a source label. \stepmark{2}~\textit{Where used} locates the output evidence for the integer-cents violation. \stepmark{3}~\textit{Enforce this next run} makes \memP{07} mandatory in the next turn's working memory; the subsequent audit reports it as \textit{shaped}.}
    \label{fig:impact}
    \Description{Three panels. Panel 1 lists injected memories with verdicts and source labels. M-07 is marked violated and audit-found, with Where used and Enforce this next run controls. Other rows show shaped, not applicable, and no visible effect. Panel 2 highlights a sentence reporting 12.50 USD above a tool call containing amount: 12.5 and currency: 'usd'. Panel 3 shows M-07 as an enforced working-memory item with a Cancel enforce control and a subsequent audit reporting it as shaped with an amount of 1250 cents.}
    \vspace{-5pt}
\end{figure*}

Ben clicks \textit{Stop} beside the citation and then \textit{Confirm}, interrupting the current agent run. The recovery card quotes the associated sentence (Figure~\ref{fig:execution}, \stepmark{2}). In its correction field, he writes ``Never trust the client's total, recompute it from stored prices before creating the session.'' He selects \textit{Enforce this run}, which makes the correction mandatory for the resumed run, and clicks send to submit it. The coding agent continues with the corrective prompt (Section~\ref{s:implementation}), recomputes the total from stored prices (\memS{16}), creates the checkout session in integer cents (\memP{07}), and runs the tests. The corrected content of \memS{16} is stored back to the Memory Board \circmark{B}.

\subsection{Impact: Auditing Memory Use and Enforcing a Rule}
\label{s:impact}

After execution, \textit{Memory Use Audit} reports one verdict for each injected item: \textit{violated} (the agent did not follow it), \textit{shaped} (it changed what the agent did), \textit{not applicable} (nothing in the turn required it), or \textit{no visible effect} (Figure~\ref{fig:impact}, \stepmark{1}). A separate conversation branch in the same coding-agent runtime reviews the full exchange against the injected memories, allowing the audit to identify violations even when the agent did not cite the item (Section~\ref{s:implementation}). Results from this check are marked \textit{audit-found}; those based on the agent's citations are marked \textit{self-reported}.

In the second turn, Ben asks the agent to charge a test order without inspecting its intermediate actions. The audit flags \memP{07} as \textit{violated} and \textit{audit-found}: the transcript shows \texttt{amount: 12.5} for an order of USD~12.50, conflicting with the stored rule's required integer-cent representation, \texttt{1250}. Ben clicks \textit{Where used} to locate this evidence (Figure~\ref{fig:impact}, \stepmark{2}), then \textit{Enforce this next run} to require \memP{07}'s inclusion in the next turn's working memory. In the third turn, only Ben can remove this enforced item; the audit reports it as \textit{shaped}, with the amount now recorded as 1250 cents (\stepmark{3}). Audit verdicts and enforcement actions enter the item's history. Repeated violations trigger a proposal to revise the stored item in \textit{Review Suggested Changes to Existing Memories}, the third review step before the next execution (Figure~\ref{fig:evolution}, \stepmark{3}).

\subsection{Memory Board and Memory Record}
\label{s:board}

\begin{figure*}[t]
    \centering
    \includegraphics[width=\textwidth]{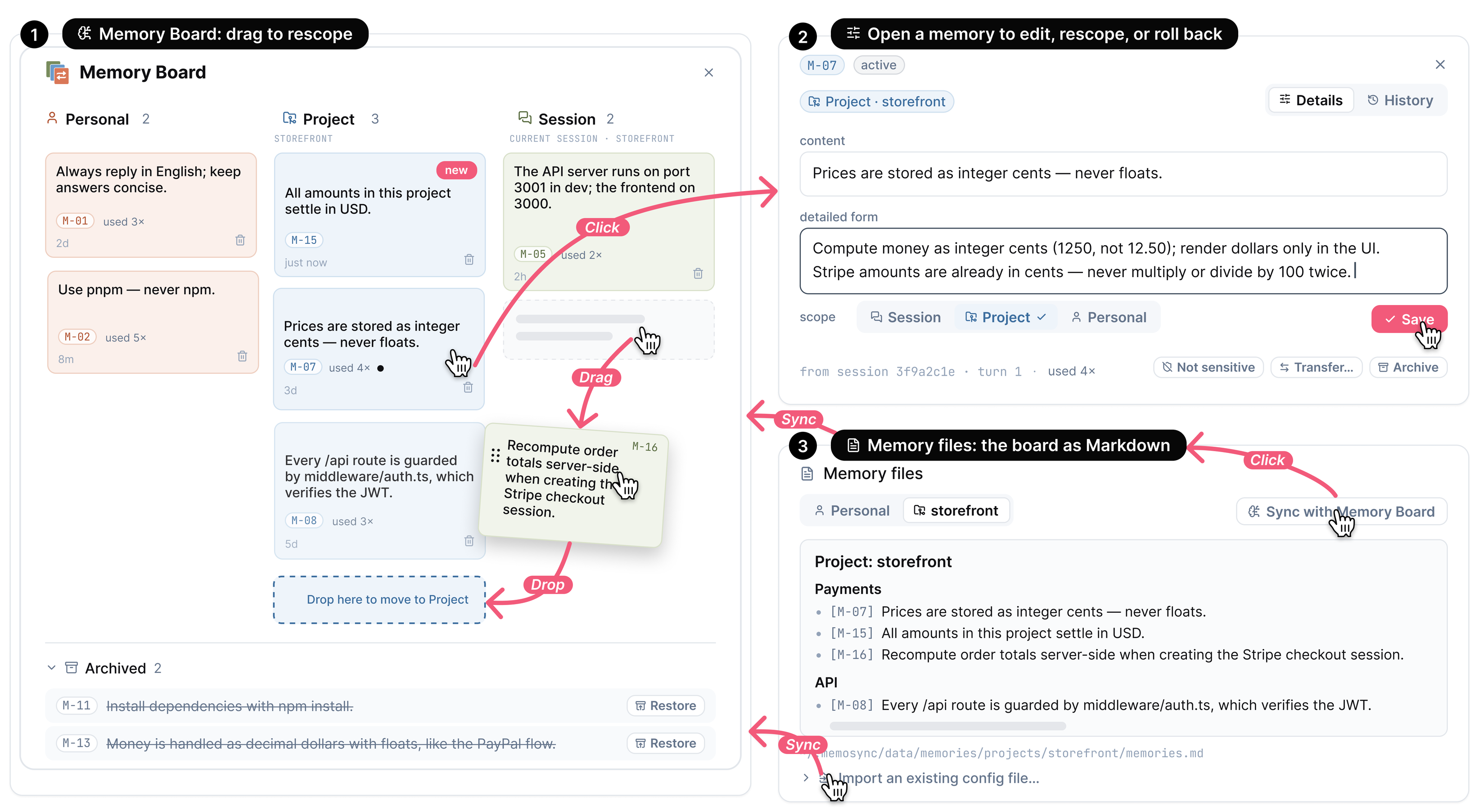}
    \caption{Managing memories across turns. \stepmark{1}~The \textit{Memory Board} groups items by scope and supports rescoping by drag and drop. Archived items remain restorable. \stepmark{2}~The detail panel supports editing an item's summary and description, changing its scope, transferring or archiving it, and reviewing its \textit{History}. \stepmark{3}~\textit{Memory files} provides a synchronized Markdown view and file import.}
    \label{fig:memory_board}
    \Description{Three panels. Panel 1 shows Personal, Project, and Session columns, with M-16 being dragged toward Project and two retired items in Archived. Panel 2 shows M-07's summary, an editable detailed description, a Save button, scope selectors, Transfer and Archive controls, and Details and History tabs. Panel 3 shows a Markdown view grouped by topic, a Sync with Memory Board button, a file path, and an import control.}
\end{figure*}

After the three turns, Ben opens the \textit{Memory Board} (Figure~\ref{fig:memory_board}, \stepmark{1}). It organizes his personal memories, \textit{storefront}'s project memories, and the active session's memories into separate columns for direct review and editing. To retain \memS{16} for future work on \textit{storefront}, he drags it from Session to Project. Ben opens \memP{07}'s detail panel and clarifies its detailed description: Stripe amounts are already in cents and should not be multiplied or divided by 100 twice. He clicks \textit{Save} (Figure~\ref{fig:memory_board}, \stepmark{2}). The \textit{History} tab traces changes to this individual item, including its conflict with \memS{13}, the audit-found violation, and the enforcement action. Retired items remain available in \textit{Archived} and can be restored.

\begin{figure*}[t]
    \centering
\newlength{\orchH}\setlength{\orchH}{0.850175\textwidth}\newcommand{\orchlink}[5]{\node[anchor=south west,inner sep=0] at (#1,#2) {\hyperref[#5]{\phantom{\rule{\dimexpr#3\textwidth\relax}{\dimexpr#4\orchH\relax}}}};}
    \begin{tikzpicture}
      \node[anchor=south west,inner sep=0] (orch) {\includegraphics[width=\textwidth]{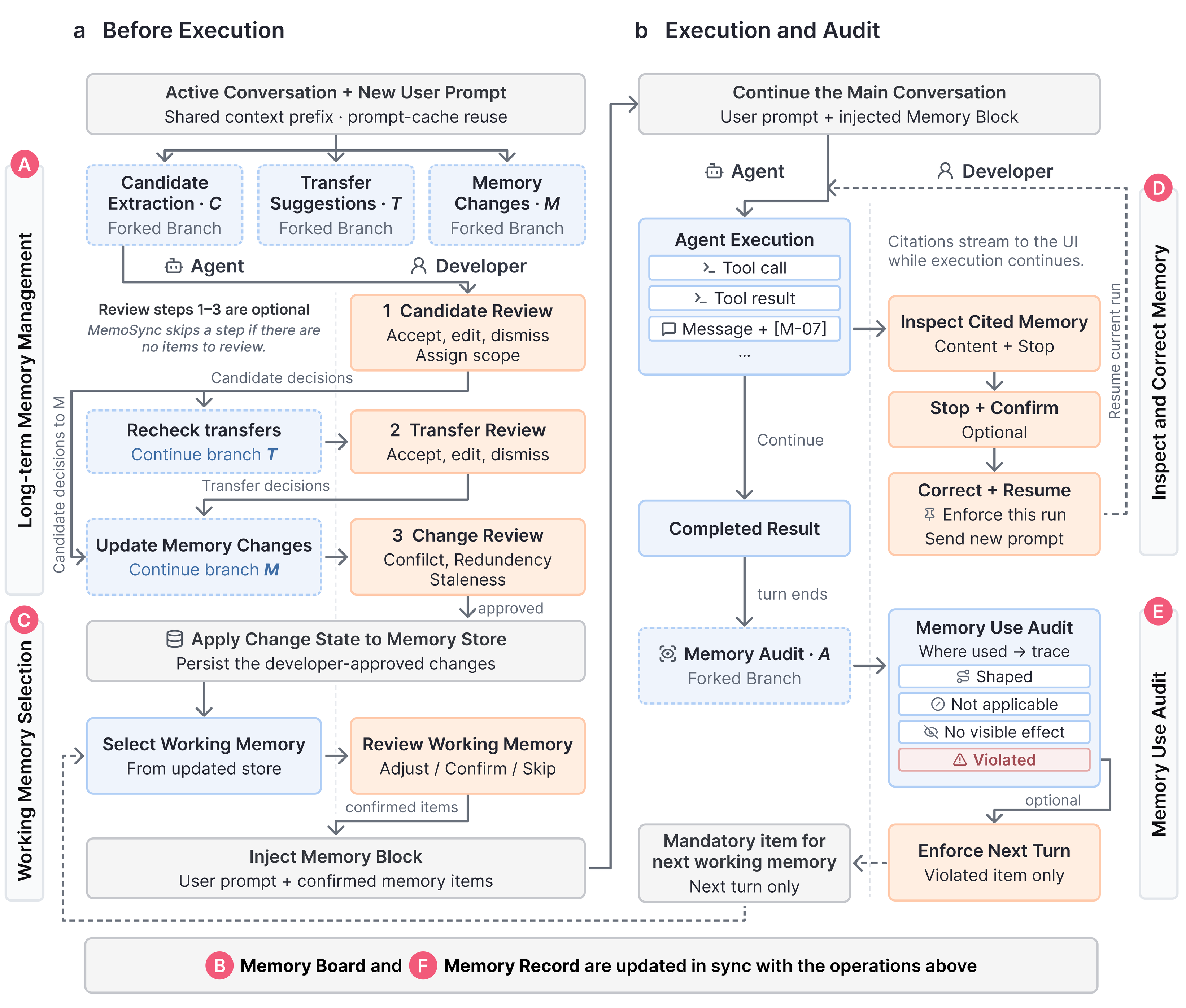}};
      \begin{scope}[x={(orch.south east)},y={(orch.north west)}]
        \orchlink{0.006}{0.411}{0.029}{0.441}{s:evolution}  \orchlink{0.006}{0.108}{0.029}{0.292}{s:selection}  \orchlink{0.962}{0.451}{0.030}{0.377}{s:execution}  \orchlink{0.962}{0.110}{0.030}{0.298}{s:impact}     \orchlink{0.171}{0.024}{0.139}{0.036}{s:board}      \orchlink{0.343}{0.024}{0.148}{0.036}{s:record}     \end{scope}
    \end{tikzpicture}
    \caption{MemoSync's memory orchestration. \textbf{(a)}~Parallel branches prepare proposals for sequential developer review; approved changes update the store before working-memory selection and injection. \textbf{(b)}~Execution supports interruption and correction, followed by an audit branch and optional next-turn enforcement. The \textit{Memory Board} and \textit{Memory Record} synchronize across both stages. Blue boxes denote agent operations or results, orange boxes denote developer actions, and dashed borders indicate forked conversations. Dashed arrows show current-run recovery and next-turn enforcement. In the PDF, lettered regions link to subsections.}
    \label{fig:implementation}
    \Description{Two-panel workflow. The left panel shows parallel branches for candidate extraction, transfer suggestions, and memory changes, followed by sequential reviews. Earlier review decisions feed into later branches. Approved changes update the store before working-memory review and injection into the main conversation. The right panel shows agent execution, an optional interruption-and-correction loop, a post-execution audit branch, and mandatory inclusion of selected violated items in the next turn's working memory. A shared Memory Record spans both panels.}
    \vspace{-8pt}
\end{figure*}

The \textit{Memory files} card provides a synchronized Markdown view of the Board's contents (Figure~\ref{fig:memory_board}, \stepmark{3}). Ben can edit either representation; after editing the file, he clicks \textit{Sync with Memory Board} to reconcile the changes. He can also import an existing configuration file, whose entries become candidates for review.

\phantomsection\label{s:record}At the session level, the \textit{Memory Record} in the right sidebar (Figure~\ref{fig:main_ui}F) organizes memory activity by turn, recording proposals, selections, citations, interruptions, and audits under a one-line turn summary. It persists outside the coding agent's context window, so Ben and the agent can consult it after the conversation has been compacted.

\subsection{Implementation}
\label{s:implementation}

MemoSync is implemented as a TypeScript web application with a React frontend, a Bun backend, and SQLite storage for memory items and their histories. Built on Kanna\footnote{\url{https://github.com/jakemor/kanna}}, an open-source agent framework, it supports both the Claude Agent SDK\footnote{\url{https://github.com/anthropics/claude-agent-sdk-typescript}} and the Codex CLI\footnote{\url{https://github.com/openai/codex}} as the underlying coding agent. All memory operations run through the existing coding-agent runtime without requiring a separate LLM API integration. These operations use model calls in forked conversations that share the active session's context prefix, retaining task context and enabling prompt-cache reuse.

Figure~\ref{fig:implementation} shows the orchestration. A new user prompt triggers parallel branches for candidate extraction, transfer suggestions, and memory-change checks, while the frontend presents their results for sequential review. Candidate-review decisions are passed to both downstream branches, and transfer-review decisions are passed to the memory-change branch. These branches continue with the decisions incorporated. Review steps with no items are skipped, and approved changes update the store before working-memory selection. The confirmed items are injected into the main conversation as a Memory Block alongside the user prompt.

During execution, memory citations stream to the frontend, and corrective prompts allow an interrupted run to continue. After completion, a separate fork audits the full exchange against the injected memories. Developer-marked violated items are retained as mandatory entries in the next turn's working memory. All stages synchronize with the \textit{Memory Board} and \textit{Memory Record}. Further implementation details appear in Appendix~\ref{app:implementation}.
 \section{User Study}

To evaluate MemoSync and explore how memory co-management affects developer-agent memory alignment and task performance in cross-context collaboration (RQ3), we conducted a between-subjects study comparing MemoSync against two baselines: an automatic memory extraction system that represents how current coding agents manage memory autonomously (referred to as the Auto condition), and a static configuration file system that represents how developers manage agent memory through shared Markdown files (referred to as the Static condition). Participants collaborated with an AI coding agent in a browser-based coding workbench to complete the task. We measured memory alignment, monitoring accuracy, control accuracy, task performance, memory application gap, interaction activity (monitoring and control action counts), perceived workload, and system usability.

\subsection{Task Setup}

We selected two full-stack web development projects from FullStack-Bench~\cite{flu2026fullstack}: an apartment-rental application and a car-rental application (Figure~\ref{fig:task-projects}). Each project used a Next.js frontend, a NestJS backend, and a PostgreSQL database. We chose full-stack development for two reasons. First, it is a prevalent use case for AI coding agents and involves multi-file, multi-layer reasoning where accumulated context directly affects output quality. Second, the benchmark provides an official scoring rubric, enabling reproducible performance measurement. To keep the study duration manageable (approximately two hours), we limited the design to two projects with two sessions each.

The two projects share a similar application structure (browsing, booking, and transaction management) but differ in business logic and data models, so that memory from the first project could apply to the second only after adaptation. Each project was divided into two consecutive sessions. The first session established the project's core functionality, while the second extended the same codebase with additional transaction and management features. Session~2 built on the code produced in Session~1: the two sessions within a project shared the same workspace directory, and Session~2 did not reset the codebase. Session~1 was scored from its own frozen snapshot, so subsequent changes in Session~2 did not retroactively affect Session~1 scores. Session~2 was evaluated on the cumulative workspace. The four sessions had different maximum scores (9, 5, 11, and 5 points, respectively; see Appendix~\ref{app:rubric} for the full rubric), but the required knowledge and task complexity are equivalent, as our task difficulty evaluation shows (Appendix~\ref{app:baseline}).

Project order was counterbalanced across participants, with Session~1 always preceding Session~2 within each project. This design provided both within-project continuity for memory accumulation and a cross-project boundary for memory transfer. In each session, participants read the task brief, described the requirements to the agent in natural language, iterated on the implementation, and verified the result through an embedded browser preview. At the end of the session, they submitted a frozen workspace snapshot as the final artifact (see Section~\ref{sec:procedure} for the detailed flow).

To calibrate task difficulty, we used the same agent configuration as in the study (DeepSeek V4 Flash, reasoning effort: high; see section~\ref{sec:condition} for the full specification) and ran it without any accumulated memory ten times per session. The mean score was approximately 50\% of the session maximum, indicating that the tasks were neither trivial nor out of reach for the agent alone (see Appendix~\ref{app:baseline} for details).

\subsection{Conditions}
\label{sec:condition}

Participants were randomly assigned to one of three conditions. All conditions used the same coding agent, browser-based workbench, and task materials, differing only in how agent memory was presented and managed. The workbench was built on the open-source Kanna framework, and each participant ran in an isolated Docker container with separate transcripts, memory data, and project workspaces. The coding agent was Claude Code (v2.1.170) via the Claude Agent SDK, using DeepSeek V4 Flash as the underlying model with high reasoning effort, and could access workspace files, the shell, and the web.

\begin{itemize}

\item \textbf{Auto (Automatic Memory).} After each successful agent turn, the agent could create, update, or reinforce memory entries, and these changes were applied automatically without participant review. Memory was stored in a SQLite database. At the start of each turn, all active memories for the current project were injected into the agent's context as a plain-text Markdown block. When participants switched projects, the first project's memory was copied once into a separate project-level store. Participants could view an LLM-generated summary of the current memory state and request memory changes through natural-language conversation, but could not directly edit individual entries (Figure~\ref{fig:baseline-conditions}a). \textbf{This condition reflects the automated memory management approach used by agents like Claude, ChatGPT, and Kimi.}\footnote{Among the major agents we examined, Kimi was the only provider that publicly documented a concrete capacity limit for persistent memory. We therefore followed Kimi's documented limits, allowing up to 50 memory entries. See \url{https://www.kimi.ai/help/features/memory-tips}.}

\item \textbf{Static (Static Configuration).} Participants and the agent shared plain-text Markdown files in the workspace root (\texttt{MEMORY.md}), similar to \texttt{CLAUDE.md}. At the start of each turn, the full contents of these files were injected into the agent's context together with a fixed maintenance instruction. Participants could edit the root \texttt{MEMORY.md} directly from the sidebar or ask the agent to edit the files through the main conversation (Figure~\ref{fig:baseline-conditions}b). \textbf{This condition represents the common practice of using configuration files such as \texttt{CLAUDE.md} to manage agent behavior across sessions and projects.}

\item \textbf{MemoSync.} Participants used the MemoSync system described in Section~\ref{s:system}, with developer review and control over memory management throughout the interaction. \textbf{This condition operationalizes monitoring and control across memory evolution, selection, execution, and impact.}

\end{itemize}

\subsection{Procedure}
\label{sec:procedure}

\begin{figure*}[t]
\centering
\includegraphics[width=\textwidth]{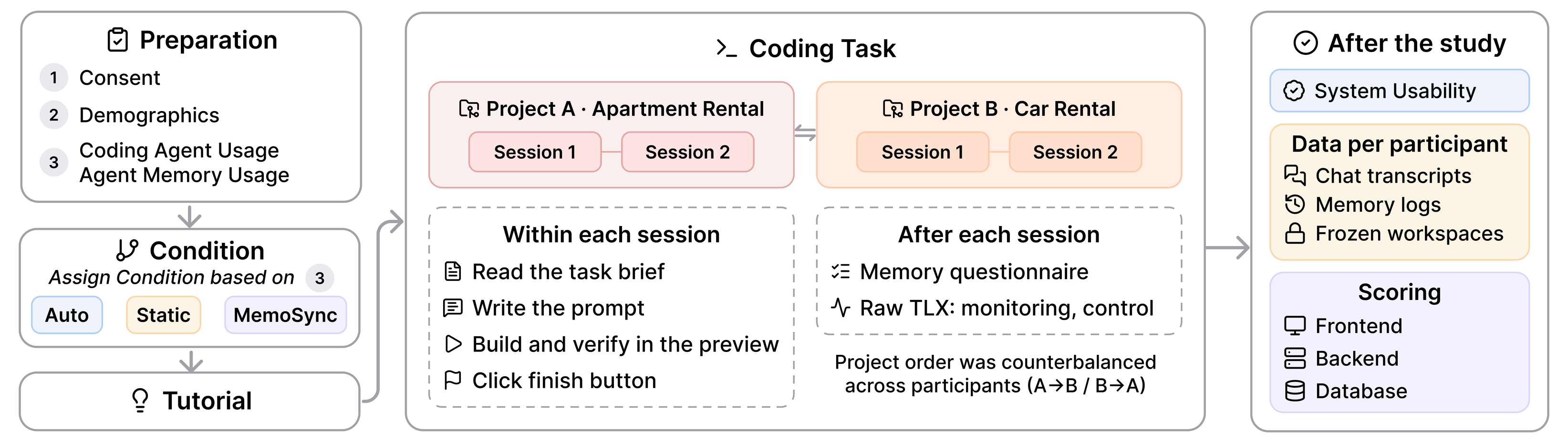}
\caption{Study procedure. Participants completed a background questionnaire, were assigned to one of three conditions, and followed a guided tutorial before the coding tasks. Each participant completed two projects (order counterbalanced), each consisting of two consecutive sessions. Within each session, participants read the task brief, prompted the agent, built and verified the result, and clicked finish. After each session, participants completed a memory questionnaire and two Raw TLX scales (monitoring, control). After all sessions, participants completed the SUS. The right panel lists collected data and scoring dimensions.}
\label{fig:procedure}
\Description{A flowchart in three panels from left to right. Left panel, Preparation: 1 Consent, 2 Demographics, 3 Coding Agent Usage and Agent Memory Usage; an arrow leads to Condition, assigned based on step 3, with three chips Auto, Static, and MemoSync; then Tutorial. Middle panel, Coding Task: Project A Apartment Rental with Session 1 and Session 2, and Project B Car Rental with Session 1 and Session 2, joined by a double arrow; below, Within each session: read the task brief, write the prompt, build and verify in the preview, click finish button; After each session: memory questionnaire and Raw TLX for monitoring and control; a note reads Project order was counterbalanced across participants (A to B / B to A). Right panel, After the study: System Usability; Data per participant: chat transcripts, memory logs, frozen workspaces; Scoring: Frontend, Backend, Database.}
\end{figure*}

Figure~\ref{fig:procedure} summarizes the study procedure. After providing informed consent, participants completed a background questionnaire collecting demographic information and assessing their experience with AI coding agents and agent memory, including AI agent usage frequency and agent memory experience on 5-point scales, as well as previously used AI agent tools. Based on these experience levels, we assigned participants to one of three conditions. Participants then completed a tutorial on the workbench, agent interaction, and the memory mechanism of their condition.

Participants completed four sessions (P1S1, P1S2, P2S1, P2S2) in sequence. Project order was counterbalanced across participants. At the start of each session, participants read a task brief describing the goals and requirements for that session (Appendix~\ref{app:task-specs}). The interface disabled copy, cut, context menus, and drag on the brief text, and the server rejected prompts with near-verbatim overlap with the brief. At the end of each session, participants clicked Finish this session, which froze the workspace code as the submission artifact and set the chat to read-only.

After each session, participants completed the Memory Questionnaire (Section~\ref{sec:measures}) and two Raw TLX questionnaires~\cite{hart2006nasa} measuring monitoring and control workload separately; each questionnaire included one instructed-response attention check; in the Memory Questionnaire, the check was formatted as a memory item so that participants who rated items without reading them would fail it. After all four sessions, participants completed the SUS~\cite{brooke1996sus}.

\subsection{Participants}

We used G*Power~\cite{faul2009statistical} to determine the required sample size for a three-condition between-subjects design, with an expected effect size of $f = 0.30$, $\alpha = 0.05$, and power $1 - \beta = 0.80$. The analysis indicated a minimum of 111 participants (37 per condition) needed. After obtaining institutional ethics approval, we recruited 136 participants through Prolific\footnote{https://www.prolific.com/}, an online participant recruitment platform. Eligibility criteria required experience with at least one AI coding assistant (GitHub Copilot, Cursor, Claude Code, Codex, Windsurf, Amazon Q Developer, Gemini Code Assist, JetBrains AI, Tabnine, or other) and English fluency, as the system and tasks were presented in English. We excluded four participants who did not pass all four attention checks or did not complete all four sessions with valid submissions (i.e., at least one participant-origin prompt and one successful agent turn per session, a frozen workspace snapshot, and completed Memory Questionnaire, Monitoring Raw TLX, and Control Raw TLX for each session). The final sample comprised 132 participants (44 per condition). Among them, 77 identified as male and 55 as female. Ages ranged from 18 to 61 ($M = 30.13$, $SD = 9.18$). Participants were assigned to conditions balanced by their AI agent usage frequency and memory experience. Kruskal--Wallis tests did not detect group differences in AI agent usage frequency ($H(2)=3.78$, $p=.151$, $\epsilon^2=.01$) or agent memory experience ($H(2)=1.06$, $p=.589$, $\epsilon^2=.00$). The study took a median of 131 minutes end-to-end. Participants received a completion code after submitting the final SUS and were compensated at approximately \$10 per hour upon approval of their Prolific submissions.

\subsection{Measures}
\label{sec:measures}

We collected three types of data: (1)~system logs recording the memory the agent actually held and participant monitoring and control action counts; (2)~participant questionnaires, including the post-session Memory Questionnaire (Q1--Q3), two Raw NASA TLX scales (monitoring and control), and the System Usability Scale; and (3)~task artifacts, the frozen workspace code snapshot from each session. From these data we derived memory measures (Table~\ref{tab:measures}), task performance measures, and behavioral and subjective measures (Table~\ref{tab:other-measures}).

\subsubsection{Preliminary}

We use memory items as the unit of analysis, considering both their content and their scope of application. A memory item is represented as a tuple $M_i=(c_i,s_i)$, where the \textbf{content} $c_i$ is the stored text (e.g., ``commit messages should be concise''), and the \textbf{scope} $s_i$ indicates where that content applies: Personal, Project, or Session. The index $i$ identifies the memory item version being evaluated. To distinguish memory content from its evaluation, we use $c$ for text and $q$ for questionnaire or evaluator ratings on a 1--5 scale. For measures that assess both content and scope, we report Content, Scope, and Content+Scope, and treat Content+Scope as the primary reported dimension.

\begin{table*}[t]
  \centering
  \small
  \setlength{\tabcolsep}{5pt}
  \renewcommand{\arraystretch}{1.15}
  \caption{Per-item quantities used by the memory measures. Q1--Q3 are the three questions of the post-session Memory Questionnaire, which displayed every currently stored memory item. Ratings $q$ use a 1--5 scale; scopes $s$ take the values Session, Project, or Personal.}
  \label{tab:quantities}
  \begin{tabularx}{\textwidth}{@{}>{\raggedright\arraybackslash}p{3.1cm} >{\raggedright\arraybackslash}p{1.5cm} >{\hsize=1.3\hsize\raggedright\arraybackslash}X >{\hsize=0.7\hsize\raggedright\arraybackslash}X@{}}
    \toprule
    \textbf{Quantity} & \textbf{Source} & \textbf{Content} & \textbf{Scope} \\
    \midrule
    Object-Level Memory\newline $O_i=(c_i^O,s_i^O)$ & System log & $c_i^O$: the stored text of the item, as the agent actually held it & $s_i^O$: the stored scope \\
    \addlinespace[4pt]
    Desired Memory\newline $D_i=(c_i^D,s_i^D)$ & Q1 & $q_i^D$: how closely $c_i^O$ matched what the participant wanted stored. A rating of 5 means $c_i^D=c_i^O$; for lower ratings, an optional free-text field specifies $c_i^D$ & $s_i^D$: the scope the participant wanted \\
    \addlinespace[4pt]
    Assessed Memory\newline $A_i=(c_i^A,s_i^A)$ & Q2 & $q_i^A$: how completely and accurately the participant believed the agent remembered $c_i^O$, with an ``I do not know'' option. A rating of 5 means $c_i^A=c_i^O$, and 1 means the item is absent; for 2 to 4, an optional field describes $c_i^A$ & $s_i^A$: the scope the participant believed the agent had assigned \\
    \addlinespace[4pt]
    Perceived Memory Application\newline $P_i=q_i^P$ & Q3 & $q_i^P$: the extent to which the participant believed the item had been applied in the session's output, with a ``Not applicable'' option & --- \\
    \addlinespace[4pt]
    Actual Memory Application\newline $R_i=(q_i^R,s_i^R)$ & LLM evaluator & $q_i^R$: the extent to which the desired content $c_i^D$ was realized in the frozen task artifact & $s_i^R$: 1 if the application conformed to $s_i^D$, else 0 \\
    \bottomrule
  \end{tabularx}
\end{table*}

\begin{figure*}[t]
  \centering
  \includegraphics[width=\textwidth]{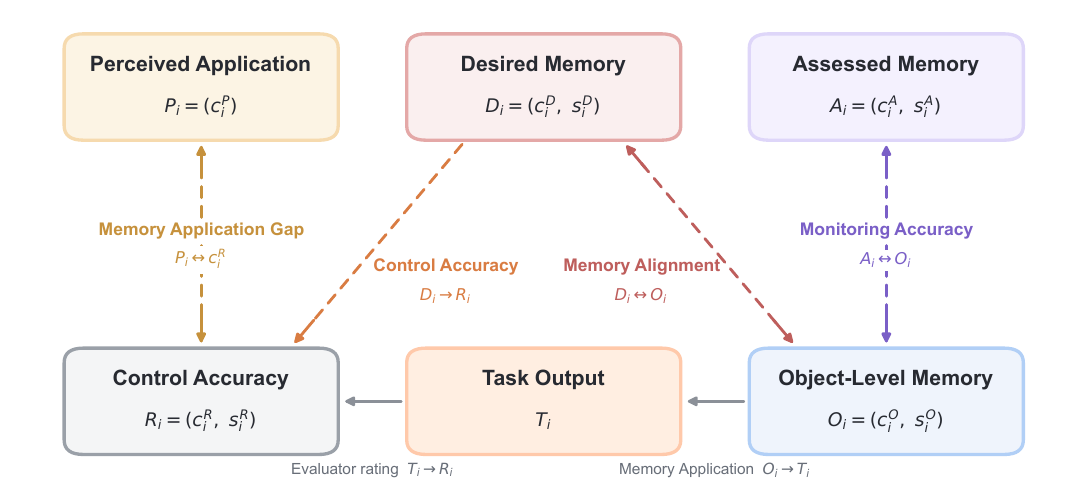}
  \caption{Measurement map. Top row: participant reports elicited by questionnaire; bottom row: observable states. Double-headed dashed arrows are computed comparisons between two ratings of the same reference; the single-headed Control arrow indicates that the evaluator produces $R_i$ by rating each desired item's realization in the task output.}
  \label{fig:measurement-map}
  \Description{A diagram of six boxes in two rows. Top row: Perceived Application P (yellow), Desired Memory D with content and scope (red), and Assessed Memory A with content and scope (blue). Bottom row: Control Accuracy R with a rating and scope (gray), Task Output T (orange), and Object-Level Memory O with content and scope (blue). Dashed double-headed arrows connect P and R (Memory Application Gap), D and O (Memory Alignment), and A and O (Monitoring Accuracy). Solid single-headed arrows run from D to R (Control Accuracy), from O to T (Memory Injection), and from T to R (Evaluator rating).}
\end{figure*}

\begin{table*}[!t]
  \centering
  \small
  \setlength{\tabcolsep}{5pt}
  \renewcommand{\arraystretch}{1.15}
  \caption{Memory measures, computed per item from the quantities in Table~\ref{tab:quantities}. Content scores are normalized to $[0,1]$, Scope scores are binary, and Content+Scope is their product. $^{a}$\,$N=132$. $^{b}$\,$N=123$ (see footnote~\ref{fn:n123}).}
  \label{tab:measures}
  \begin{tabularx}{\textwidth}{@{}>{\raggedright\arraybackslash}p{3.0cm} >{\raggedright\arraybackslash}p{3.5cm} >{\raggedright\arraybackslash}X@{}}
    \toprule
    \textbf{Measure} & \textbf{Scoring} & \textbf{Operationalization} \\
    \midrule
    Monitoring Accuracy$^{a}$\newline $A_i\leftrightarrow O_i$ & Content: $(q_i^A-1)/4$ \newline Scope: 1 if $s_i^A=s_i^O$, else 0 & How accurately the participant assessed the agent's actual memory state. All displayed items were fully stored in the logged memory, so an accurate assessment of an item's content corresponds to a Q2 rating of 5. \\
    \addlinespace[4pt]
    Memory Alignment$^{a}$\newline $D_i\leftrightarrow O_i$ & Content: $(q_i^D-1)/4$ \newline Scope: 1 if $s_i^D=s_i^O$, else 0 & The agreement between the agent's actual memory and what the participant wanted it to retain. Q1 directly assessed how closely the displayed stored content matched the participant's expectations, so its rating serves as the Content score. \\
    \addlinespace[4pt]
    Control Accuracy$^{b}$\newline $D_i\rightarrow R_i$ & Content: $(q_i^R-1)/4$ \newline Scope: $s_i^R$ & The extent to which the participant's desired memory was realized in the frozen task artifact. For each desired item $D_i$, an LLM evaluator (GPT-5.6 Sol, extra-high) examined the artifact against $c_i^D$ and $s_i^D$. Two researchers independently annotated a random sample of 200 items ($\kappa=.83$)~\cite{landis1977kappa}; disagreements were adjudicated, and the evaluator agreed with the adjudicated labels on 97.5\% of these items. \\
    \addlinespace[4pt]
    Memory Application Gap$^{b}$\newline $P_i\leftrightarrow R_i$ & Content: $|q_i^P-q_i^R|$ & The discrepancy between participant-perceived and actual memory application, on the original 1--5 scales. \\
    \bottomrule
  \end{tabularx}
\end{table*}

\begin{table*}[!t]
  \centering
  \small
  \setlength{\tabcolsep}{5pt}
  \renewcommand{\arraystretch}{1.15}
  \caption{Task performance measures (top) and behavioral and subjective measures (bottom).}
  \label{tab:other-measures}
  \begin{tabularx}{\textwidth}{@{}>{\raggedright\arraybackslash}p{2.8cm} >{\raggedright\arraybackslash}p{1.9cm} >{\raggedright\arraybackslash}p{3.0cm} >{\raggedright\arraybackslash}X@{}}
    \toprule
    \textbf{Measure} & \textbf{Data} & \textbf{Scoring} & \textbf{Operationalization} \\
    \midrule
    \tabgroup{4}{Task performance}
    Overall Performance & Frozen workspaces & Sum of the four session scores divided by the 30-point maximum & Each session's frozen workspace was evaluated with the benchmark's official pipeline and rubric, covering frontend, backend, and database criteria (Appendices~\ref{app:rubric} and~\ref{app:eval-pipeline}), with GPT-5.6 Sol (extra-high reasoning) as the evaluator model. Each rubric item was scored YES (1~point), PARTIAL (0.5, frontend only), or NO (0). \\
    \addlinespace[4pt]
    Session / Project Performance & Frozen workspaces & Session or project total as a percentage of its maximum, centered on the mean of all participants on the same benchmark session or project & Session maxima were 9 and 5~points in the apartment-rental and 11 and 5~points in the car-rental. Because project order was counterbalanced, we report deviations in percentage points from the same-task mean; changes across consecutive sessions are differences between these deviations. \\
    \midrule
    \tabgroup{4}{Behavioral and subjective measures}
    Monitoring / Control Activity & System logs & Cumulative counts across the four sessions & Monitoring actions (e.g., viewing the memory panel, expanding memory details) and control actions (e.g., approving memory proposals, editing memory content, adjusting scope). \\
    \addlinespace[4pt]
    Monitoring / Control Workload & Raw TLX, per session & Raw TLX score & Two questionnaires~\cite{hart2006nasa} targeting monitoring and control separately. \\
    \addlinespace[4pt]
    System Usability & SUS, after Session 4 & SUS score & System Usability Scale~\cite{brooke1996sus}. \\
    \bottomrule
  \end{tabularx}
\end{table*}

\subsubsection{Memory measures}

To examine how memory co-management affected developer-agent memory alignment and developers' monitoring and control of agent memory, we proposed four measures. Each compares two of the five per-item quantities listed in Table~\ref{tab:quantities}: the Object-Level Memory $O_i$ recorded in system logs, three participant reports collected by the post-session Memory Questionnaire, and the Actual Memory Application $R_i$ rated by an LLM evaluator. Figure~\ref{fig:measurement-map} illustrates the relationships between these data and the four measures. Following the metamemory framework, control refers to meta-level regulation of object-level actions~\cite{nelson1990metamemory}. In this study, we assessed the outcome of this regulation by examining whether developers' memory expectations were realized in the final task artifact, and therefore operationalized Control Accuracy as Actual Memory Application. Nine participants had no items eligible for this evaluation, leaving $N=123$ for Control Accuracy and Memory Application Gap.\footnote{\label{fn:n123}Evaluating actual memory application requires explicitly specified desired content $c_i^D$. When the Q1 rating $q_i^D=5$, the displayed content can serve directly as the evaluation target; lower ratings require the participant to supply corrected content. We kept the correction field optional because requiring it could encourage participants to rate items as~5 to avoid writing corrections. Nine participants rated every item between~1 and~4 across all four sessions without providing any corrected text, leaving the desired content unspecified for all their items and preventing actual-application evaluation. On this basis, these participants were excluded only from Control Accuracy, Memory Application Gap and Mediation Analysis.} ``I do not know'' responses in Q2 were excluded from the primary scores (2.1\% of all ratings).

\subsubsection{Task performance and behavioral measures}

To explore how memory co-management affected task performance, we scored each session's frozen workspace with the benchmark's official evaluation pipeline and rubric, and report overall performance as well as session- and project-level deviations that account for the counterbalanced project order. To address what workload and effort accompanied these outcomes, we counted monitoring and control actions in the system logs and collected the Raw TLX after each session and the SUS after all four sessions.

\subsection{Analysis}

We tested normality with Shapiro--Wilk separately within each condition (42 tests: 14 measures $\times$ 3 conditions). Thirty-one tests were significant. We therefore used non-parametric tests for group comparisons, except for the count-based activity measures modeled separately below. Per-measure $W$ and $p$ values are reported in the supplementary material. Three-group comparisons used Kruskal--Wallis $H$, followed by all three pairwise Mann--Whitney $U$ tests (MemoSync vs.\ Auto, MemoSync vs.\ Static, Auto vs.\ Static). We applied Holm correction to the three pairwise comparisons for each measure at each level of aggregation, but did not apply additional correction across measures or positions. All pairwise $p$ values reported below are Holm-corrected. Omnibus effect sizes are reported as $\epsilon^2$, and pairwise effect sizes as signed rank-biserial $r_{rb}$. The supplementary material reports the full test statistics for every outcome, the negative-binomial models for the activity counts, all components of the mediation model, and the per-measure normality tests. \textbf{Monitoring Activity and Control Activity} were overdispersed count measures, with variance-to-mean ratios of 13.7--27.4 and 2.4--6.4, respectively. We therefore modeled both using negative-binomial regression with HC3 robust contrasts. We report incidence rate ratios (IRRs) with Holm-corrected pairwise contrasts. \textbf{The exploratory mediation analysis} treated Memory Alignment (Content+Scope) as the outcome and Monitoring Accuracy and Control Accuracy as simultaneous mediators. Path coefficients were estimated with OLS and HC3 robust standard errors; indirect effects were tested with BCa 95\% CIs from 50{,}000 condition-stratified bootstrap resamples~\cite{hayes2018introduction}. This analysis used the 123 participants with complete scores (see the footnote in the Measures subsection for the derivation of $N=123$). For the two workload measures, we additionally ran equivalence tests (TOST)~\cite{lakens2017equivalence} with a margin of $d=0.5$ to test whether MemoSync's workload was equivalent to each baseline within a medium effect. All analyses used Python 3.12.7 with SciPy 1.16.2 and statsmodels 0.14.5. 

\section{Results}
\label{s:results}

To address RQ3, we organize the results around three aspects. First, how did memory co-management affect developer-agent Memory Alignment across sessions and project boundaries? Second, how did memory co-management affect Task Performance? Third, what workload and effort accompanied these outcomes? 

\begin{figure*}[t]
\centering
\includegraphics[width=\textwidth]{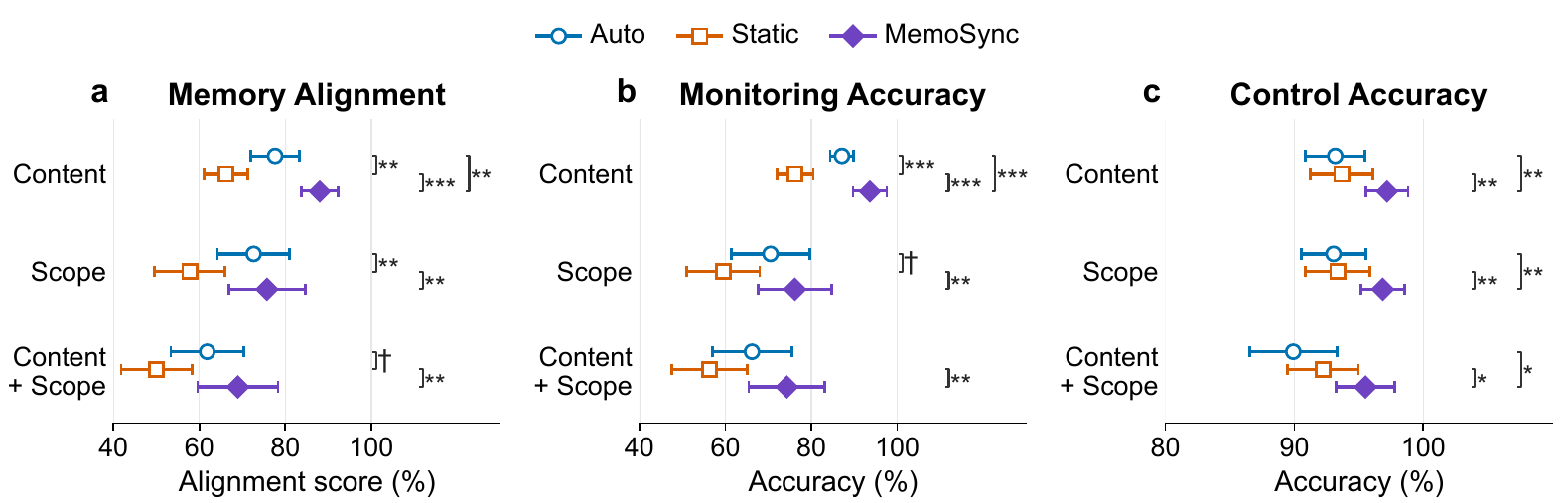}
\caption{Memory Alignment, Monitoring Accuracy, and Control Accuracy across Content, Scope, and Content+Scope dimensions. Markers show condition means, and error bars show 95\% confidence intervals. Brackets show all three pairwise Mann--Whitney $U$ tests with Holm correction within each measure. ${}^{***}p<.001$, ${}^{**}p<.01$, ${}^{*}p<.05$, ${}^{\dagger}p<.10$.}
\label{fig:three-views}
\Description{Three horizontal dot plots with 95\% confidence-interval bars, sharing a legend of Auto (blue circle), Static (orange square), and MemoSync (purple diamond). Each panel has three rows: Content, Scope, and Content + Scope. (a) Memory Alignment, x-axis 40 to 100\%: MemoSync's marker lies to the right of both baselines in every row, with Content near 87\%. (b) Monitoring Accuracy, x-axis 40 to 100\%: MemoSync's Content marker is near 95\% and clearly separated from both baselines. (c) Control Accuracy, x-axis 80 to 100\%: all three conditions are above 88\%, with MemoSync furthest right. Brackets at the right of each row mark significant pairwise differences.}
\end{figure*}

\begin{figure*}[t]
\centering
\includegraphics[width=\textwidth]{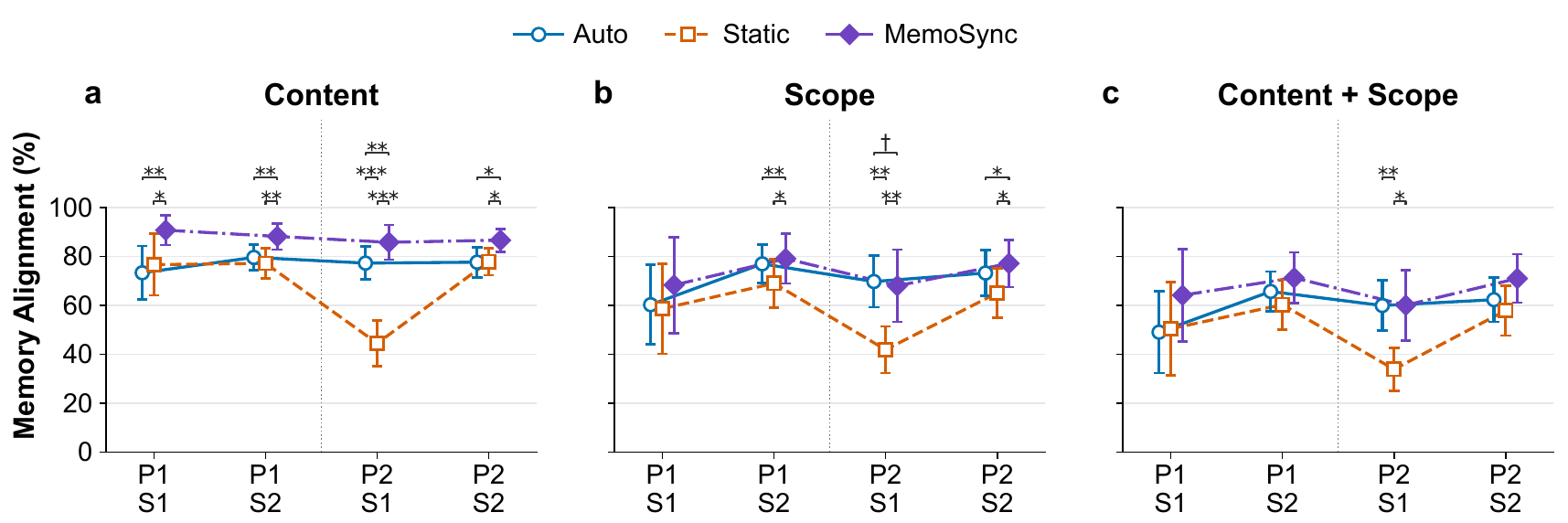}
\caption{Memory Alignment at P2S1 was lower under Static across Content, Scope, and Content+Scope. Points show condition means; lines connect the ordered session means, and error bars show 95\% confidence intervals. P1 and P2 denote the first and second projects in each participant's counterbalanced order. Brackets show all three pairwise Mann--Whitney $U$ tests with Holm correction within each dimension and session position. ${}^{***}p<.001$, ${}^{**}p<.01$, ${}^{*}p<.05$, ${}^{\dagger}p<.10$.}
\label{fig:alignment-across-sessions}
\Description{Three line panels for (a) Content, (b) Scope, and (c) Content + Scope, with Memory Alignment (\%) on the y-axis from 0 to 100 and the four sessions P1S1, P1S2, P2S1, P2S2 on the x-axis. A dotted vertical line between P1S2 and P2S1 marks the project boundary. MemoSync (purple diamonds, dashed line) stays between about 80 and 90\% in all four sessions of panel (a) and above 60\% in panels (b) and (c). Auto (blue circles) stays between about 65 and 80\%. Static (orange squares) is close to Auto in the first project and drops at P2S1 to about 40\% in (a) and (b) and about 30\% in (c), then returns to the earlier level at P2S2. Brackets above P2S1 mark significant differences between Static and the other two conditions.}
\end{figure*}

\subsection{Developer-Agent Memory Alignment and Associated Metamemory Processes}

This subsection examines how memory co-management affected developer-agent memory alignment and the metamemory processes underlying it. We first report condition differences in memory alignment, monitoring accuracy, and control accuracy across the content, scope, and content+scope dimensions, and then test through mediation analysis how monitoring and control each contributed to the observed alignment differences.

\textbf{Finding 1:} \textbf{MemoSync participants achieved the highest memory alignment, with the advantage most pronounced on the content.} A Kruskal--Wallis test showed a significant effect of condition on Memory Alignment Content, $H(2) = 35.86$, $p < .001$, $\epsilon^2 = .26$. MemoSync ($M = 88.0\%$, $SD = 13.9$) was significantly higher than Auto ($M = 77.6\%$, $SD = 18.8$; $U = 1348$, $p = .002$, $r_{rb}=.39$) and Static ($M = 66.2\%$, $SD = 16.9$; $U = 1651.5$, $p < .001$, $r_{rb}=.71$; Figure~\ref{fig:three-views}a). On Scope ($H(2) = 14.32$, $p < .001$, $\epsilon^2 = .10$) and Content+Scope ($H(2) = 11.58$, $p = .003$, $\epsilon^2 = .07$), MemoSync was significantly higher than Static (Scope: $M = 75.7$ vs.\ $57.8$, $p = .002$, $r_{rb}=.42$; Content+Scope: $M = 69.0$ vs.\ $50.0$, $p = .005$, $r_{rb}=.39$) but not than Auto (Scope: $p = .126$; Content+Scope: $p = .101$), though the mean was higher.

Session-level trajectories reveal where each dimension's differences emerged (Figure~\ref{fig:alignment-across-sessions}). On Content, MemoSync was significantly higher than both baselines at every session position (all $p \leq .031$, $r_{rb}=.31$--$.73$). On Scope, MemoSync was significantly higher than both baselines within each project's second session (P1S2: $r_{rb}=.41$ and $.29$; P2S2: $r_{rb}=.29$ and $.33$), but at the project boundary (P2S1) the advantage held only over Static ($p = .006$, $r_{rb}=.40$); the MemoSync mean ($M = 68.0\%$) was slightly below Auto's ($M = 69.7\%$; $p = .075$). On Content+Scope, the only significant pairwise difference appeared at P2S1, where both Auto and MemoSync exceeded Static ($r_{rb}=.41$ and $.36$).

The scope pattern likely reflects the structural similarity between the two projects. Both were rental applications sharing the same technology stack and application architecture. Under these conditions, carrying forward prior memory without scope distinctions introduced few conflicts, leaving little room for MemoSync's explicit scope mechanism to outperform automatic carryover. Static's lower scope alignment, by contrast, likely reflects progressive configuration expansion: as the shared Markdown file accumulated entries across sessions, participants lost track of its contents.

\textbf{Finding 2:} \textbf{Control accuracy was significantly higher under MemoSync across all dimensions; monitoring accuracy gains were concentrated on the content dimension.} Kruskal--Wallis tests were significant for Control Accuracy on all three dimensions (Content: $H(2) = 13.84$, $p < .001$, $\epsilon^2 = .10$; Scope: $H(2) = 11.96$, $p = .003$, $\epsilon^2 = .08$; Content+Scope: $H(2) = 9.61$, $p = .008$, $\epsilon^2 = .06$). MemoSync was significantly higher than both baselines on all three dimensions (Content: $p = .003$ and $.003$, $r_{rb}=.40$ and $.42$; Scope: $p = .009$ and $.004$, $r_{rb}=.36$ and $.40$; Content+Scope: $p = .014$ and $.024$, $r_{rb}=.35$ and $.32$; Figure~\ref{fig:three-views}c), although absolute levels were high in every condition ($M \geq 89.9\%$). On Monitoring Accuracy Content ($H(2) = 47.78$, $p < .001$, $\epsilon^2 = .35$), MemoSync ($M = 93.6\%$, $SD = 13.0$) was significantly higher than both Auto ($M = 87.1\%$, $SD = 9.0$; $U = 1548$, $p < .001$, $r_{rb}=.60$) and Static ($M = 76.2\%$, $SD = 13.9$; $U = 1688.5$, $p < .001$, $r_{rb}=.74$; Figure~\ref{fig:three-views}b). On Scope and Content+Scope, MemoSync means remained higher than Auto's but did not reach significance (Scope: $M = 76.2$ vs.\ $70.5$, $p = .116$; Content+Scope: $M = 74.3$ vs.\ $66.2$, $p = .110$); MemoSync was significantly higher than Static on both (Scope: $p = .003$, $r_{rb}=.40$; Content+Scope: $p = .004$, $r_{rb}=.40$). We speculate that Auto's memory summaries already provided implicit scope cues through headings such as ``Preferences \& Habits'' and ``Project Knowledge'' (Figure~\ref{fig:baseline-conditions}a). These cues may have supported scope judgments without item-level labels, which may explain why the scope comparisons with Auto were not significant despite higher means under MemoSync.

\begin{figure*}[t]
\centering
\includegraphics[width=\textwidth]{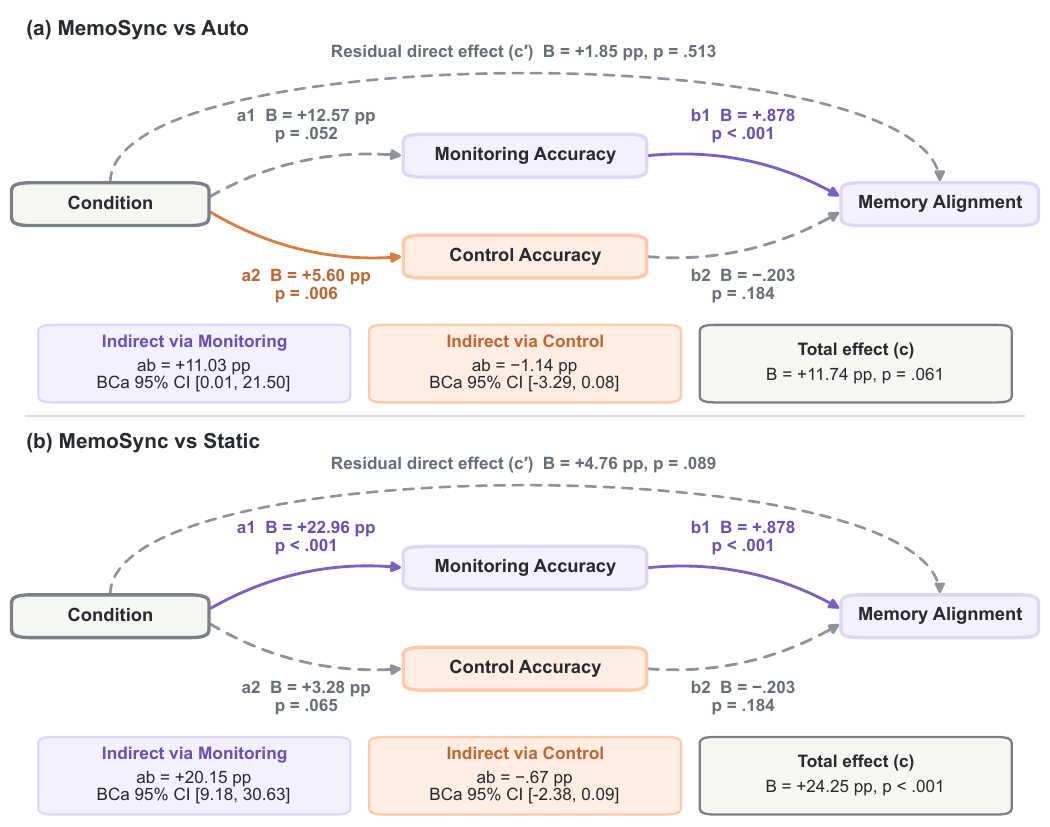}
\caption{Parallel mediation of Memory Alignment (Content+Scope; $N=123$). Panels (a) and (b) show MemoSync relative to Auto and Static, respectively; positive condition effects favor MemoSync. Arrows report unstandardized OLS coefficients with HC3 $p$ values; solid and dashed paths denote $p<.05$ and $p\geq.05$, respectively. The $b$ paths are shared across contrasts and therefore repeated in both panels. Summary boxes report specific indirect effects with BCa 95\% CIs from 50{,}000 condition-stratified participant bootstrap resamples and the total effect ($c$); $c'$ is the residual direct effect after both mediators.}
\label{fig:mediation-parallel}
\Description{Two path diagrams. In each, Condition on the left connects to Monitoring Accuracy and Control Accuracy in the middle, which connect to Memory Alignment on the right; a curved dashed arrow across the top is the residual direct effect. (a) MemoSync vs Auto: a1 = +12.57 pp, p = .052; a2 = +5.60 pp, p = .006; b1 = +.878, p < .001; b2 = -.203, p = .184; residual direct effect +1.85 pp, p = .513. Summary boxes: indirect via monitoring ab = +11.03 pp, BCa 95\% CI [.01, 21.50]; indirect via control ab = -1.14 pp, CI [-3.29, 0.08]; total effect +11.74 pp, p = .061. (b) MemoSync vs Static: a1 = +22.96 pp, p < .001; a2 = +3.28 pp, p = .065; the same b paths; residual direct effect +4.76 pp, p = .089. Summary boxes: indirect via monitoring ab = +20.15 pp, CI [9.18, 30.63]; indirect via control ab = -.67 pp, CI [-2.38, 0.09]; total effect +24.25 pp, p < .001. Solid arrows mark paths with p < .05 and dashed arrows the others.}
\end{figure*}

\textbf{Finding 3:} \textbf{Monitoring accuracy mediated the condition differences in memory alignment.}\label{sec:mechanism} In an exploratory parallel mediation analysis with Memory Alignment (Content+Scope) as the outcome and Monitoring Accuracy and Control Accuracy as simultaneous mediators ($N=123$; Figure~\ref{fig:mediation-parallel}), the specific indirect effect through Monitoring Accuracy was significant for both contrasts (vs.\ Static: $ab=+20.2$~pp, BCa 95\% CI $[9.18, 30.63]$; vs.\ Auto: $ab=+11.0$~pp, BCa 95\% CI $[0.01, 21.50]$).\footnote{Mediation inference treats the bootstrap CI for the indirect effect as the criterion~\cite{hayes2018introduction, zhao2010reconsidering, mackinnon2008introduction}, without requiring path or total-effect significance.} The corresponding indirect effects through Control Accuracy were not significant (vs.\ Auto: $ab=-1.14$, CI $[-3.29, 0.08]$; vs.\ Static: $ab=-0.67$, CI $[-2.38, 0.09]$). After both mediators were included, neither residual direct effect reached $p<.05$ (vs.\ Static: $c'=+4.76$~pp, $p=.089$; vs.\ Auto: $c'=+1.85$~pp, $p=.513$). This pattern suggests that MemoSync's alignment advantage, and thus its reduction of misalignment, operated through participants' accurate knowledge of the agent's memory state rather than through the system's automatic operations alone. Had the advantage come from better extraction and proposals by themselves, the direct effect of condition would have remained after controlling for the two mediators; instead, neither residual direct effect reached significance. The indirect effect through Control Accuracy was not significant, consistent with its near-ceiling level in all conditions: once participants knew the memory state, realizing desired memories in the output was not the bottleneck.

\begin{figure*}[t]
\centering
\includegraphics[width=0.8\textwidth]{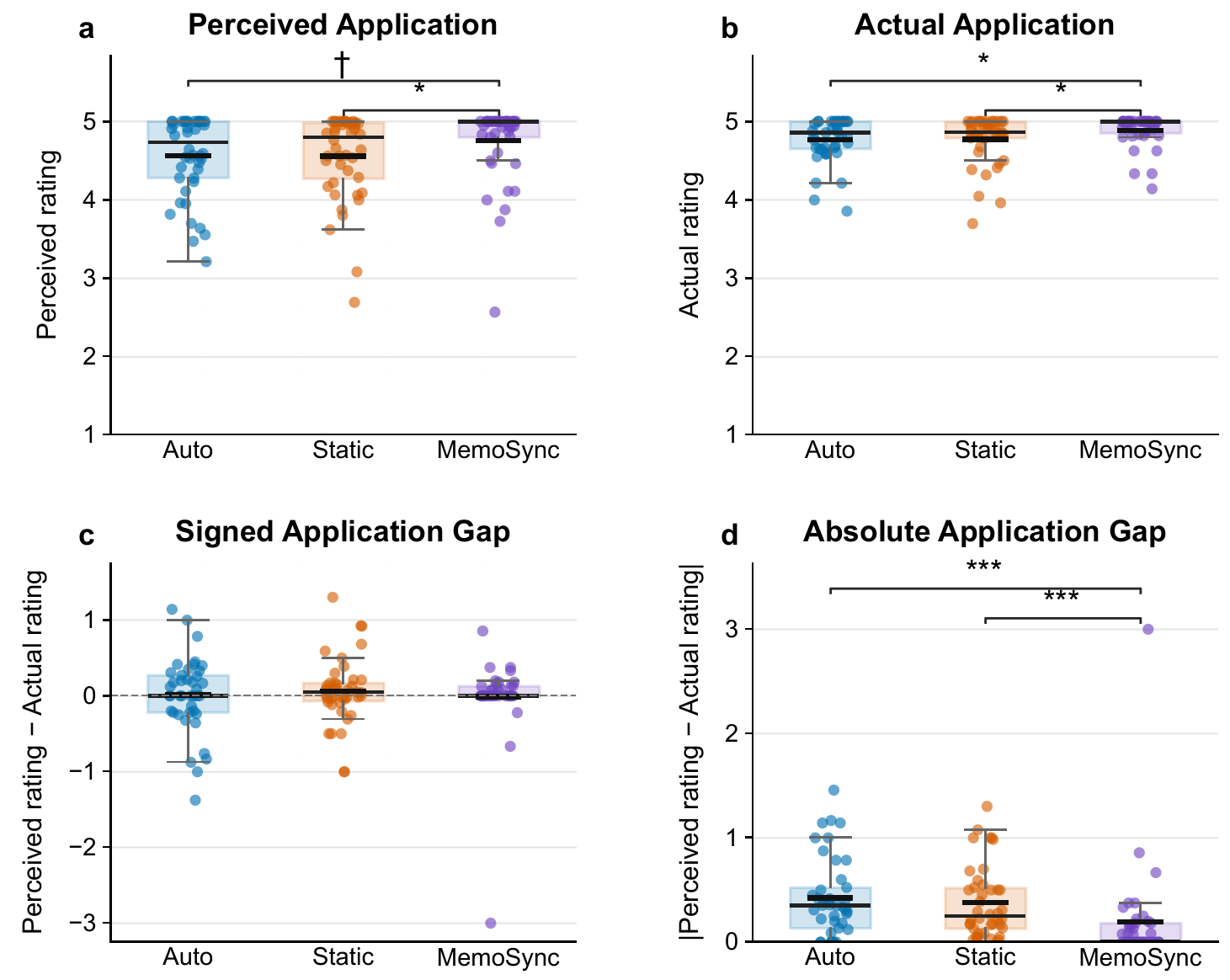}
\caption{Memory application perception. (a)~Perceived application rating. (b)~Actual application rating. (c)~Signed gap (perceived minus actual). (d)~Absolute gap. Boxes show medians and interquartile ranges, thick black bars show means, whiskers extend to the most extreme values within 1.5 IQR, and dots show individual participants. $n=41$, 42, and 40 for Auto, Static, and MemoSync, respectively. MemoSync users had the smallest gap between perceived and actual memory application. Brackets show all three pairwise Mann--Whitney $U$ tests with Holm correction within each measure. ${}^{***}p<.001$, ${}^{*}p<.05$, ${}^{\dagger}p<.10$.}
\label{fig:application-gap}
\Description{Four box plots with jittered points for Auto (blue), Static (orange), and MemoSync (purple). (a) Perceived rating, 1 to 5: all medians near 5, with MemoSync's points clustered most tightly near 5. (b) Actual rating, 1 to 5: the same pattern. (c) Signed gap, perceived minus actual: all three boxes sit at zero; Auto and Static show a wider spread of points reaching about -3. (d) Absolute gap: MemoSync's box lies near zero and below the Auto and Static boxes, with brackets marked *** between MemoSync and each baseline.}
\end{figure*}

\textbf{Finding 4:} \textbf{MemoSync reduced the gap between perceived and actual memory application.} A Kruskal--Wallis test on the absolute application gap was significant, $H(2) = 23.65$, $p < .001$, $\epsilon^2 = .18$. The gap under MemoSync ($M = 0.19$ on a 1--5 scale) was less than half of that under Auto ($M = 0.42$; $U = 395.5$, $p < .001$, $r_{rb}=-.52$) and Static ($M = 0.38$; $U = 371$, $p < .001$, $r_{rb}=-.56$; Figure~\ref{fig:application-gap}d), while no significant difference was detected between Auto and Static ($U = 904.5$, $p = .695$). Signed gaps were near zero in every condition (Figure~\ref{fig:application-gap}c), indicating no systematic over- or under-estimation. All three conditions allowed participants to inspect stored memory items, but only MemoSync showed which items the agent selected and referenced during each generation turn, together with whether each item was followed, unused, or violated. We speculate that this direct link between individual memories and evidence of their application reduced the gap between perceived and actual memory use. Inline citations located the relevant agent actions, while post-turn audits assessed whether those actions followed the corresponding memory items, helping participants judge memory application more accurately.

\subsection{Task Performance across Projects and Sessions}
\label{sec:task-performance}

This subsection examines how memory co-management affected task performance. We first report overall and per-project condition differences, then trace how performance evolved across sessions, and finally discuss the relationship between memory alignment and task performance.

\begin{figure*}[t]
\centering
\includegraphics[width=0.8\textwidth]{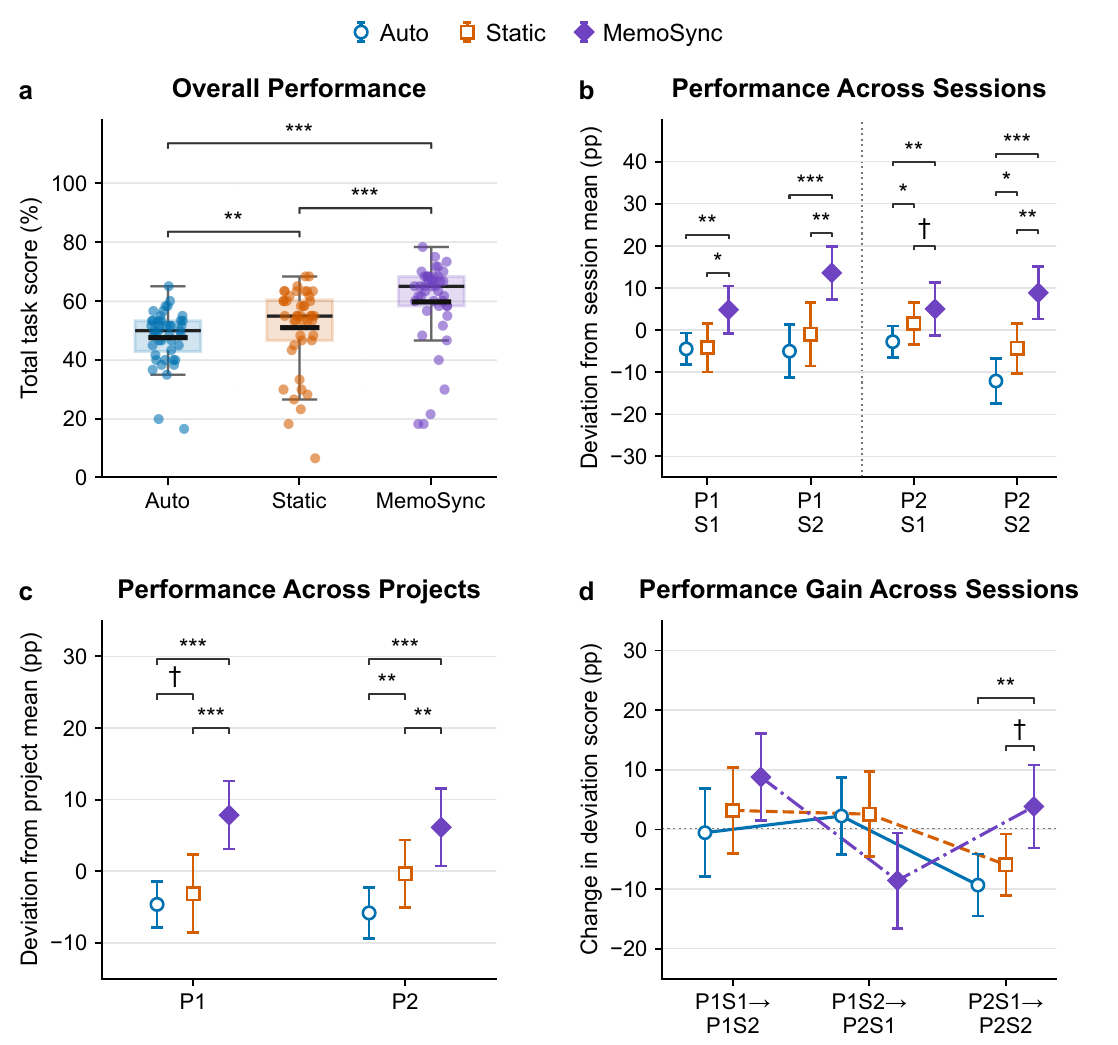}
\caption{Task performance. (a) Overall performance as a percentage of the 30-point maximum. (b) Per-session deviation from the corresponding task mean. (c) Project-level deviation from the corresponding project mean. (d) Change in session-level deviation score across each pair of consecutive sessions. In (a), boxes show medians and interquartile ranges, thick black bars show means, whiskers extend to the most extreme values within 1.5 IQR, and dots show individual participants. In (b--d), markers show condition means and error bars show 95\% CIs; lines in (d) connect the three ordered transitions within each condition. Brackets show all three pairwise Mann--Whitney $U$ tests with Holm correction within each panel, session, project, or transition position. $n=44$ per condition. $^{***}p<.001$, $^{**}p<.01$, $^{*}p<.05$, $^{\dagger}p<.10$.}
\label{fig:task-performance}
\Description{Four panels. (a) Box plots of total task score in percent: Auto centered near 48, Static near 51, MemoSync near 60, with brackets marked *** between MemoSync and each baseline and ** between Static and Auto. (b) Deviation from the session mean in percentage points at P1S1, P1S2, P2S1, and P2S2, with a dotted line at the project boundary: MemoSync's markers are above zero at all four sessions, between about +5 and +14 pp, while Auto's and Static's markers are at or below zero. (c) Deviation from the project mean: MemoSync about +8 pp at P1 and +6 pp at P2; Auto about -5 and -6 pp; Static about -3 and 0 pp. (d) Change in deviation score across the three consecutive transitions, drawn as lines: brackets appear only at the last transition, P2S1 to P2S2, where MemoSync rises by 3.8 pp, Auto changes by -9.3 pp, and Static by -5.9 pp.}
\end{figure*}

\textbf{Finding 5:} \textbf{MemoSync achieved significantly higher task performance than both baselines; Static significantly outperformed Auto.} A Kruskal--Wallis test on overall performance was significant, $H(2) = 35.57$, $p < .001$, $\epsilon^2 = .26$. MemoSync participants scored $59.8\%$ of the maximum ($SD = 14.2$), significantly higher than Auto ($M = 47.7\%$, $SD = 9.3$; $U = 1626.5$, $p < .001$, $r_{rb}=.68$) and Static ($M = 51.1\%$, $SD = 14.3$; $U = 1442.5$, $p < .001$, $r_{rb}=.49$; Figure~\ref{fig:task-performance}a). Static was also significantly higher than Auto ($U = 637$, $p = .006$, $r_{rb}=-.34$). At the session level (Figure~\ref{fig:task-performance}b), MemoSync remained above the task mean across all four sessions, significantly higher than both baselines at P1S1, P1S2, and P2S2, and significantly higher than Auto at P2S1 ($p = .004$, $r_{rb}=.40$) with a near-significant advantage over Static ($p = .073$, $r_{rb}=.22$). MemoSync's advantage was already present at P1S1, likely because its extraction and per-turn selection captured and selected task-relevant memory from the first turn onward. At the project level (Figure~\ref{fig:task-performance}c), MemoSync outperformed Auto and Static at P1 ($p < .001$, $r_{rb}=.59$ and $p < .001$, $r_{rb}=.45$, respectively) and P2 ($p < .001$, $r_{rb}=.56$ and $p = .010$, $r_{rb}=.32$, respectively). Static outperformed Auto at P2 ($p = .009$, $r_{rb}=-.35$), with a marginal difference at P1 ($p = .064$, $r_{rb}=-.23$). We speculate that Static benefited from retaining accumulated implementation context that remained useful across structurally similar tasks. The shared technology stack and application structure may have made these details reusable, whereas Auto's summarization and storage cap left fewer details in memory to reuse.

\textbf{Finding 6:} \textbf{At the final session (P2S2), MemoSync's performance increased while both baselines declined.} P2S1$\rightarrow$P2S2 was the only transition where condition differences emerged (Figure~\ref{fig:task-performance}d): MemoSync gained $+3.8$~pp, whereas Auto declined by $9.3$~pp ($p = .010$, $r_{rb}=.36$); the difference from Static ($-5.9$~pp) approached significance ($p = .075$, $r_{rb}=.26$). The first two transitions did not differ by condition (all adjusted $p \geq .204$, all $|r_{rb}| \leq .23$). By the final transition, the memory store could contain both knowledge carried across projects and decisions made during P2S1. We conjecture that extending the second project's codebase in P2S2 increased the need to bring this accumulated context into alignment with new requirements. MemoSync's review of proposed changes may have supported this process, helping preserve the usefulness of prior memory as the implementation evolved.

\textbf{Finding 7:} \textbf{Memory alignment and task performance traded off between Auto and Static; MemoSync achieved the highest on both.} Auto participants had higher memory alignment than Static (Finding~1), yet their task performance was significantly lower (Finding~5). Auto's automatic extraction mechanism imposed a storage cap, retaining high-level summaries; detailed but reusable information accumulated during sessions (e.g., environment-specific debugging approaches, concrete API call patterns) could be compressed during summarization or replaced once the cap was reached, leaving the agent without directly reusable references in later sessions. Static imposed no such cap, and developers and the agent continuously added context to the shared Markdown file, providing the agent with a richer pool of references that could be reused across structurally similar projects, even though developers gradually lost track of what the file contained (Figure~\ref{fig:alignment-across-sessions}). MemoSync showed no such trade-off, achieving the highest alignment and performance. One explanation is that continuous accumulation and cross-project transfer kept its memory rich, while the mixed-initiative proposal mechanism kept developers aware of the agent's memory and its scope.

\subsection{Interaction Activity, Workload, and Usability}

\begin{figure*}[t]
\centering
\includegraphics[width=\textwidth]{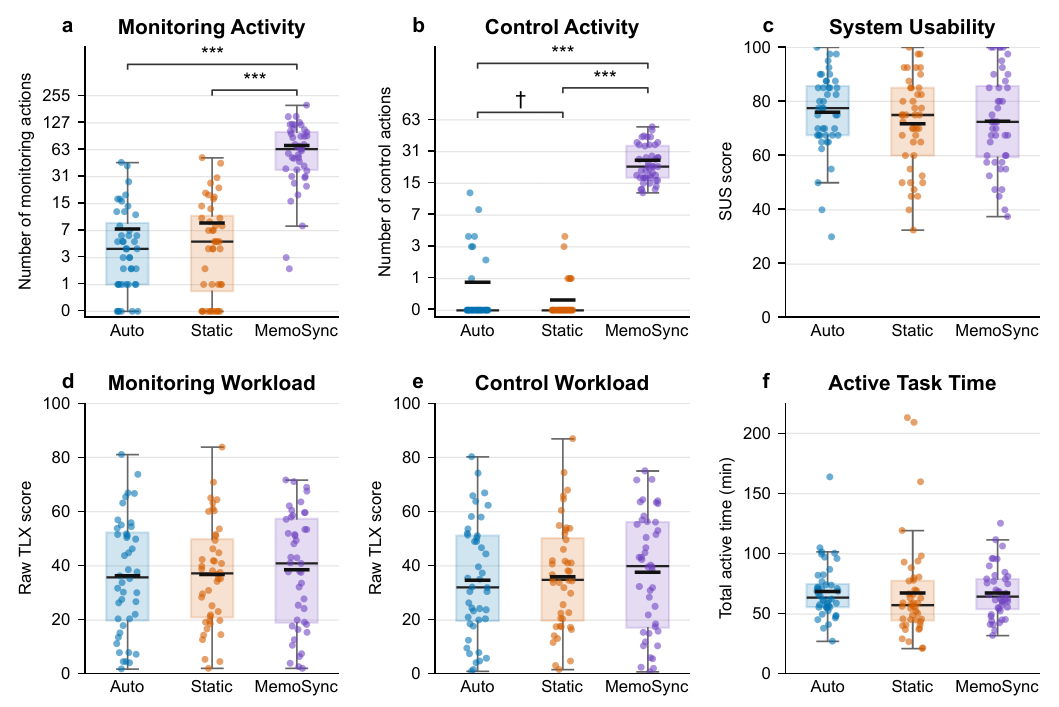}
\caption{Behavioral measures and cost. (a)~Monitoring activity (log scale). (b)~Control activity (log scale). (c)~System usability (SUS). (d)~Monitoring workload (Raw TLX). (e)~Control workload (Raw TLX). (f)~Active task time. Boxes show medians and interquartile ranges, thick black bars show means, whiskers extend to the most extreme values within 1.5 IQR, and dots show individual participants. Brackets show all three pairwise contrasts with Holm correction within each measure, from negative-binomial regression in (a,~b) and Mann--Whitney $U$ tests in (c--f). ${}^{***}p<.001$, ${}^{\dagger}p<.10$.}
\label{fig:behavior-effort}
\Description{Six box plots with jittered points for Auto (blue), Static (orange), and MemoSync (purple). (a) Number of monitoring actions on a log scale: MemoSync's box is centered around 60 to 70, Auto's and Static's below 10, with brackets marked ***. (b) Number of control actions on a log scale: MemoSync's box is centered around 25, Auto's and Static's near 0. (c) SUS score: all three boxes lie between about 70 and 80 with no brackets. (d) Monitoring Raw TLX and (e) Control Raw TLX: all three boxes lie between about 30 and 40 with no brackets. (f) Total active time in minutes: all three boxes lie between about 60 and 70 with no brackets.}
\vspace{-8pt}
\end{figure*}

\textbf{Finding 8:} \textbf{MemoSync participants performed significantly more monitoring and control actions, yet no condition differences were detected in workload, usability, or task time.} MemoSync participants performed on average 70.3 monitoring actions over the four sessions ($SD = 43.9$), compared with 7.3 under Auto ($\mathrm{IRR}=9.64$, 95\% CI $[6.20, 14.97]$, $p_{\mathrm{H}}<.001$) and 8.7 under Static ($\mathrm{IRR}=8.06$, 95\% CI $[5.25, 12.36]$, $p_{\mathrm{H}}<.001$; Figure~\ref{fig:behavior-effort}a). They also performed 25.3 control actions ($SD = 11.1$), compared with fewer than one under Auto ($\mathrm{IRR}=30.14$, 95\% CI $[13.32, 68.20]$, $p_{\mathrm{H}}<.001$) and Static ($\mathrm{IRR}=101.36$, 95\% CI $[40.32, 254.82]$, $p_{\mathrm{H}}<.001$; Figure~\ref{fig:behavior-effort}b). Kruskal--Wallis tests detected no condition differences in monitoring workload ($H(2) = 0.38$, $p = .829$, $\epsilon^2 < .01$), control workload ($H(2) = 0.52$, $p = .771$, $\epsilon^2 < .01$), system usability ($H(2) = 1.34$, $p = .511$, $\epsilon^2 < .01$), or active task time ($H(2) = 3.22$, $p = .200$, $\epsilon^2 = .01$; Figure~\ref{fig:behavior-effort}c--f). Equivalence tests supported equivalence within a medium effect for both monitoring and control workload against both baselines (all TOST $p \le .047$). As described in the Introduction, developer-agent collaboration splits the meta-level and object-level of traditional metamemory apart, delegating object-level memory operations to the agent. Under Auto and Static, users had to self-initiate checks on this external object-level state. MemoSync's mixed-initiative design surfaced object-level state changes at each interaction point, allowing the monitoring-control loop to operate without requiring users to initiate across this boundary. We conjecture that this design allowed the substantially increased activity to proceed without a perceived increase in workload.

 \section{Deployment Study}
\label{s:deployment}

In the controlled study, MemoSync's task performance advantage was already shown in the first session. Participants also performed more monitoring and control actions, although we detected no condition differences in perceived workload. We sought to understand how memory co-management supported adjustments to the current task, how these actions fit into ongoing development work, and how developers reused memory across their own projects and parallel sessions. To investigate these questions, we conducted a week-long deployment study.

\subsection{Participants and Procedure}

Nine developers participated in the deployment study (P1 to P9, demographics in Table \ref{tab:deployment_participants}), using MemoSync to work on their ongoing projects over one week. Before the study, we provided a user manual and asked participants to spend at least two hours using the system. System logs recorded participants' monitoring and control actions, agent execution, and context switches. At the end of the week, each participant completed a semi-structured interview lasting approximately 30 minutes, covering their overall experience and specific interaction episodes selected by the researchers from their logs. Participants received unlimited LLM API access to use MemoSync on their own projects throughout the study and a \$50 USD honorarium upon completing the exit interview.

\subsection{Results}

\begin{figure*}[t]
\centering
\includegraphics[width=\textwidth]{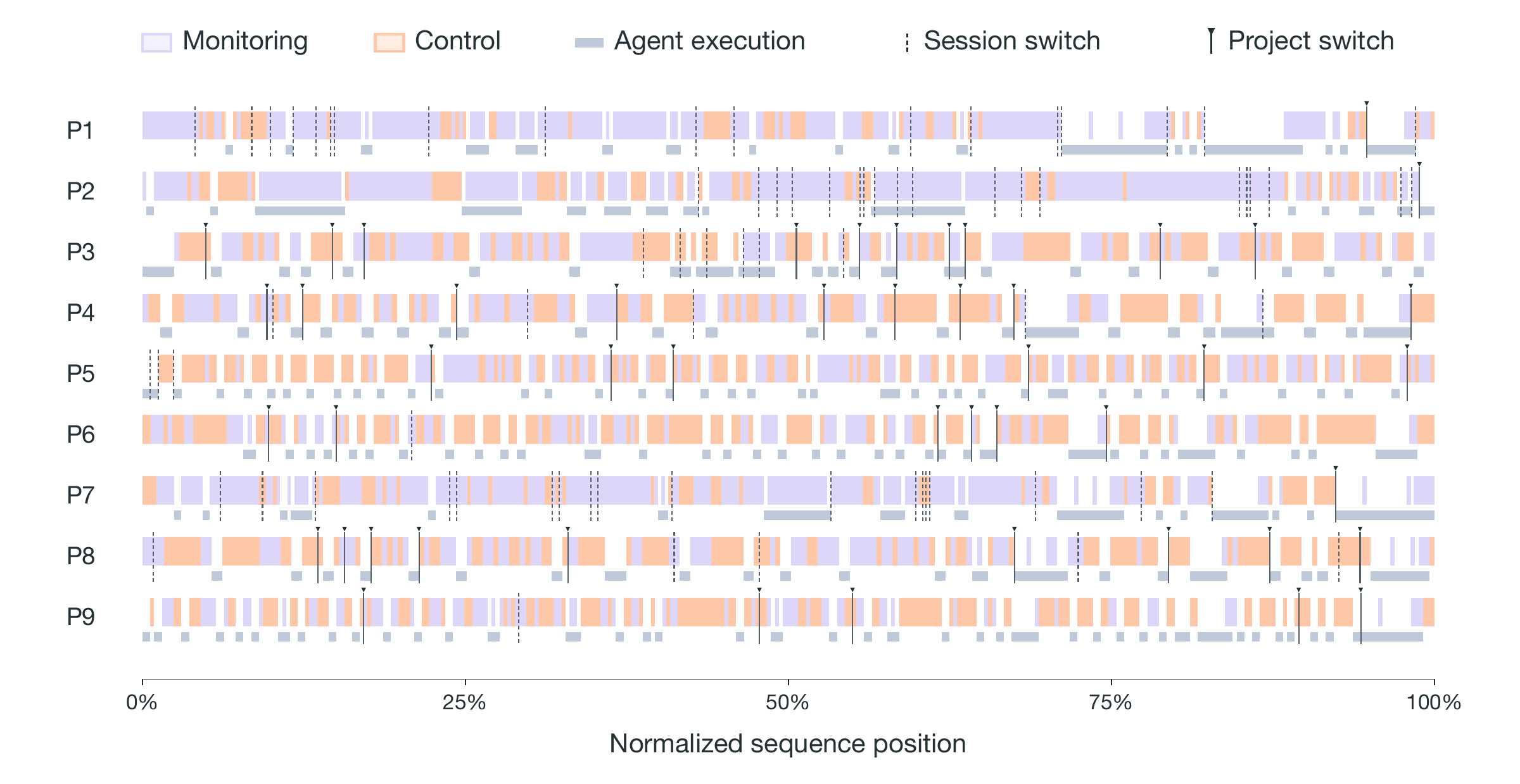}
\caption{Activity sequences from the week-long deployment study. Each row represents one participant. Purple and orange bars indicate monitoring and control actions, respectively; gray bars indicate agent execution. Dashed vertical lines mark session switches; solid lines with triangular markers mark project switches.}
\label{fig:deployment-activity}
\Description{A horizontal timeline chart with nine rows labeled P1 to P9 and an x-axis labeled Normalized sequence position from 0 to 100\%. Each row is a bar made of many short colored segments, purple for monitoring and orange for control, with thin gray segments beneath the bar marking agent execution. Dashed vertical ticks mark session switches and solid ticks with a triangular head mark project switches. P1, P2, and P7 are mostly purple with few project switches. P4, P5, P6, P8, and P9 contain more orange segments and more project switches. The later portions of P1, P4, P6, P7, P8, and P9 contain longer gray segments with fewer purple or orange segments above them.}
\end{figure*}

We transcribed the audio-recorded interviews and coded them with the same inductive procedure as the formative study (Appendix~\ref{app:analysis}): two researchers coded the transcripts independently, compared and merged their codes into a shared codebook, and grouped the codes into themes describing how participants used memory co-management in their daily work. Figure~\ref{fig:deployment-activity} presents each participant's activity sequence over the week, with the horizontal axis indicating normalized sequence position. Below, we draw on these sequences and interviews to describe how developers incorporated memory co-management into daily work.

\textbf{Finding 1: Participants used shared memory both to check execution and to make direct corrections to the current task.} Participants incorporated memory co-management between agent runs, showing two usage tendencies. P1, P2, and P7 more often began by inspecting \textit{Working Memory for This Turn} and the \textit{Memory Record}, checking the relevant memories and what the agent had done before deciding whether to make changes. For these participants, inspection itself helped them judge whether the next steps matched their expectations. P1 felt that the pre-execution expected-use statements gave them ``a greater sense of control over the task,'' while P2 found per-turn working-memory selection smoother than editing files beforehand or asking the agent to pause the current task to retrieve context. In comparison, P4, P5, P6, P8, and P9 more often directly edited memory or specified items to use between runs, using individual items to adjust subsequent execution. P5 explained that outputs often deviated from requirements only in particular places, while editing and resending the original prompt was costly. They therefore preferred to mark the relevant memory item as mandatory for the next turn to ``make small adjustments to the work.'' P3 combined both tendencies, often beginning with inspection after execution while also frequently modifying memory. System-initiated proposals further integrated these adjustments into the current task. P8 had previously checked memory only after outputs went wrong, and their Markdown file had grown beyond 2,000 lines, making it almost impossible to review in full. With MemoSync proposing new items and changes during work, P8 felt that the system ``helped me better manage my memory store.'' These uses connected memory maintenance to the task at hand: participants could inspect memory to decide whether intervention was needed or directly revise the relevant items for problems they identified.

\textbf{Finding 2: Some participants became more willing to delegate execution and used shared records to resume oversight as needed.} In the latter halves of their activity sequences, six participants (P1, P4, P6, P7, P8, and P9) showed a higher proportion of agent execution and longer continuous execution segments; P2, P3, and P5 did not show the same trend. This change occurred among participants who emphasized inspection as well as those who emphasized modification, consistent with interview accounts of becoming more willing to delegate tasks. P1 perceived improvements in the agent's adherence to their requirements and personal habits and ``gradually felt confident enough to delegate complex tasks to it.'' Shared records helped participants catch up after stepping away and identify issues that needed attention. P4 called the \textit{Memory Record} ``an anchor between me and the agent,'' explaining that it took over part of the progress tracking they had previously maintained manually in Markdown. P7 delegated a task to the agent, switched to writing a paper, and returned to find that \textit{Memory Use Audit} had flagged a memory item that had not been followed. They noted that ``in my previous workflow, I would have just moved on and missed this deviation.'' Oversight could also resume during execution: P5 stopped execution immediately after seeing the agent cite an incorrect memory item, a problem that had been difficult to notice in the lengthy execution logs of previous tools. For these participants, shared records preserved information about memory use during delegation, allowing them to find specific grounds for intervention when returning or checking in without continuously following the full execution.

\textbf{Finding 3: Cross-context memory supported both reusing existing requirements and handing off information between parallel sessions.} P3 and P4, who switched projects and used memory transfer most frequently, reused existing memories across projects and concurrent conversations. P3 frequently needed to reuse common requirements across projects. Previously, this required manually organizing the content or explaining it again in new conversations; transfer allowed them to bring in existing memory directly, reducing the need to ``repeat the same things.'' Memory reuse also occurred across parallel sessions. P3 and P4 described MemoSync proactively bringing in memory generated in other conversations, reducing the work of manually passing information between them. P4 described this as ``enabling communication between sessions to some extent.'' In these experiences, memory carried requirements that could be reused across projects and allowed information developed in one conversation to be carried into another task.

\textbf{Additional expectations: Participants wanted the system to judge when further confirmation was necessary.} Participants valued the opportunities to participate in memory co-management, but wanted fewer repeated confirmations for simple instructions, explicit preferences, and error correction. P4 and P5 wanted the system to proceed directly when instructions were simple, without asking them to confirm working memory again. P2 felt that after expressing a preference for camelCase over underscore naming, the system should store it directly in personal memory. P7 argued that ``when the audit finds an error, it should proactively fix it,'' and wanted to avoid having the system first report the error and then wait for approval to repair it. These expectations came from participants with both usage tendencies and centered on whether further confirmation was necessary: they wanted the system to proceed based on requirements they had already made explicit and request their involvement when further judgment or information was still needed.

 \section{Discussion}
\label{s:discussion}

MemoSync treats persistent memory as a shared operable representation that developers and agents jointly monitor and control. Drawing on the formative, controlled, and deployment studies, we discuss memory co-management in four parts: how it supports selective oversight of accumulated memory and the bias and accountability issues raised by the agent's selection (Section~\ref{s:discussion-metamemory}), how it builds human-AI common ground through layered memory representations (Section~\ref{s:discussion-scope}), how it re-establishes memory across sessions and projects (Section~\ref{s:discussion-cross-context}), and how it lets people shape their extended cognition (Section~\ref{s:discussion-extended-cognition}).

\subsection{From Memory Visibility to Selective Oversight}
\label{s:discussion-metamemory}

\subsubsection{Monitoring as the Bottleneck of Metamemory Regulation}

Metamemory connects people's understanding of memory with memory processes through monitoring and control: monitoring informs the meta level about object-level states, and control guides storage and retrieval accordingly~\cite{nelson1990metamemory}. When the agent carries out memory operations, this regulation crosses the human-agent interaction boundary. Developers judge what should be retained and where it applies, while the agent stores, selects, and applies that information. In this loop, control is guided by monitoring; without accurate monitoring, control lacks an informed basis.

Our findings indicate that monitoring is the main bottleneck in developer-agent metamemory regulation. In the controlled study, Control Accuracy was high across all three conditions, whereas Monitoring Accuracy varied much more. The exploratory mediation analysis also identified significant indirect effects through Monitoring Accuracy. Static participants could inspect and edit the full memory file, yet their Monitoring Accuracy was the lowest. In the formative study, as configuration files grew, participants rarely read them in full and struggled to confirm which rules the agent actually followed. This finding suggests that the accumulated information made it difficult for developers to identify which memories needed attention in the current task. Formative participants also had difficulty identifying which memories the agent selected for each turn and often restated requirements in prompts after errors appeared. These findings suggest a need for selective oversight: developers need to identify task-relevant content within accumulated memory and judge whether its use matches their expectations.

\subsubsection{Agent-Supported Selective Oversight}

Our work suggests that agents can proactively support this selective oversight, helping developers judge which memories are relevant to the current task and whether their use matches expectations. Prior interactive memory systems support users in retrieving, organizing, reusing, and revising memory~\cite{huang2023memory,yen2024memolet}, but users still largely initiate reviews to identify which memories need attention in the current task. In MemoSync, the agent proposes which items to use and explains how it expects to use each one. Developers check for missing or irrelevant items, while citations and audits connect those items to their actual application. Developers and agents jointly identify which memories to monitor and check how they are used in the output.

Selective oversight can help developers monitor memory more accurately with little difference in perceived workload. In the controlled study, MemoSync participants' judgments of memory application were closer to actual application. They also performed significantly more monitoring and control actions, with little difference in perceived workload compared with the baselines. Deployment participants' feedback (Finding 1 in Section~\ref{s:deployment}) suggests that the system organized accumulated information around the current task, leaving developers to make judgments about specific items. This means that monitoring support involves both presenting information and helping people identify what currently needs attention. AI can help direct attention to relevant memories, connecting people's judgments with specific control actions in the current task.

\subsubsection{Flexible Control across Collaboration Stages}

Monitoring and control in metamemory do not occur at a single point in time. Nelson and Narens describe monitoring judgments and the corresponding control across the acquisition, retention, and retrieval of memory, such as allocating study time, selecting a search strategy, and deciding when to stop a search~\cite{nelson1990metamemory}. In developer-agent collaboration, regulation is likewise distributed before, during, and after each turn: developers can intervene when reviewing the working memory for the turn, when seeing citations during execution, or after the audit results appear. In the deployment study, some participants resumed oversight through shared records after delegating execution, and a participant stopped execution immediately after seeing the agent cite an incorrect memory item (Finding 2 in Section~\ref{s:deployment}). This suggests that selective oversight also involves choosing when to intervene. Developers do not need to continuously follow execution; they can adjust how they intervene according to how much they delegate and still find grounds for intervention when needed.

\subsubsection{Bias and Accountability}

Selective oversight also allows the agent's selection to shape what developers see. Which memories the agent selects depends on the judgment of the underlying model, and current language models tend to assign uniform priority to instructions, which makes it difficult to distinguish their importance~\cite{wallace2024instruction, sharma2024towards}. People who rely on automated aids also tend to miss problems that the aids do not flag~\cite{parasuraman2010complacency}. In the deployment study, participants wanted the system to proceed directly when instructions were simple, with fewer repeated confirmations (Section~\ref{s:deployment}). As delegation increases, developers' confirmations cover only part of the agent's decisions, and the model's biases are more likely to enter human judgment through its selection.

This influence also changes the distribution of responsibility. In human-machine systems, responsibility for an action may be attributed to a human operator who had limited control over the behavior of the automated system~\cite{elish2019moral}. In selective oversight, the agent shapes which memories developers can see and act on, while developers make the approvals. Selective oversight needs to make the selection itself open to inspection, showing which memories were not selected and why, and distinguishing decisions made by the agent from those made by developers, so that responsibility corresponds to actual influence.

\textbf{Implications.} From this metamemory perspective, human oversight of AI involves people and agents jointly identifying which information needs attention, when to intervene, and who is responsible for these decisions. Agents with persistent memory can propose which memories need attention in each turn and explain how they expect to use them, keep shared records that let developers resume oversight during and after execution, and show which memories were selected and which were not. Selective oversight offers a way to organize human-AI interaction around the current task, connecting relevant memories, human judgments, and direct control so that developers and agents can continue to regulate how accumulated experience is used.

\subsection{Building Human-AI Common Ground through Layered Memory Representations}
\label{s:discussion-scope}

People and agents differ in their capacity to process information. Human reading and judgment are limited by working memory capacity and time~\cite{cowan2001magical,rayner2016reading}, while LLMs can process large amounts of text within their context windows and use specific details to complete tasks. Effective communication is based on a shared understanding sufficient for the current task~\cite{clark1991grounding}. In memory co-management, developers need to understand the requirements expressed in memory and their scopes, while the agent uses detailed content to carry out those requirements. Developers and agents can build common ground around the same requirement using representations at different levels of detail.

Existing automatic memory and static configuration practices typically use a single memory representation to support both developer understanding and agent execution. Our controlled study included Auto and Static as baselines representing these two practices. The results showed a trade-off: Auto participants had higher Monitoring Accuracy and Memory Alignment for Content than Static participants, whereas Static participants achieved significantly higher task performance than Auto participants. Prior work has found that LLM agents can improve their performance as they accumulate experience~\cite{zhao2024expel}, whereas compressing contexts into concise summaries or repeatedly rewriting them loses domain-specific details needed for execution~\cite{zhang2026agentic}. Together with these findings, our results suggest that Auto's summaries and storage cap helped developers maintain an overview of memory while limiting the details available for the agent to reuse, affecting task performance. Static retained implementation details that may have supported subsequent tasks, but the growing content exceeded what developers could review, making it harder for them to monitor and understand what the memory file contained.

MemoSync uses linked, layered representations of the same memory: developers review summaries and scopes, while the agent loads task-relevant details on demand. Links between these levels allow developers to review specific details when needed and inspect their application through citations and audits. This design builds common ground around the same requirement while allowing developers and agents to use information at different levels of detail. In the controlled study, MemoSync achieved the highest Memory Alignment and task performance (Finding 7 in Section~\ref{s:results}). As memory accumulates, developers can understand requirements through concise representations, while the agent retains the implementation details needed for execution.

\textbf{Implications.} Layered memory representations let people understand and regulate accumulating knowledge around task requirements, allowing human-AI common ground to expand in ways that accommodate human cognitive capacity and agents’ ability to process detailed information. Systems can organize information for people according to the purpose of collaboration, show implementation details when people need to review them, and apply developers' revisions to the memory used in execution. Future work could explore how the level of detail can change with tasks and collaboration experience, so that people and agents can understand and jointly use the growing knowledge stored in memory.

\subsection{Re-establishing Memory across Contexts}
\label{s:discussion-cross-context}

To reuse memory in another context, it needs to be separated from the context in which it was formed and re-established in the context where it is used~\cite{ackerman2000reexamining}. Agents with persistent memory can take part in extracting, rewriting, and applying memory, making it easier for memory to move across sessions and projects. A practice from one project can be adapted for another project, and a requirement can extend from the current session to the whole project. Memory reuse thus involves changes in both its content and the scope in which it applies.

\subsubsection{Abstracting and Localizing Memory across Projects}

Cross-project transfer shows how content is re-established in a new context. In the formative study, participants typically copied existing context documents with little revision when switching projects and assumed that the agent would interpret them correctly, but this often led to unexpected behavior. They wanted guidance on which memories could transfer directly, which needed rewriting, and which should remain in the original project. MemoSync first abstracts a source memory into a principle, removing implementation details specific to the source project, then rewrites it for the destination project, and presents the source, abstract, and localized forms for developer review. In the controlled study, MemoSync participants' Memory Alignment for Content was significantly higher than both baselines in all four sessions across the two projects.

\subsubsection{Refining Scope through Later Use}

Scope determines how widely memory content can continue to be used. Crossing a boundary can extend the use of a memory item or limit it to a more specific scope: MemoSync allows developers to move a session memory to the project scope, and to keep content transferred from another project in the current session only. These decisions require anticipating where a memory will be used in later work; Ackerman and Halverson note that such anticipation remains difficult, although people do it every day in their work~\cite{ackerman2000reexamining}. In the controlled study, at the project boundary, MemoSync participants' Memory Alignment for Content remained significantly higher than both baselines, while their Memory Alignment for Scope was close to Auto's and was significantly higher than both baselines in the second session of each project. The different trajectories of the two dimensions suggest that content can be re-established during transfer, whereas scope continues to become clearer through later use after memory enters a new context. Systems can connect scope adjustment with how memory is actually used, allowing developers to keep, extend, or narrow its scope according to new task requirements. The re-establishment of memory in the destination context also continues into later work as the project progresses.

\subsubsection{From Individual to Group Memory}

Our study focused on individual developers reusing memory across their own sessions and projects. In practice, much development work is carried out by multiple people. Contributors to open-source repositories follow project conventions when submitting pull requests, and development in companies needs to meet team coding standards. As coding agents become part of everyday workflows, these conventions are increasingly written as Markdown files or packaged as skills that agents read and follow. These conventions become group memory shared by team members and their agents, and the scope of memory extends from individuals to teams. Group memory crosses boundaries in two directions. Experience that a member develops in a task needs to be separated from the details of that member's work before it can become a team convention; when team conventions enter each member's specific task, they need to be re-established for that task, a process increasingly carried out by agents. As development progresses, where a convention applies also becomes clearer through members' own use. Some conventions extend to the whole team, while requirements set only for a particular task are retired when the task ends.

\textbf{Implications.} Agent memory should be continually re-established in new contexts as collaboration progresses, allowing past experience to adapt to changing task requirements. Systems can let memory move across sessions, projects, and team members as collaboration continues, re-establishing its content and adjusting its scope each time it crosses a boundary, while preserving links to where it came from and where it has been used, so that the experience individuals develop in collaboration can be shared and developed more widely.

\subsection{Supporting Extended Cognition through Metamemory Co-Regulation}
\label{s:discussion-extended-cognition}

People have long relied on notes, documents, and digital tools to record past experience, organize thought, and coordinate action. Through continued use, these external resources become part of people's cognitive activity, extending what they can remember, understand, and accomplish~\cite{clark1998extended,salomon1991partners}. AI agents with persistent memory take on a more active role in this relationship: they retain requirements and decisions from collaboration, proactively organize and retrieve relevant information, and apply it to subsequent tasks. As people increasingly rely on this accumulated information in their work, AI memory gradually becomes part of their extended cognition.

Metamemory co-regulation lets people shape this part of their extended cognition. Developers use monitoring to understand what information influences their work and control to decide how it should be retained, revised, and applied. These judgments guide the cognitive system formed by people and AI as people's goals and contexts change. In this process, people also learn to organize and use external cognitive resources: judging when to rely on what the agent remembers, connecting knowledge across tasks, and adjusting the division of cognitive work as new needs arise. Collaboration with AI includes people's active participation in shaping their own cognition.

The deployment study showed how developers participated in this process (Section~\ref{s:deployment}). Participants used shared records to resume oversight after delegating execution (Finding 2) and used memory to reuse and pass information across projects and parallel sessions (Finding 3). Developers reorganized some of their record-keeping and coordination around AI memory and used the information preserved by the system to keep track of that work. Shared memory and records of its use supported both ongoing work and developers' judgments, allowing them to regulate these external resources while using them in their everyday development work.

Salomon et al. note that working with intelligent technologies can affect people beyond their performance during the collaboration, leaving lasting effects on their abilities~\cite{salomon1991partners}. Building on this view, our work suggests that people's ability to shape their extended cognition provides a basis for further human augmentation. As collaboration continues, people can connect knowledge across tasks to develop new understanding and revisit past decisions and their outcomes to revise their judgments and ways of working.

\textbf{Implications.} Future AI memory systems can support people in continuing to shape their extended cognition, bringing metamemory co-regulation into learning, problem-solving, and creative work. People can benefit from the information agents retain and use while continuing to decide how these resources are used in their thinking and actions, allowing their extended cognition to develop with their goals and needs.

\subsection{Limitations and Future Work}
\label{s:discussion-limitations}

First, we selected automatic memory and static configuration as baselines grounded in existing practices to examine the overall value of MemoSync as a complete memory co-management approach in collaboration. Memory updating, selection, application, and auditing are interconnected stages, and feedback from audits can also influence memory management in subsequent turns. Our comparison therefore cannot disentangle the contributions of individual mechanisms and their interactions to task performance and memory alignment. The deployment study provided specific insights into how these mechanisms were used in practice, but could not establish their independent causal contributions. Future work could combine stage-specific comparisons with interaction tracing across turns to examine how memory mismatches are introduced, detected, and corrected, and how these processes affect subsequent memory use.

Second, our cross-context findings are bounded by the task settings and study duration. The two projects in the controlled study were chosen to share a technology stack and overlapping features so that transfer could occur; this leaves untested how the system behaves when projects differ enough that transfer should be suppressed. The overall difference in scope alignment between MemoSync and Auto also did not reach statistical significance. The week-long deployment with nine developers primarily provided accounts of everyday memory reuse, maintenance, and oversight. Future work should examine more heterogeneous projects and longer collaborations to investigate whether systems can continue to preserve useful experience, limit inappropriate transfer, and sustain developers' understanding of memory.

Third, MemoSync's candidate extraction, transfer rewriting, and auditing all depend on the judgment of the underlying model. Current language models tend to assign uniform priority to instructions and to comply with them~\cite{wallace2024instruction, sharma2024towards}, which is consistent with deployment participants' feedback: to avoid errors, the system requested confirmation even for simple instructions. This limits the system's ability to judge the scope and severity of a memory change, and such judgment is a prerequisite for adapting the division of initiative to developer feedback. Future work could explore letting the model or a separate mechanism estimate a change's impact and decide which operations can proceed directly. \section{Conclusion}

In this paper, we present MemoSync, a system for co-managing developer-agent metamemory across sessions and projects. Informed by a formative study with 12 developers, MemoSync treats agent memory as a shared, operable representation that pairs monitoring with control across evolution, selection, execution, and impact, with explicit scopes to support cross-context reuse. A between-subjects study ($N = 132$) found that MemoSync participants achieved significantly higher task performance than participants using automatic memory or static configuration and the highest mean memory alignment, with no detected condition differences in perceived workload despite increased monitoring and control activity. A week-long deployment with nine developers further illustrated how co-management supported ongoing memory maintenance, cross-context reuse, and the resumption of oversight after stepping away. Together, these findings highlight memory as a continuing site of human-agent collaboration, where developers can understand and shape how accumulated experience informs subsequent work. As agents become more capable and autonomous, people may become less directly involved in task execution. We hope this work serves as an exploratory step toward preserving human agency over what agents remember and how those memories are applied across contexts. 

\bibliographystyle{ACM-Reference-Format}

\appendix
\renewcommand{\thetable}{\Alph{section}\arabic{table}}
\renewcommand{\thefigure}{\Alph{section}\arabic{figure}}

\section{Baseline Interfaces}
\label{app:baselines}
\setcounter{table}{0}
\setcounter{figure}{0}

\begin{figure}[!htbp]
  \centering
  \includegraphics[height=0.27\textwidth]{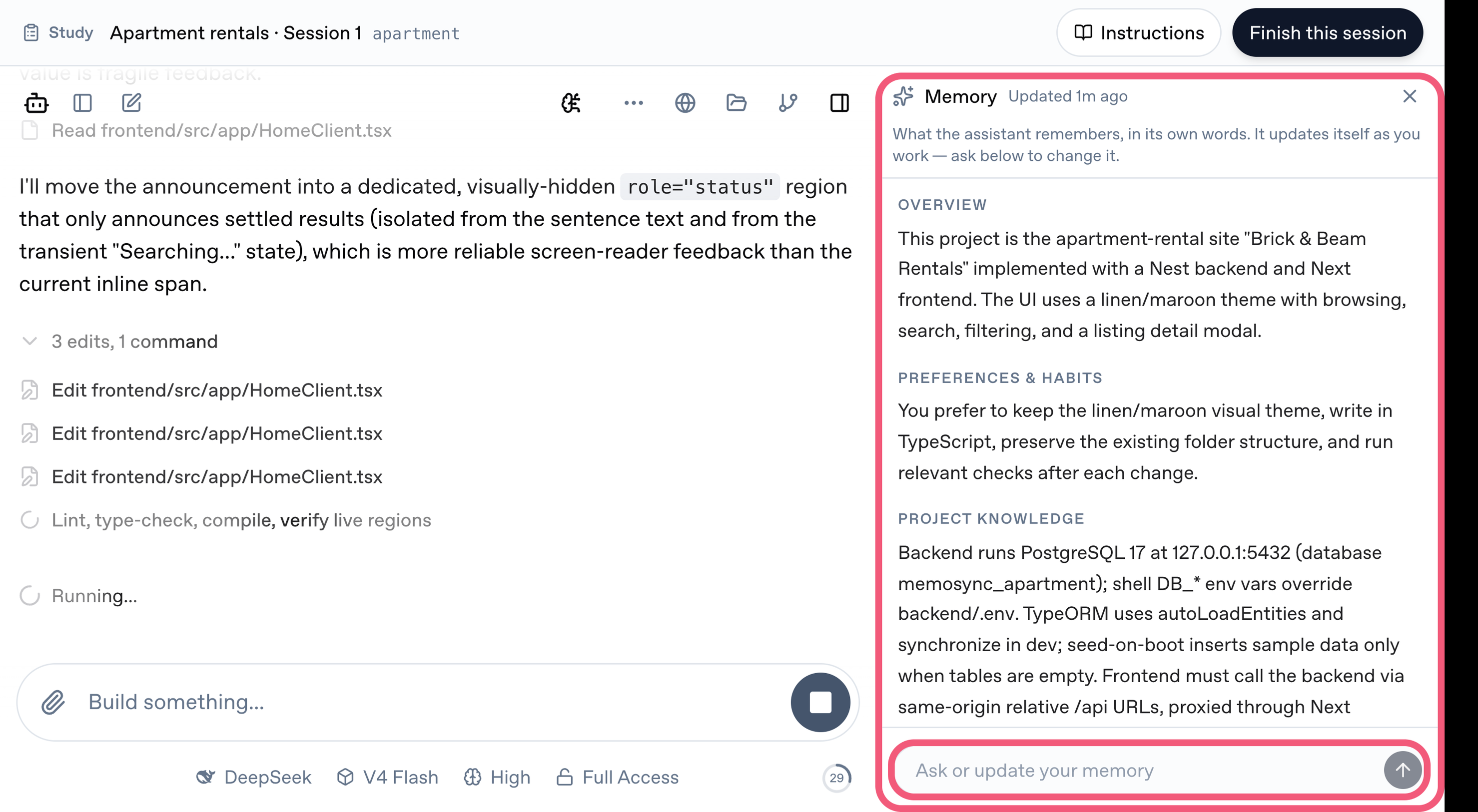}\hfill
  \includegraphics[height=0.27\textwidth]{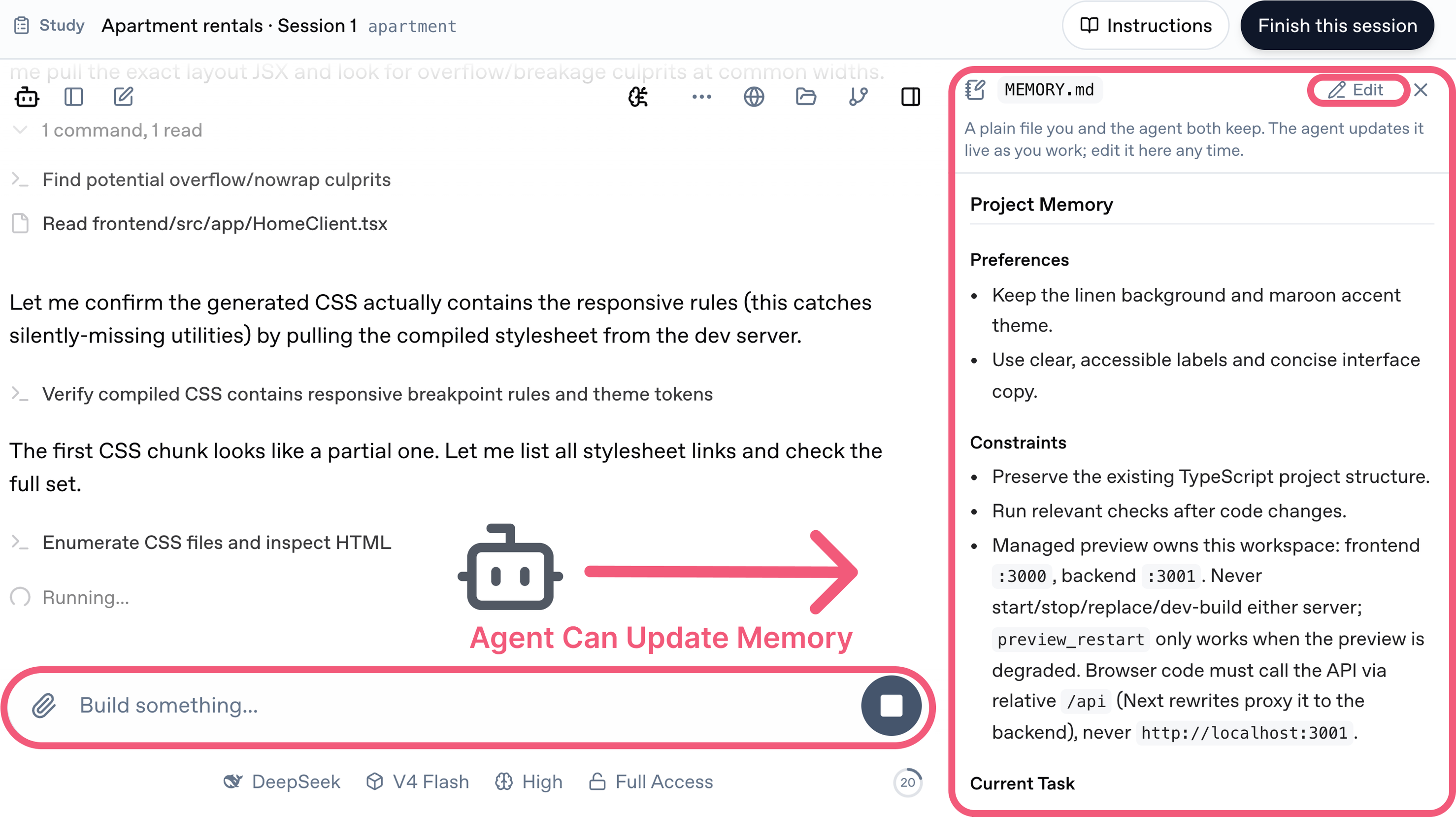}
  \caption{The two baseline conditions. \textbf{(a)}~Auto (left): the agent's memory is stored and displayed in the Memory sidebar as a summary. Participants cannot edit the memory directly; they can revise it only through the sidebar composer, which operates without the main conversation context. \textbf{(b)}~Static (right): the agent's memory is stored as a Markdown file and displayed in the Memory sidebar. Participants can edit the file directly or ask the agent in the main conversation to revise it.}
  \label{fig:baseline-conditions}
  \Description{Two screenshots of the study workbench. Each shows a header reading Apartment rentals, Session 1 with Instructions and Finish this session buttons, a chat panel on the left in which the agent reads and edits frontend files and runs commands, a prompt box reading Build something, and a Memory sidebar on the right outlined in red. (a) Auto: the sidebar is titled Memory, Updated 1m ago, and shows an agent-written summary under the headings Overview, Preferences and Habits, and Project Knowledge, with a text box at the bottom reading Ask or update your memory. (b) Static: the sidebar shows a file MEMORY.md with an Edit button and Markdown sections Project Memory, Preferences, Constraints, and Current Task; an annotation with a robot icon and a pink arrow reads Agent Can Update Memory. The footer of both screenshots shows DeepSeek V4 Flash, High, and Full Access.}
\end{figure}

Figure~\ref{fig:baseline-conditions} shows the workbench under the two baseline conditions described in Section~\ref{sec:condition}. Both provide a Memory sidebar for viewing the agent's memory; only the mechanism for changing it differs.

\section{Formative Study}
\label{app:formative}
\setcounter{table}{0}
\setcounter{figure}{0}
\subsection{Participant Demographics}
Table~\ref{tab:appendix_participants} lists the 12 participants in the formative study, with their programming experience, agent use, and project context.
\begin{table}[htbp]
\centering
\small
\setlength{\tabcolsep}{3.5pt}
\renewcommand{\arraystretch}{1.2}
\caption{Participants in our formative study ($N=12$): Coding gives years of programming experience, and Agent Use gives how long the participant has used AI coding agents. Projects is the number of projects in which the participant uses an agent on a sustained basis, and Sessions is the approximate number of agent sessions on their most-used project. Languages lists each participant's main programming languages, and Project Context gives the domain of their recent agent-assisted projects.}
\label{tab:appendix_participants}
\begin{tabularx}{\textwidth}{@{} l >{\raggedright\arraybackslash}p{2.1cm} c c c c c >{\hsize=1.18\hsize\raggedright\arraybackslash}X >{\hsize=0.82\hsize\raggedright\arraybackslash}X @{}}
\toprule
P\# & Role & Age & Coding & Agent Use & Projects & Sessions & Languages & Project Context \\
\midrule
P1  & PhD Student        & 24 & 3--5  & 3--6 months  & 2       & 2--3  & Python, Shell                        & Visualization \\
P2  & Master's Student   & 23 & 3--5  & $>$1 year    & 3       & $>$10 & Python, JS, C\#, SQL         & Interactive Tools \\
P3  & Agent Engineer     & 28 & $>$10 & 6--12 months & $\geq$4 & $>$10 & JS, TS, Shell, Scala & Compilers, Agent \\
P4  & PhD Student        & 23 & 6--10 & $>$1 year    & 2       & $>$10 & Python, JS, C/C++            & 3D Generation \\
P5  & PhD Student        & 25 & 6--10 & $>$1 year    & $\geq$4 & $>$10 & Python                               & Super-Resolution \\
P6  & Master's Student   & 23 & 3--5  & 6--12 months & 2       & 4--10 & Python, Shell                        & Graphics \\
P7  & Software Engineer  & 30 & 6--10 & $>$1 year    & $\geq$4 & 4--10 & Python, TS, SQL       & Full-Stack System \\
P8  & LLM Engineer     & 23 & 3--5  & $<$3 months  & $\geq$4 & $>$10 & Python                               & LLM Decoding \\
P9  & Undergraduate    & 21 & 1--2  & 3--6 months  & 3       & 4--10 & Python, Java                         & RAG \\
P10 & Agent Engineer  & 24 & 3--5  & 6--12 months & 2       & $>$10 & Python                               & Agent Workflow \\
P11 & Software Engineer  & 27 & $>$10 & $>$1 year    & 3       & $>$10 & Python, JS, TS       & LLM Inference \\
P12 & ML Researcher        & 29 & $>$10 & 6--12 months & $\geq$4 & $>$10 & Python, SQL, Shell                   & Marketing Agent \\
\bottomrule
\end{tabularx}
\end{table}

\subsection{Formative Study Procedure Details}
\label{app:procedure}

\textbf{Part A: Semi-structured Interview.} We asked how a typical agent session starts and ends, what context participants provide, how much autonomy the agent has, and what prompts a new session. We then asked about work across multiple projects, including how participants switch between projects and understand memory boundaries, and how they currently preserve information the agent should keep following, including conversation history, configuration files such as \texttt{CLAUDE.md} and \texttt{AGENT.md}, and system-managed memory: how it is created and updated, how often it is inspected, and how participants determine what the agent currently remembers and uses.

\textbf{Part B: Artifact Walkthrough.} Participants opened conversation logs, configuration files, automatically extracted memory, and the codebase from a recent agent-assisted project, and walked through them with the researcher. Starting from concrete examples, we traced what information they wanted to persist and how it was maintained, selected, used, and reflected in the output, covering five aspects. \textit{Representation}: which information was worth persisting or should not be retained, where it was stored, whether participants could inspect and modify what the agent actually used, and at which level (personal, project, or session) and granularity it should apply. \textit{Evolution}: how items were created and changed over a project, how stale, duplicated, or contradictory content was detected, which information could be reused across projects directly, rewritten, or kept local, and what confirmation participants wanted when memory was written, modified, or transferred. \textit{Selection}: for concrete turns, which memories the agent might have accessed, how participants verified the selected context, how they adjusted it when relevant items were missing or irrelevant ones included, and what they would want to see before generation. \textit{Execution}: whether participants could tell which memory the agent was acting on, when they discovered outdated or incorrect memory in use, how they handled such deviations, and what real-time information and intervention they wanted. \textit{Impact}: for completed outputs, whether a given item had influenced the output, been ignored, or been applied unexpectedly, what evidence supported that judgment, how deviations were repaired, and how the system should connect output behaviors to memory items.

\textbf{Part C: Design Expectations Interview.} Based on the episodes discussed, participants described what memory information they wanted to see, when they wanted to intervene, which operations the agent could perform proactively, how to recover from memory-related deviations, and how memory management should integrate with configuration files and existing workflows. We showed examples of open-source agent memory tools and asked which mechanisms were useful and what integration costs would discourage adoption, and closed by asking participants to rank visibility and control needs and to reflect on how proactive prevention, intervention during execution, and post-hoc correction should combine.

\subsection{Formative Study Data Analysis}
\label{app:analysis}

All sessions were audio-recorded and transcribed verbatim. Two researchers independently coded the transcripts using an inductive approach~\cite{thomas2006general}. In the first pass, both researchers conducted open coding, generating initial codes from participants' descriptions of their actual practices and artifacts without imposing predefined categories of memory problems. Example codes included `repeatedly re-explains project context,'' `does not know which stored rule the agent follows,'' `wants confirmation before memory write,'' `asks agent what it remembers,'' `cannot see memory use during generation,'' and `cannot tell whether a memory affected the output.''

After independently coding the first four transcripts, the two researchers compared their codes, discussed differences in interpretation, and merged overlapping codes into a shared codebook. They then applied this codebook to the remaining transcripts, adding new codes when new phenomena emerged and refining existing definitions through continued discussion. Recruitment and analysis proceeded iteratively; we stopped recruitment after 12 participants when subsequent interviews no longer produced new codes.

After coding was complete, we grouped the codes by the stage of memory use at which the reported issue occurred; these stages were also how the walkthrough was organized. This yielded five themes. The first concerned the representation of memory itself, including the sources from which developers currently infer the agent's memory state and how they want memory to be organized, inspected, and modified (\textrightarrow~F1). The remaining four corresponded to four stages of memory in collaboration: how memory is created, updated, becomes stale, and transfers across contexts, or \textit{evolution} (\textrightarrow~F2); how the agent determines which memories to use for the current turn, or \textit{selection} (\textrightarrow~F3); how memory guides agent behavior during generation, or \textit{execution} (\textrightarrow~F4); and how developers determine whether memory was applied correctly in the output, or \textit{impact} (\textrightarrow~F5).

Within each theme, we further distinguished what developers could currently observe or adjust, the resulting visibility and control gaps in existing practices, and the forms of support participants wanted. This analysis produced the five findings reported in the main text and informed the corresponding design goals. Participant counts reported in the paper (e.g., N=12) indicate the number of participants whose transcripts contained at least one code supporting that observation.

\section{Agent Baseline and Performance Analysis}
\label{app:baseline}
\setcounter{table}{0}
\setcounter{figure}{0}

\subsection{Agent-Only Calibration}
\label{app:calibration}

Before the controlled study, we ran the coding agent alone on the four study sessions (task briefs, rubric items, and the evaluation pipeline are described in Appendix~\ref{app:tasks}). These runs serve two purposes: they calibrate task difficulty, and they evaluate the capability of the coding agent itself. We ran the four sessions in a fixed order with no accumulated memory, repeated independently ten times (40 sessions in total). The agent was Claude Code (v2.1.170) with DeepSeek V4 Flash at the high reasoning setting. After each session, the workspace was frozen and scored following the session scoring protocol in Appendix~\ref{app:eval-pipeline}. Scores sum the Frontend, Backend, and Database items. Table~\ref{tab:agent-baseline} reports the mean score per session. The mean scoring rate across the four sessions was approximately 50\%. The tasks were thus neither trivial nor beyond the agent's reach, leaving room to observe both gains and losses from memory management.

\begin{table}[htbp]
\caption{Agent-only calibration scores (ten runs per session, no accumulated memory).}
\label{tab:agent-baseline}
\small
\begin{tabular}{l r r}
\toprule
\textbf{Session} & \textbf{Mean score / max} & \textbf{Scoring rate} \\
\midrule
Apartment-rental, Session 1 & 4.30 / 9 & 47.8\% \\
Apartment-rental, Session 2 & 2.40 / 5 & 48.0\% \\
Car-rental, Session 1 & 5.75 / 11 & 52.3\% \\
Car-rental, Session 2 & 2.50 / 5 & 50.0\% \\
\bottomrule
\end{tabular}
\end{table}

\subsection{Condition Performance Relative to the Baseline}
\label{app:condition-vs-baseline}

The agent-only runs also provide a reference point for the three conditions. Averaged over the four sessions, the agent alone scored about 50\% of the maximum with no memory and no developer involvement. Participant outputs were scored with the same evaluator and judge, so the overall scores in Table~\ref{tab:token-usage} can be read against this reference. Auto participants scored 47.7\%, slightly below the agent-only runs; Static participants scored 51.1\%, on par with them; and MemoSync participants scored 59.8\%, about ten percentage points above.

Auto participants could iterate over multiple turns, yet automatic memory brought no net gain. The memory injected into each session consisted of high-level summaries produced by the extraction pipeline, and the storage cap replaced accumulated detail as new items arrived. Participants could neither see nor edit what was injected, so stale decisions from earlier sessions could be carried forward uncorrected. Static preserved the full detail that developers and the agent had written into the Markdown file, but offered no mechanism for deciding which of it applied to the current turn, and its benefit stayed close to the agent-only level. MemoSync's advantage came from both sides of this trade-off: retained memory kept its detail because developers confirmed the content and scope, and each turn injected only the items selected for the current task. The agent-only runs were single-shot and involved no developer, so this comparison is descriptive and we did not test it statistically.

\subsection{Token Usage}
\label{app:tokens}

Table~\ref{tab:token-usage} also reports the coding agent's token usage, summed over the four sessions and averaged per participant within each condition. The counts cover the main agent session and every session it spawned. For MemoSync, they therefore include the forked sessions that perform candidate extraction, transfer encoding, memory-change checks, and post-turn auditing (Appendix~\ref{app:implementation}). Cache reads dominate in all three conditions because spawned sessions share the main session's context prefix. Despite these additional operations, MemoSync participants used the fewest tokens: 30\% fewer than Auto and 19\% fewer than Static in total, with 53\% fewer input tokens than Auto and the lowest output token count as well. MemoSync thus achieved the highest task performance at the lowest token cost.

\begin{table}[htbp]
  \centering
  \small
  \caption{Task performance and coding-agent token usage in the controlled study. Performance is the overall task score (\%) with $SD$ in parentheses. Token counts are averaged per participant over the four study sessions and include the main agent and agent-spawned subagents. Total is the sum of the three token columns. The DeepSeek API does not return cache-write token counts, so they are not included.}
  \label{tab:token-usage}
  \begin{tabular}{l r rrrr}
    \toprule
    & & \multicolumn{4}{c}{\textbf{Tokens (mean per participant)}} \\
    \cmidrule(lr){3-6}
    \textbf{Condition} & \textbf{Performance (\%)} ($SD$) & \textbf{Input} & \textbf{Output} & \textbf{Cache Read} & \textbf{Total} \\
    \midrule
    Auto     & 47.7 (9.3)  & 658{,}057 & 271{,}874 & 23{,}046{,}921 & 23{,}976{,}852 \\
    Static   & 51.1 (14.3) & 515{,}525 & 246{,}507 & 20{,}096{,}497 & 20{,}858{,}530 \\
    MemoSync & 59.8 (14.2) & 306{,}370 & 229{,}767 & 16{,}314{,}060 & 16{,}850{,}196 \\
    \bottomrule
  \end{tabular}
\end{table}

\section{Task Specifications and Evaluation Rubrics}
\label{app:tasks}
\setcounter{table}{0}
\setcounter{figure}{0}

\subsection{Task Specifications}
\label{app:task-specs}

The two study projects are tasks 000038 (apartment rental) and 000098 (car rental) from FullStack-Bench~\cite{flu2026fullstack}.\footnote{\url{https://github.com/mnluzimu/FullStack-Bench/tree/9e4cbb593e803309275cc642c5530f2092d7832b}} Each benchmark task provides a natural-language instruction and three sets of official test items: UI test tasks, each consisting of a user action and an expected result; backend API test cases; and data structures that the database must contain. We did not rewrite the tasks. We split each official instruction by functionality into two session briefs. In the apartment-rental project, Session~1 covers apartment browsing, search, and filtering, and Session~2 adds booking and booking records. In the car-rental project, Session~1 covers car browsing and booking, and Session~2 adds payment and order management. The theme-color requirement of each task was placed in the Session~1 brief, and each Session~2 brief asked participants to extend the existing implementation while preserving its functionality and visual design.

The following are the verbatim task briefs that participants read at the start of each session. Task order was counterbalanced across participants; within each project, Session~1 always preceded Session~2.

\begin{tcbraster}[raster columns=2, raster column skip=8pt, raster row skip=8pt, raster equal height=rows]
\begin{taskbox}{aptFrame}{aptFill}{Apartment-rental, Session 1}
Please implement the browsing and discovery features of a short-term apartment rental website. Users should be able to browse different types of apartments, search for apartments, and filter apartments based on their criteria. Use linen as the page background color and maroon as the accent color for components. These are the company's theme colors.
\end{taskbox}
\begin{taskbox}{aptFrame}{aptFill}{Apartment-rental, Session 2}
Continue working on the same short-term apartment rental website. Add the ability for users to book selected apartments and view their booking records. Integrate these features into the existing website while preserving its current functionality and visual design.
\end{taskbox}
\begin{taskbox}{carFrame}{carFill}{Car-rental, Session 1}
Please implement the core browsing and booking features of a comprehensive car rental platform with a user-friendly interface. Users should be able to browse different types of cars, select rental dates and times, book selected cars, and view their booking records. Set the page background color to papaya whip and use dark orange for components. These are the company's theme colors.
\end{taskbox}
\begin{taskbox}{carFrame}{carFill}{Car-rental, Session 2}
Continue working on the same car rental platform. Add support for multiple payment methods and order management. Integrate these features into the existing platform while preserving its current functionality and visual design.
\end{taskbox}
\end{tcbraster}

\begin{figure}[htbp]
\centering
\includegraphics[width=0.48\textwidth]{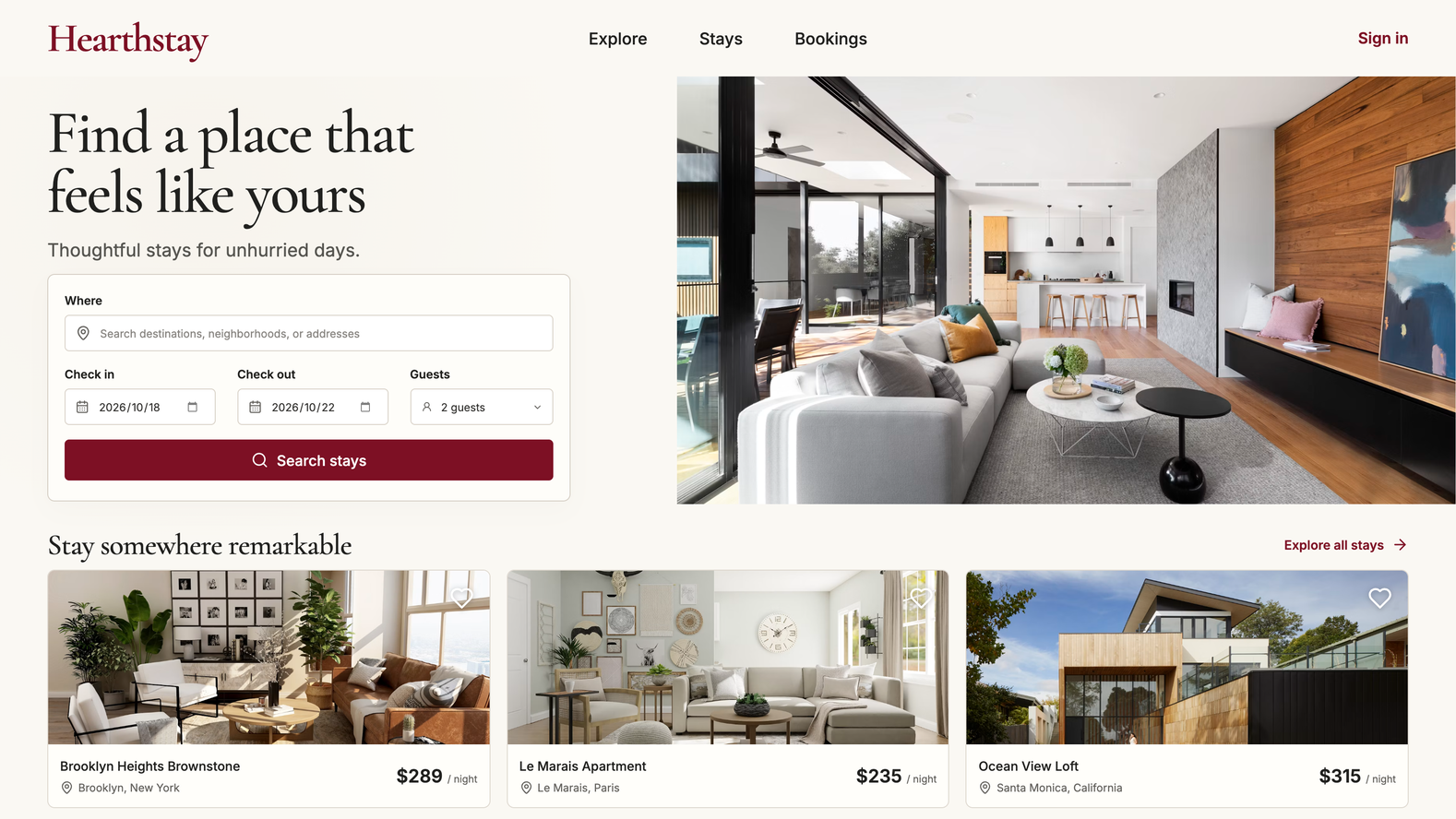}
\includegraphics[width=0.48\textwidth]{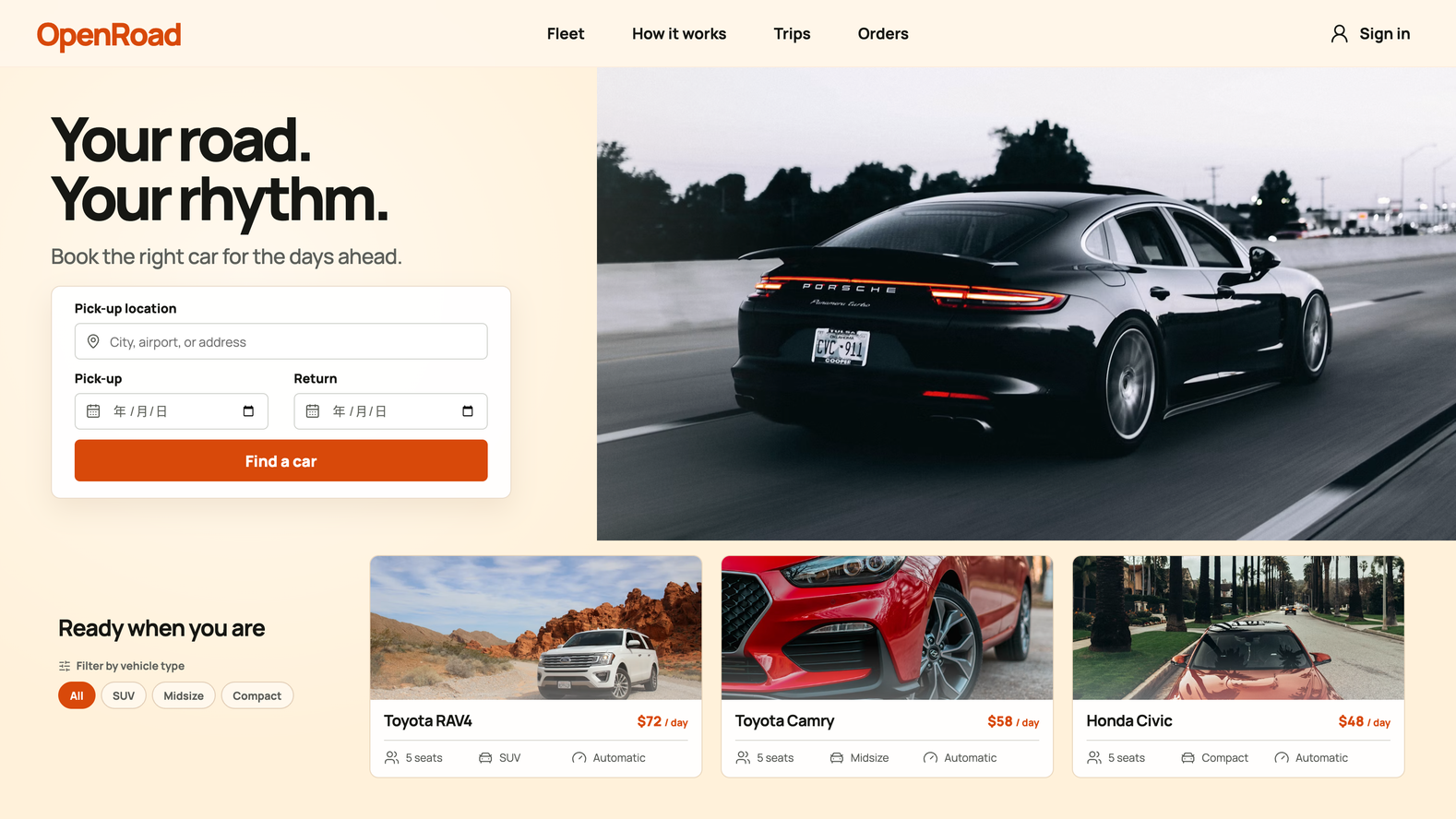}
\caption{Reference outputs generated by the coding agent (Claude Code with DeepSeek V4 Flash) given the complete task descriptions. \textbf{(a)}~The apartment-rental project (Hearthstay): location search, date selection, and property listings. \textbf{(b)}~The car-rental project (OpenRoad): pick-up search, date selection, and vehicle listings filtered by type. The two projects share a similar application structure but differ in business logic and data models.}
\label{fig:task-projects}
\Description{Two website screenshots. (a) Hearthstay: a header with Explore, Stays, Bookings, and Sign in; the headline Find a place that feels like yours; a search form with Where, Check in, Check out, and Guests fields and a maroon Search stays button; a photo of a living room; and three listing cards (Brooklyn Heights Brownstone at 289 dollars per night, Le Marais Apartment at 235, Ocean View Loft at 315), each with a photo and location. The page uses a linen background with maroon accents. (b) OpenRoad: a header with Fleet, How it works, Trips, Orders, and Sign in; the headline Your road. Your rhythm.; a form with pick-up location, pick-up and return dates and a dark orange Find a car button; a photo of a car on a road; filter chips All, SUV, Midsize, and Compact; and three car cards (Toyota RAV4 at 72 dollars per day, Toyota Camry at 58, Honda Civic at 48) listing seats, type, and transmission. The page uses a papaya whip background with dark orange accents.}
\end{figure}

\subsection{Evaluation Rubric}
\label{app:rubric}

The rubric items are the official test items of the two benchmark tasks: 14 items for the apartment-rental task (6 UI, 5 backend, 3 database) and 16 items for the car-rental task (5 UI, 7 backend, 4 database). We did not add or rewrite any item. Each item was assigned to the session in which its functionality is first requested. Each rubric item is scored as 1 (YES), 0.5 (PARTIAL, frontend items only), or 0 (NO). The four sessions have maximum scores of 9, 5, 11, and 5 points.

Table~\ref{tab:rubric} lists the rubric items for each session, grouped by Frontend/UI, Backend, and Database.

\begin{table}[htbp]
\centering
\small
\setlength{\tabcolsep}{5pt}
\caption{Evaluation rubric for the four sessions (9, 5, 11, and 5 points), grouped by Frontend/UI, Backend, and Database.}
\label{tab:rubric}
\begin{tabularx}{\textwidth}{@{}l c X@{}}
\toprule
\textbf{Category} & \textbf{\#} & \textbf{Criterion} \\
\midrule
\multicolumn{3}{@{}l}{\textbf{Apartment-rental, Session 1 (9 points)}} \\
\addlinespace[1pt]
Frontend/UI & 1 & Search apartments by location keyword and display results. \\
 & 2 & Filter to one-bedroom apartments. \\
 & 3 & Navigate back to the fully loaded homepage. \\
 & 4 & Use linen background and maroon highlights. \\
\addlinespace[2pt]
Backend & 1 & Search API returns matching apartments. \\
 & 2 & Filter API handles price, bedrooms, and amenities. \\
 & 3 & Details API returns apartment details. \\
\addlinespace[2pt]
Database & 1 & Apartment details including location and price. \\
 & 2 & User account information. \\
\midrule
\multicolumn{3}{@{}l}{\textbf{Apartment-rental, Session 2 (5 points)}} \\
\addlinespace[1pt]
Frontend/UI & 1 & Book an apartment and show confirmation or unavailability. \\
 & 2 & Show booking records. \\
\addlinespace[2pt]
Backend & 1 & Booking API creates a confirmed record. \\
 & 2 & Records API returns user bookings. \\
\addlinespace[2pt]
Database & 1 & Booking records linking users, apartments, and dates. \\
\midrule
\multicolumn{3}{@{}l}{\textbf{Car-rental, Session 1 (11 points)}} \\
\addlinespace[1pt]
Frontend/UI & 1 & Browse cars with type and daily price. \\
 & 2 & Select car and dates with availability. \\
 & 3 & Use papaya whip background and dark orange highlights. \\
\addlinespace[2pt]
Backend & 1 & Browse API returns cars. \\
 & 2 & Search/filter API handles type, dates, and availability. \\
 & 3 & Booking API confirms and stores. \\
 & 4 & Records API returns user bookings. \\
 & 5 & Invalid date ranges return an error. \\
\addlinespace[2pt]
Database & 1 & User accounts. \\
 & 2 & Car inventory and availability. \\
 & 3 & Booking records. \\
\midrule
\multicolumn{3}{@{}l}{\textbf{Car-rental, Session 2 (5 points)}} \\
\addlinespace[1pt]
Frontend/UI & 1 & Show booking confirmation with payment/edit options. \\
 & 2 & Complete booking with payment and unique reference. \\
\addlinespace[2pt]
Backend & 1 & Process and record payment. \\
 & 2 & Retrieve, update, or cancel orders. \\
\addlinespace[2pt]
Database & 1 & Payment transaction details. \\
\bottomrule
\end{tabularx}
\end{table}

\subsection{Evaluation Pipeline and Session Scoring}
\label{app:eval-pipeline}

We ran the evaluation implementation released with the benchmark without modification, replacing only the evaluator model with GPT-5.6 Sol (extra-high reasoning) for both text and vision judgments. For each frozen workspace, the pipeline starts a PostgreSQL container and the application's frontend and backend services. An information-gathering agent first reads the codebase to identify API routes, request formats, and authentication requirements. For each backend test case, a testing agent then sends requests to the running backend and reports YES or NO against the expected result. For each UI test task, a browser agent (WebVoyager) operates the running frontend for up to 15 interactions, with page screenshots passed to the evaluator model, and reports YES, PARTIAL, or NO. Finally, the pipeline dumps the database schema with up to five sample rows per table, and the evaluator model judges whether the schema covers each required data structure, reporting YES or NO. Scores are YES = 1, PARTIAL = 0.5 (UI items only), and NO = 0.

Session scoring followed the same protocol for participants and for the agent-only runs. At the end of Session~1, the workspace was frozen and evaluated on the Session~1 items only. Session~2 continued in the same workspace. At the end of Session~2, the workspace was frozen again and the full official evaluation was run over the whole project, that is, over all items of the task. The Session~2 score counted only the Session~2 items. Session~1 items were excluded from the Session~2 score regardless of their outcome in the second evaluation, and the Session~1 score remained the one obtained from the Session~1 snapshot, unaffected by later fixes or regressions. Each session score therefore reflects only the functionality requested in that session, while Session~2 functionality that depends on Session~1 was tested in the complete project.

\section{Implementation Details}
\label{app:implementation}
\setcounter{table}{0}
\setcounter{figure}{0}

Section~\ref{s:implementation} describes the orchestration in Figure~\ref{fig:implementation}. This appendix adds the implementation details behind it.

\subsection{Technology Stack}

The frontend uses React 19 and TypeScript 5.8, and the backend runs on Bun (v1.3.5+). The main agent session is created through the Claude Agent SDK's \texttt{query()} interface, which accepts an asynchronous prompt queue, session fork handles, system prompts, MCP server configurations, and tool permissions. For the Codex backend, memory tools are converted into Codex dynamic tools, and the Memory Block is injected into \texttt{developer\_instructions} at the start of each turn. Both backends support session forking.

\subsection{Forked Branches and Speculative Precomputation}

Beyond prompt-cache reuse, running memory operations as forked branches has two further effects. A branch sees the full project context accumulated in the main session, and it keeps the coding agent's tools, so it can read files or run commands to verify a proposal against the actual codebase instead of reasoning only from what is already in context.

The three branches in Figure~\ref{fig:implementation}a start at the same time, and downstream branches do not wait for upstream reviews. While the developer reviews candidates, the transfer and memory-change branches have already completed their analyses. Each review decision reaches the downstream branches as a short follow-up message in their own conversations. Because a branch already holds its analysis in context, it only needs to output what changed, so each review step is followed by one brief model call rather than a fresh analysis.

\subsection{Working-Memory Selection and Context Isolation}

After the approved changes are persisted, a further branch selects the working memory for the current turn from the updated store and produces the expected-use statement for each selected item. All memory reasoning thus happens in branches. The main session grows only by the user prompt, the Memory Block, and the agent's reply, so memory co-management adds no reasoning transcripts to the main context.

\subsection{Citations and Audit}

Memory citations are not tool calls. The system prompt instructs the coding agent to cite a memory item by writing its identifier in square brackets, such as \texttt{[M-07]}, whenever it acts on that item. The frontend parses the streamed reply for this fixed pattern and renders an inline citation only when the identifier matches an item in the store. When the developer interrupts a run, the corrective prompt continues in the same main session, with any item marked \textit{Enforce this run} attached.

The audit branch forked after the turn serves context isolation only: it runs after execution, so it gains nothing in latency, but its output stays out of the main session.

\section{Deployment Study Details}
\label{app:deployment}
\setcounter{table}{0}
\setcounter{figure}{0}

\subsection{Participant Demographics}
Table~\ref{tab:deployment_participants} lists the nine participants in the deployment study, with their programming experience, agent use, and project context.
\begin{table}[htbp]
\centering
\small
\setlength{\tabcolsep}{5pt}
\renewcommand{\arraystretch}{1.2}
\caption{Participants in our deployment study ($N=9$). Coding gives years of programming experience, and Agent Use gives how long the participant has used AI coding agents. Languages lists each participant's main programming languages, and Project Context gives the domain of their recent agent-assisted projects.}
\label{tab:deployment_participants}
\begin{tabularx}{\textwidth}{@{} 
l 
>{\raggedright\arraybackslash}p{2.2cm} 
c 
c 
c 
>{\hsize=1.05\hsize\raggedright\arraybackslash}X 
>{\hsize=0.95\hsize\raggedright\arraybackslash}X 
@{}}
\toprule
P\# & Role & Age & Coding & Agent Use & Languages & Project Context \\
\midrule
P1 & PhD Student       & 25 & 6--10  & $>$1 year & Python, JS, C++          & Webapp, Hardware \\
P2 & PhD Student    & 26 & 6--10 & 3--6 months    & Python, C++, C\#, JS, TS, Java       & Webapp, Data Analytics \\

P3 & LLM Engineer      & 30 & 6--10 & $>$1 year    & Python, C++, CUDA       & Model Serving \\
P4 & Master's Student  & 24 & 3--5  & 3--6 months  & C\#, Unity, Python      & VR/AR Applications \\
P5 & Software Engineer & 35 & $>$10 & $>$1 year    & Python, Go, TS          & Cloud Platform \\
P6 & PhD Student       & 26 & 6--10 & 6--12 months & Python, C/C++           & Robotics \\
P7 & Agent Engineer    & 29 & 6--10 & $>$1 year    & Python, TS, SQL         & Enterprise Workflow \\
P8 & Master's Student  & 25 & 3--5  & 6--12 months & Python, Java, JS        & Data Analytics \\
P9 & ML Engineer       & 32 & $>$10 & $>$1 year    & Python, C++, Shell      & Recommendation Systems \\
\bottomrule
\end{tabularx}
\end{table}

\subsection{Log Processing and Analysis}
\label{app:deployment-analysis}

System logs recorded participants' monitoring and control actions, agent execution turns, and switches between sessions and projects. To produce Figure~\ref{fig:deployment-activity}, we ordered each participant's events by time and placed each event at its index divided by the participant's total event count, so that the horizontal axis reflects relative position within the week rather than wall-clock time. Interview transcripts were analyzed as in the formative study (Appendix~\ref{app:analysis}).

\end{document}